%% file: main.tex
\documentclass[letterpaper,12pt]{article}
\usepackage[utf8]{inputenc}
\usepackage[left=1in, top=1in, right=1in, bottom=1in]{geometry}
\usepackage{graphicx}
\usepackage{subcaption}
\usepackage{soul}
\usepackage{setspace}
\usepackage{amsthm, amsfonts, amsmath, amssymb}
\usepackage{xcolor}
\usepackage{bm, bbm}
\usepackage{mathtools}
\usepackage[ruled, linesnumbered]{algorithm2e}
\usepackage{booktabs}
\usepackage{pgfplots}
\pgfplotsset{compat=1.18}
\usepackage[table]{xcolor}

\usepackage[
  style=authoryear,
  natbib=true,
  uniquename=false,   % don't add initials for disambiguation
  uniquelist=false    % don't expand author lists for disambiguation
]{biblatex}
\definecolor{prune}{rgb}{0.44, 0.11, 0.11}
\usepackage[colorlinks=true, citecolor=blue, linkcolor=prune, urlcolor=prune]{hyperref}
\usepackage{times}
\usepackage{titlesec}

\titleformat{\section} 
  {\normalfont\normalsize\scshape\centering} % 12pt, small caps, centered
  {\thesection}{1em}{}

\titleformat{\subsection}
  {\normalfont\normalsize\scshape\centering} % 12pt, small caps, centered
  {\thesubsection}{1em}{}

\titleformat{\subsubsection}
  {\normalfont\normalsize\scshape\centering} % 12pt, small caps, centered
  {\thesubsubsection}{1em}{}

\input{notations}

\title{Learning to Design Information\thanks{We are especially grateful to Bart Lipman for his detailed comments on this paper. We also thank Gerdus Benadè, Modibo Camara, Yiling Chen, Krishna Dasaratha, Maryam Farboodi, Hu Fu, Kevin He, Nicole Immorlica, Xudong Li, Yingkai Li, Chiara Margaria, Jawwad Noor, Juan Ortner, Roi Orzach, Pengyu Qian, Eitan Sapiro-Gheiler, David Simchi-Levi, Wei Tang, Zhihao Tang, Haifeng Xu, Kai Hao Yang, and Jiawei Zhang for valuable discussion. This paper also benefits from audiences at EC'25 Workshop: Information Economics and Large Language Models, the World Congress of the Econometric Society, the INFORMS Annual Meeting, Conference on Information Systems and Technology, and Stony Brook Game Theory Conference.
}
}

\author{
Ce Li\thanks{Department of Economics, Boston University, email: \texttt{celi@bu.edu}}
\and
Tao Lin\thanks{Microsoft Research, New England and the School of Engineering and Applied Sciences, Harvard University, email: \texttt{tlin@g.harvard.edu}}
}
\date{November 23, 2025}
\begin{document}

\maketitle 
\onehalfspacing

%{\centering\textbf{Job Market Paper} \par}

{\centering Latest version available \href{https://www.dropbox.com/scl/fo/s0n81cwinsjrjj2sahayx/AFPFBgBL8-l2lbCVmpwX5sw?rlkey=x20hrreh818woxzcvptzsy1ms&st=9yd5tef6&dl=0}{here}.
\par}
\bigskip

\begin{abstract}
%Recent advances in large language models call for the information design under unknown beliefs. 
%Information designers, such as large language models and online platforms, often do not have access to the subjective beliefs of their receivers or users. We design learning algorithms so that the designer can learn the receiver's belief through repeated interactions. Our learning algorithms are further robust to the receiver’s strategic manipulation on the learning process of the designer. We study regrets relative to two benchmarks, respectively, to measure the performance of the algorithms. The static benchmark is $T$ times the single-period optimality for the designer under the known belief. The dynamic benchmark, which is stronger, characterizes the global optimality for the designer under the known belief. Our learning algorithms make the designer achieve no regret against both benchmarks at fast speeds of $O(\log^2 T)$. 

Information designers, such as large language models and online platforms, often do not know the subjective beliefs of their receivers or users. We construct learning algorithms enabling the designer to learn the belief of a strategic receiver through repeated interactions. Our learning algorithms are robust to the receiver’s strategic manipulation of the learning process of the designer. We study regret relative to two benchmarks to measure the performance of the algorithms. The static benchmark is $T$ times the single-period optimum for the designer under a known belief. The dynamic benchmark, which is stronger, characterizes global dynamic optimality for the designer under a known belief. Our learning algorithms allow the designer to achieve no regret against both benchmarks at fast convergence speeds of $O(\frac{\log^2 T}{T})$, significantly faster than other approaches in the literature.

\end{abstract}

\newpage 

\tableofcontents              % <-- generates ToC with page numbers
\clearpage 
\onehalfspacing

\section{Introduction}

An information designer (the designer; she) sends signals about an unknown state to steer the action of a receiver (he) in her favor. Designing the optimal signaling strategy requires the designer to know the receiver’s prior belief over states. Canonical economic models (e.g., \cite{kamenica_bayesian_2011}) assume the parties share a common prior. In many applications, however, the designer has limited knowledge of the receiver’s belief. \textcolor{black}{We study how she can learn that belief through repeated interactions and design approximately optimal signaling strategies when the receiver acts strategically.}
%In this paper, we design learning algorithms for the designer such that the designer learns the belief of a {strategic} receiver through repeated interactions and designs approximately optimal signaling schemes.

For example, consider a large language model (LLM) API platform (the designer) that provides a business user (the receiver) with access to LLMs through Application Programming Interfaces (APIs). The LLM API platform charges the user based on the amount of tokens consumed to handle the user’s request.
After the user submits a request, the platform has an internal evaluation about how likely the request will be successfully handled, for example, high confidence and low confidence (the state). Based on that internal assessment of the success probability, the platform then decides how to signal to the user, e.g., recommending the user to proceed for more details or admitting its inability to provide a reliable solution. 
%The request may be solvable or unsolvable (the state), which can only be observed by the platform. After the user submits a request, the platform observes its condition and decides how to signal to the user, e.g., recommending the user to proceed for more details or admitting its inability to solve the request. 
Upon receiving the signal, the user updates his belief about the success probability of his request, and then takes an action to maximize his payoff, such as obeying the platform's recommendation to proceed or stopping the interaction. 
The platform always prefers that the user proceed to gain payment from the user for the token consumption, regardless of the state. On the user's side, his payoff is the highest when he pays for tokens that truly solve his request and the lowest when he pays for tokens that generate unwanted content, such as hallucinations.

%Regardless of the state, the LLM API platform always prefers that the receiver choose to proceed to
The user has a \textit{subjective} prior belief about the states, but the platform may not know that belief. As a result, the limited knowledge of the user's belief fundamentally \textit{constrains} the platform from designing the optimal signaling scheme. As implied by our previous work \citep{li_information_2025}, the platform may learn the user's belief
\textcolor{black}{in the case where the user's belief equals the true distribution}
%, though limited to an objective distribution, 
through the user's actions. However, the user may choose deceptive actions in some periods on purpose %after receiving the signal from the LLM API platform 
to \textit{manipulate} the platform's learning of his prior belief and further the platform's design of the signaling scheme in favor of the user himself.

%Motivated by such scenarios, we study an information-design problem in which the receiver’s belief is unknown to the designer and the receiver is strategic. Standard approaches to unknown beliefs endow the designer with a belief over the receiver’s prior \citep{alonso_bayesian_2016, kolotilin_persuasion_2017}, but computing the resulting optimal scheme is generally $\mathsf{NP}$-hard \citep{hossain_multi-sender_2024}. Robust information design instead optimizes against the worst-case prior in a candidate set \citep{hu_robust_2021, Piotr_2022}, which can be unduly conservative, yielding arbitrarily poor utility when the realized prior is not the worst case. Closer to our perspective, \citet{li_information_2025} study fast learning of beliefs from sequences of short-lived receivers; yet once the receiver becomes long-lived, he may purposely take myopically suboptimal actions to mislead the platform’s learning.

Motivated by such scenarios, we study an information design problem where the subjective prior belief of the receiver is unknown to the designer and the receiver is strategic.

\textcolor{black}{Standard economic models}, when tackling the problem of an unknown belief, would assume that the designer has a \textcolor{black}{distribution} over possible prior beliefs of the receiver (e.g., \cite{alonso_bayesian_2016, kolotilin_persuasion_2017}). {However, such an approach suffers from several concerns, such as computational hardness. A detailed discussion of the drawbacks of such an approach is presented in the literature review in detail.
%However, such an approach suffers from several concerns. %For one thing, it is computationally infeasible (NP-hard) to directly compute the optimal signaling scheme for a belief over beliefs \citep{hossain_multi-sender_2024}. For another, a line of robustness works considers a designer optimizing her signaling scheme for the worst case over a set of candidate prior beliefs (e.g., \cite{hu_robust_2021, Piotr_2022, ball_2025}). But that may result in a signaling scheme that makes the designer's payoff arbitrarily bad when the receiver's prior belief is not the worst-case considered. Standing out of the standard approach, \citet{li_information_2025} study learning the unknown prior belief 
%of a sequence of short-lived receivers at fast speeds 
%from the receivers' actions for a sequence of short-lived receivers who share the same prior belief. However, a receiver will be incentivized to be \textit{strategic} once he becomes long-lived: while interacting with the designer multiple times, the receiver may take suboptimal actions on purpose to convey \textit{deceptive} information about his belief, manipulating the designer’s learning in a direction favorable to him but unfavorable to the designer.
%interacts with the platform multiple times, thus becoming long-lived, : he may take suboptimal actions to convey \textit{deceptive} information about his belief, manipulating the platform’s learning in a direction favorable to him but unfavorable to the platform.

Are there near-optimal signaling strategies for the designer when she does not know the subjective prior belief of a \textit{strategic} receiver?
Our main contribution is to construct \textit{approximately optimal} signaling schemes which enable the designer to learn that subjective prior from repeated interactions with a strategic receiver. Moreover, our designer achieves such approximate optimality at efficient speeds. In addition, the designer's learning is robust to the receiver's strategic manipulation. Finally, such fast learning speeds are achieved against both a static benchmark and a dynamic benchmark.

Specifically, we study a \emph{learning} model where an information designer (e.g., the LLM platform) learns to design signaling schemes over $T$ periods based on historical interactions with a long-lived receiver (e.g., a business user). The designer does not know the belief of the receiver. In each period, the designer chooses a signaling scheme, which maps a state i.i.d.~sampled from a true distribution over the states to a randomized signal. Upon receiving the signal, the receiver updates his belief about the state and takes an action. The payoffs of the two players are then realized. We design learning algorithms for the designer to improve the signaling schemes over time, such that (1) the designer's time-averaged payoff converges to \textit{{optimum benchmarks}} and that (2) \textcolor{black}{the convergence is \textit{fast}}, while (3) being robust to \textit{strategic manipulation} by the receiver.

The rate at which the designer's payoff converges to the optimal objective depends on the learning algorithm she employs.
% \st{In particular, we care not only how \textit{close} the designer’s payoff gets to the optimal payoff she could have achieved with full knowledge of the prior, but also how \textit{quickly} such a convergence occurs.}
We measure the performance of the learning algorithm by a general notion of \emph{regret}:
the difference between the designer’s time-cumulative actual payoffs and an optimality benchmark, based on what the designer could achieve if she knew the receiver's belief.
We first characterize different regret benchmarks for the problem of an unknown prior belief in a learning model. The \textit{static} benchmark is $T$ times the single-period optimum for the designer had the designer known the receiver's belief. The \textit{static regret} is defined \textcolor{black}{relative to} the static benchmark. This allows us to relate our results to the results of the canonical information design models in \citet{kamenica_bayesian_2011}.

In addition, we consider a \textit{dynamic} benchmark that measures the designer's global dynamic optimality under a known belief, which is stronger than the static benchmark.  The \textit{dynamic regret} is then defined \textcolor{black}{relative to} the dynamic benchmark. A learning algorithm is called \textit{no-regret} if its regret grows sublinearly with time (alternatively, its time-averaged regret converges to zero). The rate at which the time-averaged regret declines reflects the learning algorithm’s efficiency. We aim to design learning algorithms that not only achieve no regret against the two benchmarks, but also \textcolor{black}{yield fast convergence to optimality under the unknown belief while being robust to the receiver's strategic manipulation.}

%We build an online learning model, in which an information designer uses an online learning algorithm to repeatedly interact with a strategic receiver through persuasion while learning the receiver's belief. In each period, the learning algorithm determines a signaling scheme, which is a mapping from states of the world to the distribution over signals. The information designer then announces that signaling scheme. A state is drawn i.i.d. from the true distribution of the states. Then, a signal is realized according to the state and the signaling scheme and is sent to the receiver. After receiving the signal, the receiver takes an action based on his rationale, which may or may not be optimal for his prior in that period. Both players' payoffs are realized based on the receiver's action and the realized state. The designer's learning algorithm is then able to elicit useful information about the receiver's belief from the receiver's action.

% \vspace{0.5em} 

Our first main result is an $O(\log^2 T)$ upper bound on the \textit{static regret}. Our second main result is also an $O(\log^2 T)$ regret upper bound on the \textit{dynamic regret}. We design two learning algorithms to ensure that, for any problem instance, the designer's regrets for $T$ periods, relative to the \textit{static benchmark} and \textit{the dynamic benchmark}, respectively, are both no more than a constant times $\log^2 T$ for all $T$. Both regret upper bounds imply convergence rates of $O(\frac{\log^2 T}{T})$ for the designer's regret to converge to zero, which are significantly faster than other approaches in the literature, as discussed later. 

Both of our two learning algorithms contain two phases. The first phase is the learning of the receiver's belief, which has the same learning logic under \textcolor{black}{the} two regret benchmarks. The learning of the belief is based on \textit{binary search}. The key idea is to learn the unknown belief from the receivers' actions. To see the intuition, consider two states and two actions, where the receiver prefers different actions in the two states. For the moment, suppose the receiver is myopic in the sense that he does not consider the future in choosing his action. Then, the designer can compute a cutoff belief such that the receiver is indifferent between the two actions.  
%when best responding. The designer has a guess about the receiver's prior belief by designing a signaling scheme such that the receiver is indifferent between the two actions under the posterior belief formed by the guessed prior belief.
The actual action chosen by the receiver then reveals whether the receiver's belief is below or above {the cutoff}. {By observing the receiver's action}, the designer gradually narrows down the range of possible values for the prior. That process does not stop until the designer obtains a sufficiently accurate estimate of the prior belief. Our formal analysis shows how to generalize that intuition to multiple states and multiple actions. 

\textcolor{black}{How can the designer learn the receiver's belief when the receiver is not myopic? The algorithm limits the strategic manipulation from the receiver by delaying the effect of the receiver's action on future signaling schemes. In particular, in each iteration, the algorithm records the receiver's action when the designated signal first occurs. It then waits a fixed number of additional periods before using that recorded action to update the signaling scheme, ignoring actions taken during the waiting periods. Thus, the recorded action can influence future information only after a delay. Since the receiver discounts future payoffs, a sufficiently long delay makes the potential gain from manipulation small relative to any substantial immediate payoff loss. The receiver may still deviate from a myopic best response when the immediate loss is small, but the resulting error in the belief estimate is small. By choosing the delay appropriately, the designer can therefore learn the receiver's belief to the desired accuracy despite the receiver's strategic manipulation.}

The second phase, where our learning algorithms for {the two benchmarks} differ, is \emph{robustification}. The signaling scheme that is optimal for the designer under the belief estimate is converted into another signaling scheme that is approximately optimal under the receiver's belief. The motivation is that even if two prior beliefs are very close, the signaling scheme optimal under one belief may {perform} arbitrarily badly under the other, leaving no guarantee on the designer's payoff. However, the robustification varies {across benchmarks}. Our first learning algorithm uses \textit{static} robustification, where the robustification process is performed on the signaling scheme optimal for the belief estimate \textit{once} to obtain a robustified signaling scheme for being repeatedly used for future periods. %only for the per-period signaling scheme to ensure the receiver's participation in each period.

% \vspace{0.5em}

%Our second main result is an $O(\log^2 T)$ regret upper bound on the \textit{dynamic} regret. We design the second learning algorithm to ensure that, for any problem instance, the designer's regret {relative to the dynamic benchmark} for $T$ periods is no more than a constant times $\log^2 T$ for all $T$ against the dynamic benchmark. 

%The first phase, the learning of the receiver's belief, follows the same logic as the first learning algorithm, while the second phase, robustification, differs. 

{Our second learning algorithm applies a \textit{dynamic robustification} to the designer's globally optimal strategy, where the signaling schemes in different periods are chosen so that the receiver at any period is still incentivized to participate in the future.} %Dynamic robustification keeps recommendations persuasive time-sensitively. 
The {challenge} is that the static robustification may backfire: the static robustification only increases the \textit{persuasiveness} of the signaling scheme, namely, increasing the {positive gap} of the expected payoff from the recommended action relative to that from any other actions. However, the static robustification may lower the receiver’s \textit{ex ante} (before a recommendation is sent) expected payoff from obeying the recommendations in the future. Thus, after an action is recommended by a statically robustified signaling scheme, the receiver’s incentive to obey that current recommendation becomes stronger, but his \textit{ex ante} payoff from obeying recommendations in \textit{future periods} may get worse, potentially violating the receiver's participation constraint. 

%\textcolor{red}{This is the first reference to a "new type of robustification."  As we discussed, it's not clear exactly how new this is.}
To overcome this, we
propose \textcolor{black}{\textit{payoff-improving robustification},} which weakly increases the receiver’s \textit{ex ante} payoff from being obedient. We then take the designer's optimal dynamic signaling strategy, given her belief estimate, and apply the payoff-improving robustification procedure to each period's signaling scheme in that dynamic strategy. We show that the resulting dynamic strategy is persuasive for the receiver and approximately optimal for the designer.  %leads to a strategy in the designer's optimal dynamic  signaling strategy. for the belief estimate, which finally leads to a signaling strategy dynamically persuasive for the receiver under his belief, as well as the approximate optimality for the designer.
%Our dynamic robustification instead plans across time: we make tiny but budgeted adjustments to today’s signaling scheme and make explicit promises about the future's gain from obeying the recommendation. Coordinating over the whole horizon, our dynamic robustification keeps the incentives of the designer and the receiver aligned each round while making the designer lose only a small amount of performance, which cannot be achieved by the static robustification.

% \vspace{0.5em}

The \textcolor{black}{efficacy}
%\textcolor{red}{ "Superiority" is a tricky word.  Superior to what?  Maybe something like "efficacy"?} 
of both of our learning algorithms is reflected by the logarithmic upper bounds on their regrets: $O (\log^2 T)$. That means that the designer's average payoff converges to the optimality benchmarks at a fast speed, in the order of $O(\frac{\log^2 T}{T})$. In contrast, an intuitive method is to use an \emph{empirical estimation}: \textcolor{black}{using the empirical frequency to estimate the state distribution}.
%counting the state revealed at the end of each period to obtain an empirical estimate of the state distribution. 
Empirical estimation does not work in our problem \textcolor{black}{as the receiver's belief is subjective and may not equal the true distribution over states.}
%, since the true distribution over the states is not the same as the receiver's subjective belief. 
Even if empirical estimation could be applied, it would suffer a regret of at least $O(\sqrt{T})$ due to sampling error, which is exponentially worse than the $O(\log^2 T)$ regret bound of our learning algorithms. 
Our algorithms circumvent the limitation of empirical estimation and achieve an exponential improvement by actively learning the receiver's belief from his actions.

\subsection{Related Work}

First, this paper contributes to the literature on robust information design under uncertainty, especially when the designer lacks knowledge of the receiver’s prior belief.
Classical models such as Bayesian persuasion \citep{kamenica_bayesian_2011, BergemannMorris2016} and cheap talk \citep{crawford_strategic_1982} typically assume a common prior shared by the designer and the receiver. \textcolor{black}{Different from the canonical models, \citet{alonso_bayesian_2016} study persuasion with heterogeneous priors. Recent work relaxes this by introducing uncertainty over the prior. One approach assumes the designer holds a belief over the unknown prior (e.g., \citet{alonso_bayesian_2016}), but computing the optimal signaling scheme is generally intractable \citep{hossain_multi-sender_2024}. Another approach, rooted in robust {mechanism} design, has the designer optimize against a worst-case prior in a set of possible priors. \citet{Kosterina2022} studies a designer who chooses a signaling scheme to maximize her payoff under the least favorable receiver prior in a given set.  \citet{hu_robust_2021, Piotr_2022} consider uncertainty about the receiver's private information or additional information sources. These papers typically study a single-shot game, limiting the designer’s ability to learn over time. {Of course, when the designer has the opportunity to learn, one expects she may achieve a much higher payoff.}
In contrast, we design learning algorithms that infer the prior, achieving near optimality quickly while avoiding computational hardness. 
}

%Recent work relaxes this by introducing uncertainty over the prior. \textcolor{red}{the Kolotilin et al paper you cite here is not one in which the designer has a belief about the unknown prior of the receiver.  They have a common prior in this paper.}
%One approach assumes the designer holds a belief over the unknown prior (e.g., \cite{kolotilin_persuasion_2017}), but computing the optimal signaling scheme is generally intractable \citep{hossain_multi-sender_2024}.
%Another approach, rooted in robust {mechanism} design, has the designer optimize against a worst-case prior in a given set (e.g., \cite{hu_robust_2021, Piotr_2022, Kosterina2022, Dworczak_Kolotilin_2024, ball_2025}).
%These papers typically study a single-shot game, limiting the designer’s ability to learn over time. {Of course, when the designer has the opportunity to learn, one expects she may achieve a much higher payoff.}
%In contrast, we design learning algorithms that infer the prior, achieving near optimality quickly while avoiding computational hardness. 

Departing from the standard approach for addressing the robustness, 
{\citet{li_information_2025} study how to \textit{learn} the belief of a sequence of \textit{short-lived} receivers from their actions.
However, if a receiver interacts with the platform repeatedly, thus becoming long-lived, then he will be incentivized to be strategic. Thus, he may take actions to convey deceptive information about his belief, manipulating the platform’s learning in a direction favorable to them but unfavorable to the platform.

%\textcolor{red}{The papers discussed at the bottom of this page, continuing to the top of the next, study issues with multiple receivers.  These paper all have common priors.  I'd suggest either not citing them or being more explicit that the issue they consider is about multiple receivers, not unknown priors.} There are also papers that address robustness from other perspectives. \citet{Mathevet2020, morris2024, Ziegler2020} study the designer’s strategy from an adversarial perspective in one-shot games. \citet{li_norman_2021} analyze sequential games from the adversarial perspective but maintain a common prior. In these papers, either a common prior is assumed or the model is one-shot, whereas our paper tackles both a learning process and an unknown prior. 
\textcolor{black}{Our work also relates to dynamic information provision. \citet{ian2023} consider a dynamic information design problem in which the beliefs of the designer and the receiver are different. However, their model uses persistent states while we consider i.i.d. states. More importantly, they consider both players to be equally patient, while we consider a patient designer and an impatient receiver. The different patience levels in our paper have crucial influence on their dynamic strategies and the designer's regret benchmarks. Besides, they assume more structure on the dynamic process and the players' payoffs, while our work keep the dynamic process and the payoff functions general without further structure.}

Our work also contributes to the literature on {online Bayesian persuasion}, which studies learning when some game parameters are unknown.
\citet{castiglioni_online_2020, castiglioni_multi-receiver_2021, feng_online_2022} consider unknown receiver utilities; \citet{zu_learning_2021, wu_learning_2022} focus on unknown priors; and \citet{bacchiocchi_markov_2024} allow both utilities and priors to be unknown.
{Previous} approaches to unknown priors rely on empirical estimation and incur $\Omega(\sqrt T)$ regret due to sampling error. {This option is not available here as the designer needs to learn the receiver's belief, not the true distribution.}
We bypass this by using best responses to infer the prior more efficiently, achieving $O(\log^2 T)$ regret.
\citet{Camara2020MechanismsFA} also study information design with learning agents but drop distributional assumptions and allow adversarial state sequences; their regret notion therefore differs from ours, which assumes i.i.d.\ states from an unknown objective prior.

Our work also connects to the line of works on misspecified learning %fudenburg2017,
\citep{ba2023robustmisspecifiedmodelsparadigm, Bohren2021, frick2023, ESPONDA2021105260, fudenburg2021, fudenburg2017} and on the long-run outcomes of misspecified Bayesian agents who cannot obtain the true distribution over states (e.g., \citet{TVERSKY1973207, rabin1999, Fudenberg2023, he2021}). Those works study belief convergence and actions in general learning environments. 
\textcolor{black}{One can view our model as studying a receiver whose belief is the limiting outcome of his misspecified learning.  We do not analyze the process by which this belief is formed,}
%However, we consider a receiver whose belief is already the convergence outcome of his misspecified learning. 
%{We neglect the process by which such a belief is formed, 
focusing instead on how the designer interacts with a receiver whose belief has already converged}. %In particular, \citet{he2021} also consider a strategic setting. But besides the aforementioned difference, \citet{he2021} show a society is vulnerable to agents' strategically beneficial misinferences about the states of the world, while our designer can achieve approximate optimality while being robust to the strategic receiver's manipulation, and the receiver's payoff is also ensured by the persuasiveness of our designer's signaling strategies.

%Finally, we study how quickly regret converges, connecting to work on convergence rates for information \citep{LIANG_MU_2020, frick2024monitoringrichdata}, beliefs \citep{iakovlev_liang_2025}, and attribute values \citep{liang2025}. Unlike those, our focus is regret, the performance measure for learning algorithms, whose convergence here is driven by our algorithm design rather than subjective beliefs or exogenous information sources.

\section{Models and Prelimaries}
\label{sec:model}

\subsection{Models}
\label{sec:prelim-BP}

%{\color{red} (change a bit!!!)}
%We consider a learning environment with repeated persuasion games between an information designer (designer; she) and a receiver (he).

\paragraph{\bf Single-period game.}
We {first} define a single-period game of information design (Bayesian persuasion) with subjective prior beliefs. 
There is a finite set of states of the world $\Omega$. The information designer and the receiver have subjective prior beliefs $\dprior, \rprior \in \Delta(\Omega)$ over the states, respectively, where $\Delta(\Omega)$ denotes the set of probability distributions over $\Omega$. {We think of $\dprior$ as the true distribution over the states and $\rprior$ as a subjective prior belief over the states, which may not be the same as the true distribution. The designer is better informed about the true distribution from access to sufficient state observation data, so the designer is assumed to know the true distribution.
The receiver's prior $\rprior$ can be different from the true distribution, which may be the convergence outcome of the learning process of a misspecified belief (e.g., \citet{fudenburg2017}).}
% We think of $\rprior$ as the convergence outcome of the receiver's misspecified learning (i.e., the receiver starts with some belief and his belief converges to a certain belief), and such a belief $\rprior$ is different from the true distribution (or, $\dprior$) due to the receiver's misspecified learning.

The receiver has a finite set of actions $A$. The designer and the receiver have payoff functions $u, v: A \times \Omega \rightarrow [0,1]$, respectively.
This single-period game is summarized by an \emph{instance} $\mathcal{I}=(\Omega, A, u, v, \dprior, \rprior)$.

At the beginning of the game, the designer announces a \emph{signaling scheme} $\pi: \Omega \rightarrow \Delta(S)$, which maps each state to a probability distribution over a finite set of signals $S$.
Let $\Pi = \Omega \rightarrow \Delta(S)$ be the set of signaling schemes.
The receiver believes that the state is drawn according to the distribution $\rprior$. Thus, after receiving a signal $s \sim \pi(\cdot | \omega)$ sampled according to state $\omega$ and the signaling scheme $\pi$, the receiver obtains his posterior belief $\frac{\rprior(\omega) \pi(s|\omega)}{\sum_{\omega'\in \Omega}\rprior(\omega')\pi(s|\omega')}, \forall \omega \in \Omega$, and takes an action in response.
Let $\rho : S \to \Delta(A)$ be a possibly randomized single-period strategy of the receiver. Under a pair of strategies $(\pi, \rho)$, the expected payoffs of the designer and the receiver in the single-period game are, respectively,
\begin{align} \label{eq:single-period-expected-utility-definition}
    & U(\pi, \rho; \dprior)  = \sum_{\omega \in \Omega} \dprior(\omega) \sum_{s\in S} \pi(s|\omega) \sum_{a\in A} \rho(a|s) u(a, \omega), \\
    & V(\pi, \rho; \rprior) = \sum_{\omega \in \Omega} \rprior(\omega) \sum_{s\in S} \pi(s|\omega) \sum_{a\in A} \rho(a|s) v(a, \omega).    
\end{align}
%For simplicity, we neglect $\rho$ until randomization is discussed.

%We say the receiver \emph{myopically best responds}\footnote{We use ``myopic best response'' to distinguish from the best response of a strategy receiver in the $T$-period game. } if he takes an action $a_s^*$ that is optimal for his posterior belief (breaking ties in favor of the designer):  
%\begin{align}
%    a^*_s(\pi, \rprior) ~ \in ~ \argmax_{a\in A} \sum_{\omega \in \Omega} \rprior(\omega) \pi(s|\omega) v(a, \omega).
%\end{align}

We say the receiver \emph{myopically best responds}\footnote{We use ``myopic best response'' to distinguish from the best response of a strategic receiver in the $T$-period game. } if he maximizes his single-period payoff by choosing a strategy $$\rho^* \in \argmax_{\rho: S\to\Delta(A)} V(\pi, \rho)$$ or, equivalently, an action $\rho^*(s) \in \argmax_{a\in A} \rprior(\omega) \pi(s|\omega) v(a, \omega)$ given each signal $s\in S$ based on his posterior belief. 
% 
% At the beginning of each period $t\in [T]$, the receiver arrives, and the information designer then picks the signaling scheme $\pi^{(t)}$ from the learning algorithm for Nature to use. The state in that period $\omega^{(t)}$ is i.i.d. sampled from the true distribution $\mu^*$. After that, Nature sends a signal $s^{(t)} \sim \pi^{(t)}(\cdot | \omega^{(t)})$ to the receiver. The receiver then takes an action $a^{(t)}$ based on $\rprior$, $\pi^{(t)}$, and $s^{(t)}$. Both players' payoffs $u,v$ are realized after the receiver takes an action. 
%
%
% Upon receiving signal $s \in S$ from a signaling scheme $\pi$, the receiver may or may not take the action optimal for his belief $\rprior$. We denote by $a_s^*$ the receiver's action optimal for his belief $\rprior$, i.e.,
% 
% \begin{align}
%     a^*_s (\rprior) ~ \in ~ \argmax_{a\in A} \sum_{\omega \in \Omega} \rprior(\omega) \pi(s|\omega) v(a, \omega).
% \end{align}
% 
With a myopically best-responding receiver, the optimal utility of the designer is
\begin{align}
    U_\BP^*(\mathcal I) := \max_{\pi \in \Pi} \max_{\rho \in \argmax_{S\to \Delta(A)} V(\pi, \rho; \rprior) } U(\pi, \rho; \dprior).  
\end{align}
We call $U_\BP^*(\mathcal I)$ the \emph{Bayesian persuasion benchmark}.

\paragraph{\bf $T$-period learning process.}
% The $T$-period learning process is as follows.
% The information designer has no access to the belief of the strategic receiver and is therefore unable to approach her optimality. Thus, the designer aims at learning the receiver's belief through repeated interactions with the receiver.

% Finding the optimal signaling scheme $\pi^*$, however, requires knowledge of the receiver's prior belief $\rprior$, which is often unavailable in practice.
We consider the $T$-period repetitions of the above single-period game with an instance $\mathcal I = (\Omega, A, u, v, \dprior, \rprior)$.  Importantly, the designer now does not know the receiver's prior belief $\rprior$ but she knows other parameters $u, v, \dprior$. 
In particular, the designer \emph{learns} to design signaling schemes
%{\color{blue}{Specifically, an \emph{instance} $\mathcal{I}$ for the $T$-period learning process is $\mathcal{I}=\{u, v, \rprior, \dprior\}$}}.\footnote{{\color{blue}{For a slight abuse of notation, only part of the instance $\{u, v, \rprior, \dprior\}$ is the input for the designer's learning algorithm $\Alg$.}}}
using an \textit{algorithm} $\Alg$ which, at every period $t$, maps the history $H^{(t-1)} \in \Pi^{(t-1)} \times S^{(t-1)} \times A^{(t-1)}$ 
%(consisting of past signaling schemes, realized signals, and the receiver's actions) 
to a signaling scheme $\pi^{(t)} \in \Pi$ for the current period.
% The receiver's strategy is a mapping $\phi$ that at every period $t$, maps the history to the current action.
In response to the learning algorithm $\Alg$, the receiver chooses a $T$-period strategy $\phi = (\phi^{(t)})_{t=1}^T$ where each $\phi^{(t)}$ maps the history $H^{(t-1)} \in \Pi^{(t-1)} \times S^{(t-1)} \times A^{(t-1)}$ and the current period's signaling scheme $\pi^{(t)} \in \Pi$ to a strategy $\rho^{(t)}: S \to \Delta(A)$ that specifies a randomized action for each possible signal. 
Then, the pair of strategies $(\Alg, \phi)$ will generate a stochastic sequence of $(\pi^{(t)}, \rho^{(t)})_{t=1}^T$ of the designer's signaling schemes and the receiver's single-period strategies.
% is a stochastic process generated by $\Alg$ and $\phi^*(\Alg, \rprior)$.
We assume that the receiver is \textit{impatient} in the sense that he has a discount factor $\gamma \in (0, 1)$, 
% a weakly decreasing discount factor sequence $(\gamma^t)_{t=1}^T$, e.g., the exponential sequence $\gamma^t = \gamma^t$ for some $0< \gamma < 1$.
so his total discounted expected payoff is 
% has a commonly known discount factor $\gamma \in (0, 1)$. 
% So, the receiver's total expected utility over the $T$ periods is
% \begin{equation}
%     \mathcal{V}(\Alg, \phi; \rprior) ~ = ~ \E_{\Alg, \phi}\bigg[ \sum_{t=1}^T \gamma^t  \sum_{\omega \in \Omega^{(t)}} \rprior(\omega^{(t)}) \cdot v(a^{(t)}, \omega^{(t)}) \bigg]. 
% \end{equation}
\begin{equation}
    \mathcal{V}_T(\Alg, \phi; \rprior) ~ = ~ \E_{(\pi^{(t)}, \rho^{(t)})_{t=1}^T \sim (\Alg, \phi)}\bigg[ \sum_{t=1}^T \gamma^t \cdot   V(\pi^{(t)}, \rho^{(t)}; \rprior) \bigg].  
\end{equation}
% {\color{blue}{The $T$-period learning process is summarized by an \emph{instance} $\mathcal{I}_T=(\Omega, A, u, v, \dprior, \rprior, (\gamma^t)_{t=1}^T, T)$.}}
The designer, on the other hand, is \textit{patient}, that is, having a discount factor of one. 
% We assume the designer to be patient, so she aims to maximize her
The designer aims to maximize her total expected payoff
% \begin{equation}
%     \mathcal{U}(\Alg, \phi; \dprior) ~ = ~ \E_{\Alg, \phi}\bigg[ \sum_{t=1}^T \sum_{\omega^{(t)} \in \Omega} \dprior(\omega^{(t)}) \cdot u(a^{(t)}, \omega^{(t)}) \bigg], 
% \end{equation}
\begin{equation}
    \mathcal{U}_T(\Alg, \phi; \dprior) ~ = ~ \E_{(\pi^{(t)}, \rho^{(t)})_{t=1}^T \sim (\Alg, \phi)}\bigg[ \sum_{t=1}^T  U(\pi^{(t)}, \rho^{(t)}; \dprior) \bigg]. 
\end{equation}

%{\bf Remark} 
% In particular??
We consider a \textit{strategic} (or long-lived) receiver who best responds to the designer's learning algorithm $\Alg$.
Formally, given the algorithm $\Alg$ and the belief $\rprior$, the receiver chooses a globally optimal strategy $$\phi^*(\Alg, \rprior) = \argmax_{\phi} \mathcal{V}_T(\Alg, \phi; \rprior).$$ Then, the designer obtains her total expected payoff $$\mathcal U_T(\Alg, \phi^*(\Alg, \rprior); \dprior).$$

\begin{remark}
%The assumption that the designer is more patient than the receiver is commonly seen in real-world problems. As a high-level intuition, the designer represents a {principal}, a platform, or a social planner, etc. Thus, the designer is long-lived {and} patient. The receiver may be an agent or a user, who can be long-lived but impatient. {We discuss how the results change if the designer is impatient in Section \ref{sec:lower-bound}.} 
Technically,
% %and well-motivated theoretically.
if the designer is impatient (\textcolor{black}{i.e., has} a discount factor less than one), then she will care \textcolor{black}{more} about the early periods \textcolor{black}{than later ones and may obtain a total payoff sublinear in $T$, thus trivially making her regret sublinear in $T$. 
%(to be formally defined below) trivially. 
What's even worse, the designer may have already started to neglect her continuing payoffs before finishing learning the receiver's belief.
If the receiver is patient (with discount factor $\gamma = 1$), then the designer cannot} achieve sublinear regret regardless of the learning algorithm, as we show in Section \ref{sec:lower-bound}. 
%If the receiver is patient, then their payoffs under the designer's dynamic benchmark will not oscillate over time, as shown in Section \ref{sec:characterization-dynamic-benchmark}.

% For one thing, the designer needs to be either patient or nearly patient with a discount factor close to one such that her no regret (to be defined below) is obtained non-trivially. Otherwise, the designer will neglect her utilities in the tail periods.
% For another, we will show in Section \ref{sec:lower-bound} that if the receiver is patient with a discount factor $\gamma = 1$, then it is impossible for the designer to achieve no regret in any learning algorithm. \st{[Regret is not defined yet.]} \ce{our introduction has introduced regret and no regret, so it should be ok to mention them simply here.} 
\end{remark}

%Against a best-responding strategic receiver, we aim to design learning algorithms for the designer to achieve sub-linear regret.

\paragraph{\bf Benchmarks and regrets.}

The \textit{regret} of the designer under instance $\mathcal{I} = \{\Omega, A, u, v, \dprior, \rprior\}$ is the difference between {a} benchmark $\mathcal U^*_T(\mathcal I)$ and the actual payoff $\mathcal{U}_T$ that the designer obtains by using the algorithm $\mathcal{G}$ over the $T$ periods. In particular, we will consider two benchmarks for defining regret.

The first benchmark is the \textbf{static benchmark} denoted by $U^*_{\mathrm{BP}}(\mathcal I)$, which is $T$ times the single-period Bayesian persuasion benchmark at a problem instance $\mathcal{I}$. 
\begin{align*}
    \mathcal{U}^{*}_{T}(\mathcal I) := T \cdot U^*_{\mathrm{BP}}(\mathcal I).
\end{align*}
The static benchmark is a natural benchmark to consider, especially for measuring how good the performance of our learning algorithm is compared to the Bayesian persuasion optimality in the canonical work \citep{kamenica_bayesian_2011}, in which the designer knows the receiver's prior $\rprior$. 
%The static benchmark is also natural for an unambitious designer (less ambitious than the designer introduced in the second benchmark)  who cares about her repeated optimality period by period.\footnote{There are decision-making rationales for such a less ambitious designer, for example, satisficing.}

We call the regret against the {static benchmark} \textbf{static regret}:  
\begin{equation}
        \Reg^{\mathrm{static}}(T; \Alg, \mathcal I) ~ = ~  \mathcal{U}^{*}_{T}(\mathcal I) ~ - ~ \mathcal{U}_T(\Alg, \phi^*(\Alg, \rprior); \dprior).
        %\Reg(T; \Alg, \mathcal I) ~ = ~ U^{*}_{T}(\mathcal I) ~ - ~ \mathcal{U}_T(\Alg, \phi^*(\Alg, \rprior); \dprior).
\end{equation}

The second benchmark{, the \textbf{dynamic benchmark},} is the maximal $T$-period payoff the designer could obtain had the designer known the receiver's belief $\rprior$ and the receiver best responded over the $T$ periods{. We denote this} by $U^{**}_{T}(\mathcal I)$ at a problem instance $\mathcal{I}$.
\begin{align*}
    \mathcal U^{**}_{T}(\mathcal I) := \max_{\text{$T$-period strategy $\sigma$}} \, \mathcal U_T(\sigma, \phi^*(\sigma, \rprior); \dprior),
\end{align*}
where $\sigma$ is a $T$-period strategy of the designer that designs signaling schemes based on historical information and the receiver's belief $\rprior$.\footnote{{We assume tie-breaking in favor of the designer if the receiver has multiple best responses. However, our results guarantee that obedience is the unique best response for the receiver.}} %The dynamic benchmark is especially suitable for an ambitious designer who focuses on her global optimality over the whole learning process.

We call the regret against the {dynamic benchmark} \textbf{dynamic regret}:  
\begin{equation}
    \Reg^{\mathrm{dynamic}}(T; \Alg, \mathcal I) ~ = ~ \mathcal U^{**}_{T}(\mathcal I) ~ - ~ \mathcal{U}_T(\Alg, \phi^*(\Alg, \rprior); \dprior). 
    %\Reg(T; \Alg, \mathcal I) ~ = ~ \mathcal U^{**}_{T}(\mathcal I) ~ - ~ \mathcal{U}_T(\Alg, \phi^*(\Alg, \rprior); \dprior).
\end{equation}
% \begin{itemize}
%     \item \textbf{Static benchmark:} $T$ times the single-period Bayesian persuasion benchmark $U^*_{\mathrm{BP}}(\mathcal I)$:
%     \begin{equation}
%         \Reg^{\mathrm{static}}(T; \Alg, \mathcal I) ~ = ~ T \cdot U^*_{\mathrm{BP}}(\mathcal I) ~ - ~ \mathcal{U}_T(\Alg, \phi^*(\Alg, \rprior); \dprior).
%     \end{equation}
%     \item \textbf{Dynamic benchmark:} The best utility obtainable by the designer if the designer knows the receiver's prior and the receiver best responds globally: 
%     \begin{equation}
%          \Reg^{\mathrm{dynamic}}(T; \Alg, \mathcal I) ~ = ~ \underbrace{\max_{\text{$T$-period strategy $\sigma$}} \, \mathcal U_T(\sigma, \phi^*(\sigma, \rprior); \dprior)}_{\text{denoted by $\mathcal U^{**}_{T}(\mathcal I)$}} ~ - ~ \mathcal{U}_T(\Alg, \phi^*(\Alg, \rprior); \dprior). 
%     \end{equation}
%     Here, $\sigma$ is a $T$-period strategy of the information designer that designs signaling schemes based on historical information and the receiver's prior $\rprior$. 
% \end{itemize}
%\textcolor{red}{you should comment here on the speed issue.  I think you want to highlight the surprise that you can achieve the dynamic benchmark as quickly as the static one.}
\textcolor{black}{Not surprisingly, as} we will show in Section \ref{sec:characterization-dynamic-benchmark}, the dynamic benchmark $\mathcal U^{**}_T (\mathcal I)$ is \textcolor{black}{always weakly larger than the static benchmark $\mathcal U^{*}_{T}(\mathcal I)$. and can be strictly larger. Consequently, any upper bound on
dynamic regret also bounds static regret. We remark that our results show
the designer can approach this stronger benchmark at the same
asymptotic rate. In particular, we construct algorithms that achieve
$O(\log^2 T)$ regret against the static and dynamic benchmarks,
respectively. Thus, in both cases, the designer's average
expected payoff falls below the corresponding per-period
benchmark by at most $O(\log^2 T/T)$. Learning to attain the
stronger dynamic benchmark therefore does not require a slower
rate of convergence.} 
%Thus, achieving no regret against the dynamic benchmark automatically implies no regret against the static benchmark. 
%\tao{Here or in Section \ref{sec:static-benchmark}, explain why we are still considering the static benchmark given that it is weaker than the dynamic benchmark. }

A regret \textit{upper bound} is a performance guarantee of an algorithm $\Alg$ in the following \textcolor{black}{sense}: for any problem instance $\mathcal I$, that is, for any unknown belief $\rprior$, the regret incurred by the algorithm $\Alg$ grows no faster than some function of $T$, i.e.,
\begin{equation*}
    \Reg(T; \Alg, \mathcal I) ~ \leq ~ C_{\mathcal{I}} \cdot f(T) ~ = ~ O(f(T)), \quad \forall T, ~ \forall \mathcal I,
\end{equation*}
where $C_{\mathcal{I}}$ is a constant depending on $\mathcal I$ but not on $T$. 
Here, $f(T)$ represents the \emph{worst-case} rate at which the regret can grow. % {\color{red} For example, the empirical estimation has $\Reg(T; \Alg, \mathcal I) = O(\sqrt{T})$.} 
An algorithm $\Alg$ with regret upper bound $O(f(T))$ is \textit{no-regret} if $f(T)$ grows at a rate that is sublinear with $T$, that is,
$$\lim_{T \rightarrow \infty} \frac{f(T)}{T}\to 0.$$
For example, the worst-case rate  $f(T)$ can be $f(T)=\sqrt{T}$ or  $f(T)=\log{T}$.
%is sub-linear in $T$: $\frac{f(T)}{T}\to 0$.  

%\paragraph{Regularity assumption on priors}

\subsection{Regularity Assumptions}

We present two regularity assumptions {used throughout the paper}. First, we assume that the beliefs $\dprior, \rprior$ have full support and the designer knows the minimum probability $p_0>0$ of any state. {This} probability $p_0$ will be used by the designer for designing the algorithms later.

\begin{assumption}[\textcolor{black}{Known lower bound on prior}
%Full-support priors
]
% [Priors have full support]
%[Regularity on priors]
\label{ass:prior-p0}
Both players' prior beliefs $\dprior, \rprior \in \Delta(\Omega)$ have full support and
there exists $p_0 \in (0,1)$ such that the designer knows $\min_{\omega \in \Omega} \rprior(\omega) \ge p_0 > 0.$
%$\dprior(\omega), \rprior(\omega) \ge p_0, \forall \omega \in \Omega$. The designer knows $p_0$ but not $\rprior$. 
\end{assumption}

%\textcolor{red}{I would not call this "full-support priors."  That makes it too easy for a referee to miss what's really important:  namely your assumption that $p_0$ is known to the designer.}

\begin{assumption}[Regularity of receiver's utility] \label{ass:D}
There is no weakly dominated action for the receiver.  
\end{assumption}

%Then, we assume that there exists

Assumption \ref{ass:D} implies that there exists a positive number $D > 0$
such that, for every action $a \in A$ of the receiver, there exists a belief $\eta_a \in \Delta(\Omega)$ on which action $a$ is \textit{strictly} better than any other action by a margin of $D$:
\begin{equation}
\label{eq:D}
    \E_{\omega \sim \eta_a}[v(a, \omega)] ~\ge ~ \E_{\omega \sim \eta_a}[v(a', \omega)] + D, \quad \forall a' \in A\setminus\{a\}.
\end{equation}
The designer knows $D$ since she knows the receiver's utility function $v$.

\textcolor{black}{Denote by $a^*_\omega = \argmax_{a \in A} v(a, \omega)$ the receiver's optimal action in state $\omega \in \Omega$. We impose two further assumptions on the receiver's utility function.}

\begin{assumption}[Unique optimal action]
\begingroup\color{black}
 \label{ass:receiver-optimal-action}
The receiver has a unique optimal action $a^*_\omega$ in each state $\omega\in\Omega$.
\par\endgroup

\end{assumption}

\begingroup\color{black}
Since the state and action sets are finite, Assumption \ref{ass:receiver-optimal-action} immediately implies that there exists a constant $G > 0$ such that $$\forall \omega \in \Omega, \forall a \in A\setminus \{ a^*_\omega\}, v(a^*_\omega, \omega) - v(a, \omega) > G > 0.$$ {Since the designer knows the utility functions, she knows this constant $G$.}
% there exists a constant $G>0$ such that
% \[
%  v(a^*_\omega,\omega)-v(a,\omega)>G,
%  \qquad \omega\in\Omega,\quad a\ne a^*_\omega.
% \]
% The designer can compute such a constant from the known utility function $v$.
\par\endgroup

\begingroup\color{black}
We call two states \emph{distinguishable} if their receiver-optimal actions differ.
\begin{assumption}[Distinguishable states] \label{ass:two-states-different-actions}
There exist at least one pair of states $\omega_i, \omega_j \in \Omega$ whose corresponding receiver-optimal actions are different: $a^*_{\omega_i} \ne a^*_{\omega_j}$.
\end{assumption}
\par\endgroup

\begingroup\color{black}
% Assumption \ref{ass:two-states-different-actions} makes the designer's learning problem non-trivial. If Assumption \ref{ass:two-states-different-actions} does not hold, then the receiver would have the same optimal action $a^*$ across all states. Thus, regardless of the signaling scheme and the belief, the receiver would always take action $a^*$, and the expected payoff of the designer would be a constant, leaving the designer zero regret trivially.

Assumption \ref{ass:two-states-different-actions} makes the designer's learning problem non-trivial. If Assumption \ref{ass:two-states-different-actions} does not hold, then the receiver would have the same optimal action $a^*$ across all states regardless of the signal sent. The designer then cannot influence the receiver's action through signals. The expected payoff of the designer would be a constant, leaving the designer zero regret trivially from not knowing the receiver's prior.
\par\endgroup

\section{Learning to Achieve No Regret Against the Static Benchmark}
\label{sec:static-benchmark}

%This section studies the $T$-period learning model of information design with unknown subjective belief. 
We will design a learning algorithm for the designer to achieve a regret upper bounded by $O(\log^2 T)$ against the static benchmark. The learning algorithm contains two parts\textcolor{black}{. The first is the learning. The second} 
%of the receiver's belief, which will be used again in the later section for dynamic regret. The other 
is a static robustification process for obtaining a near-optimal signaling scheme from the learned belief.

% Thus, the learning algorithm we design is efficient up to a constant factor.

% We design a learning algorithm for the information designer based on the idea of first estimating the prior $\rprior$ and then computing a robust and approximately optimal signaling scheme from the estimated prior.

\subsection{Efficient Learning of the Strategic Receiver's Belief}
\label{subsec:pair}

% Algorithm \ref{alg:any-pair-of-states}, together with its core part Algorithm \ref{alg:binary-search}, shows how the information designer learns the unknown prior $\mu^*$ over time.  The design of the two algorithms makes use of the receivers' knowledge about the prior $\mu^*$. Specifically, by knowing the prior $\mu^*$ and being able to do Bayesian updating, it is clear to the receiver which action is optimal for him after a signal is received. Observing the action the receiver takes, the information designer then infers information about the prior $\mu^*$. Thus, the information designer will gradually acquire information enough to learn the prior accurately with multiple different signaling schemes. Moreover, such a process only requires one natural and mild assumption on the receiver's utility function $v(\cdot, \cdot)$.

% How to learn the unknown belief $\rprior$? 
% %Instead of using the empirical distribution of the state of the world, we design a much more efficient algorithm that uses the actions taken by the receivers to infer $\rprior$.  
% A receiver takes an action based on his belief $\rprior$ after receiving a signal. Such an action, whether it is optimal for the receiver or not, contains information about his belief $\rprior$. %By using multiple different signaling schemes, the designer gradually acquires enough information to learn the prior accurately. 
% Our idea is to infer the receiver's belief $\rprior$ through his actions.
\textcolor{black}{The designer learns the receiver's prior belief $\rprior$ by observing his actions under suitably chosen signaling schemes. Our approach is to learn the prior belief ratio
$\rprior(\omega_i)/\rprior(\omega_j)$ for any pair of states to recover the receiver's prior belief $\rprior$.
We first consider pairs of {distinguishable} states, for which the receiver's optimal actions differ, and develop a \textit{binary search procedure} for learning their ratios under the prior belief. We show that how delaying the effect of the receiver's actions on future
signaling schemes limits his incentive to manipulate the designer's learning of his prior. We then extend the procedure to arbitrary pairs of states and construct an estimate of the receiver's prior belief $\rprior$.}

Algorithm~\ref{alg:binary-search} shows how to estimate the receiver's prior belief ratio $\frac{\rprior(\omega_i)}{\rprior(\omega_j)}$ for any pair of distinguishable states $\omega_1$ and $\omega_2$ %We describe the main idea assuming that the receiver best responds in each period the case of a strategic receiver will be discussed later. 
% 
% For a fixed pair of distinguishable states $\omega_1, \omega_2$, we want to estimate the receiver's prior belief ratio $\frac{\rprior(\omega_1)}{\rprior(\omega_2)}$.
% The main idea is a binary search.
% Assume the receiver to be myopic for now; we will discuss the strategic receiver later.  
% Suppose the information designer uses a signaling scheme $\pi$ that sends some signal $s_0$ only under states $\omega_1$ and $\omega_2$ (namely, $\pi(s_0 | \omega) = 0$ for $\omega \ne \omega_1, \omega_2$), then the receiver will believe that the state must be $\omega_1$ or $\omega_2$ whenever he receives signal $s_0$. If the signaling ratio $\frac{\pi(s_0 | \omega_2)}{\pi(s_0 | \omega_1)}$ is zero, then the myopic receiver will know for sure that the state is $\omega_1$, thereby taking the optimal action $a_1^*$ for state $\omega_1$.
% On the other hand, if the signaling ratio $\frac{\pi(s_0 | \omega_2)}{\pi(s_0 | \omega_1)}$ is large (say, $\ge \frac{1}{Gp_0}$), then the myopic receiver will believe that the state is $\omega_2$ with a high probability and take the optimal action $a^*_2$, which is different from $a_1^*$.
% ====================
based on \textit{binary search}. 
%First, the receiver's action reflects his inference about the state he is currently in. 
The designer uses a signaling scheme $\pi$ that sends a certain signal $s_0$ only under states $\omega_i$ and $\omega_j$, that is, {$\pi(s_0 | \omega) > 0$ for $\omega = \omega_i, \omega_j$ and} $\pi(s_0 | \omega) = 0$ for $\omega \ne \omega_i, \omega_j$. Then the receiver will believe that the state must be {either} $\omega_i$ or $\omega_j$ upon receiving signal $s_0$.
In particular, the ratio of the receiver's posterior belief about those two states is
\begin{align}
\label{eq:posterior-ratio}
    \frac{\mu(\omega_j|s_0)}{\mu(\omega_i|s_0)} = \frac{\rprior(\omega_j)}{\rprior(\omega_i)} \frac{\pi(s_0|\omega_j)}{\pi(s_0|\omega_i)}, 
\end{align}
which depends on the receiver's prior belief $\rprior$ and the signaling scheme $\pi$ used by the designer. The receiver takes an action based on this posterior belief to maximize his total payoff from the current and future periods. Based on the receiver's action, the designer then updates the signaling scheme for the next period. That process iterates until the designer obtains an estimate of the receiver's prior belief ratio $\frac{\rprior(\omega_i)}{\rprior(\omega_j)}$ with a desired accuracy. 

%It is useful to first understand the decision making of a receiver who \textit{{myopically} best responds} to the signal. 
To learn the belief ratio of a strategic receiver, it is useful to first understand how to learn that ratio if the receiver best responds to the signaling scheme $\pi$ \emph{myopically}, that is, if the receiver maximizes his payoff only in the current period.
If the receiver's prior belief ratio $\frac{\rprior(\omega_j)}{\rprior(\omega_i)}$ is sufficiently small, then his posterior belief ratio \eqref{eq:posterior-ratio} will also be small. Thus, the receiver will believe that the state is $\omega_i$ with high probability and take the corresponding optimal action $a_{\omega_i}^*$.
On the other hand, if the receiver's prior belief ratio $\frac{\rprior(\omega_j)}{\rprior(\omega_i)}$ is very large, then he will believe that the state is $\omega_j$ with high probability and take the corresponding optimal action $a_{\omega_j}^*$. In between, there is a range for the ratio $\frac{\rprior(\omega_j)}{\rprior(\omega_i)}$, where the receiver might take neither action but a completely different action. 

Therefore, given the signaling scheme $\pi$  and a pair of distinguishable states, there exists a \textit{cutoff} $c(\pi, \omega_i, \omega_j)$ on the receiver's prior belief ratio %at which
such that the receiver is indifferent in taking action $a^*_{\omega_i}$ versus some other action $\tilde a_{\omega_j, \omega_i}$, which might or might not be $a^*_{\omega_j}$. Thus, the receiver
takes action $\tilde a_{\omega_j, \omega_i}$ when 
his prior belief ratio $\frac{\rprior(\omega_j)}{\rprior(\omega_i)}$ is above the cutoff $c(\pi, \omega_i, \omega_j)$.  
As explained above, $\tilde a_{\omega_j, \omega_i}$ is not necessarily $a^*_{\omega_j}$.
We call such an action $\tilde a_{\omega_j, \omega_i}$ the \emph{first indifferent action} for the receiver under the signaling scheme $\pi$ and a pair of distinguishable states $\omega_i, \omega_j$. 
The cutoff $c(\pi, \omega_i, \omega_j)$ is the prior belief ratio $\frac{\mu(\omega_j)}{\mu(\omega_i)}$ that solves the following indifference equation under the experimented signaling scheme $\pi$:
\begin{align*}
    & \mu(\omega_j) \pi(s_0 | \omega_j) \Big( v(a_{\omega_i}^*, \omega_j) - v(\tilde a_{\omega_j, \omega_i}, \omega_j) \Big) + \mu(\omega_i) \pi(s_0|\omega_i) \Big( v(a_{\omega_i}^*, \omega_i) - v(\tilde a_{\omega_j, \omega_i}, \omega_i) \Big) = 0  \\
    & \implies ~ c(\pi, \omega_i, \omega_j) = \frac{\mu(\omega_j)}{\mu(\omega_i)} = \frac{\pi(s_0 | \omega_i)}{\pi(s_0|\omega_j)} \cdot \frac{v(a^*_{\omega_i}, \omega_i) - v(\tilde a_{\omega_j, \omega_i}, \omega_i)}{v(\tilde a_{\omega_j, \omega_i}, \omega_j) - v(a^*_{\omega_i}, \omega_j)}. 
\end{align*}
The {first indifferent action} $\tilde a_{\omega_j, \omega_i}$ is formally defined as follows. Define by $\zeta_{\omega_j,\omega_i}^a := \frac{v(a_{\omega_i}^*, \omega_i)-v(a, \omega_i)}{v(a, \omega_j)-v(a_{\omega_i}^*, \omega_j)}$ the value for the 
%\textit{posterior belief} 
ratio $\frac{\mu(\omega_j)\pi(s_0|\omega_j)}{\mu(\omega_i)\pi(s_0|\omega_i)}$ under which the receiver, when best responding myopically, is indifferent in taking actions $a$ and $a_{\omega_i}^*$.
\begin{definition}[First indifferent action]
    \label{def:a-tilde}
    For a pair of distinguishable states $\omega_i, \omega_j$, for each action $a \in A \setminus \{a_{\omega_i}^*\}$,
    the first indifferent action $\tilde{a}_{\omega_j, \omega_i}$ for the receiver is 
    %\begin{align*}
        $$\tilde{a}_{\omega_j, \omega_i} := \argmin \big\{ \zeta_{\omega_j,\omega_i}^a ~|~ a \in A ~ \text{s.t.} ~ v(a,\omega_j) > v(a_{\omega_i}^*, \omega_j) \big\}.$$
    If the $\argmin$ set is not a singleton, choose $\tilde{a}_{\omega_j, \omega_i}$ to be the action in the set that maximizes $v(a, \omega_j)$. 
    %\end{align*}
    %where $\zeta_{\omega_j, \omega_i}^a := \frac{\rprior(\omega_j)}{\rprior(\omega_i)} = \frac{v(a_{\omega_i}^*, \omega_i) - v(a, \omega_i)}{v(a, \omega_j) - v(a_{\omega_i}^*, \omega_j)}$.
\end{definition}

The designer knows the cutoff $c(\cdot, \cdot, \cdot)$ for the prior belief ratio $\frac{\mu(\omega_j)}{\mu(\omega_i)}$ based on her knowledge about the receiver's payoff function $v$.
That enables the designer to use a \textit{binary search} algorithm to identify a precise range containing the receiver's prior belief ratio $\frac{\rprior(\omega_j)}{\rprior(\omega_i)}$. In particular, by observing whether the receiver takes action $a^*_{\omega_i}$ or not, the designer is able to tell whether the receiver's actual prior belief ratio $\frac{\rprior(\omega_j)}{\rprior(\omega_i)}$ is below or above the cutoff $c(\pi, \omega_i, \omega_j)$. %For example, if the designer observes that the receiver takes action $a^*_{\omega_i}$ under the signaling scheme $\pi^{(k)}$ at iteration $k$, then the designer will know that the receiver's prior belief ratio $\frac{\rprior(\omega_j)}{\rprior(\omega_i)}$ is below the threshold $\tau(\pi^{(k)}, \omega_i, \omega_j)$.
%In that way, the designer will cut by half the range of the possible values of the receiver's prior belief ratio $\frac{\rprior(\omega_j)}{\rprior(\omega_i)}$ at that iteration. 
That enables the designer to adjust the signaling scheme and the corresponding cutoff threshold through iterations over time.
By multiple iterations, the designer narrows down the range of $\frac{\rprior(\omega_j)}{\rprior(\omega_i)}$ to length $\eps$ in $O(\log\frac{1}{\eps})$ steps.

\textcolor{black}{We now turn to the case when the receiver is strategic. 
%Being long-lived may incentivize the receiver to sacrifice his current payoff to take deceptive actions to manipulate the designer’s learning of his belief $\rprior$.
A long-lived receiver may sacrifice his current payoff to take deceptive actions (i.e., deviating from a myopic best response), in order to manipulate the designer's learning of his belief $\rprior$ to obtain more favorable information in future periods. 
%Our learning algorithm limits this incentive through two features.
Our learning algorithm 
%has two crucial steps to 
makes the designer's learning \textit{robust} to such strategic manipulation. 
First, the algorithm delays the influence of the receiver's current action on the update of the signaling scheme. In particular, in each iteration $k$, after recording the action when signal $s_0$ first occurs, the algorithm keeps the signaling scheme $\pi^{(k)}$ fixed for an additional $m$ periods before updating the signaling scheme for the next iteration $k+1$. Thus, the recorded action cannot affect the signaling scheme during these $m$ periods. With discounting, the receiver's potential gain from a manipulation
%manipulating subsequent signaling schemes 
is at most $\gamma^{m+1}/(1-\gamma)$.
%, evaluated at the time of the recorded action. 
So any manipulation that incurs an immediate loss larger than that gain is not attractive to the receiver.
}

\textcolor{black}{Second, the algorithm uses only the receiver's action when signal $s_0$ first occurs in each iteration and ignores subsequent actions during the additional $m$ periods. 
%This ensures that every action used to update the signaling scheme is separated from its effect on future information by the prescribed delay. 
%If subsequent actions were also taken into account, they could affect the signaling scheme updating, especially for those actions taken near the end of the delay, which would give the receiver a stronger incentive to manipulate. 
If the algorithm also took subsequent actions into account, actions taken near the end of the delay could affect the signaling scheme updating almost immediately, giving the receiver a stronger incentive to manipulate the update.
Together, these two features make the receiver's actions relevant for future signaling scheme updating only at sufficiently separated periods. The algorithm need not induce exact myopic best responses when the receiver is nearly indifferent between manipulation and obedience (i.e., when deviating from the myopic best response incurs a small immediate loss), but the resulting error in belief estimate can be bounded and made small by a delay length $m$ chosen to be sufficiently large.}

Algorithm~\ref{alg:binary-search}, based on binary search, formally shows how to learn the receiver's prior belief ratio $\frac{\rprior(\omega_i)}{\rprior(\omega_j)}$ for any pair of distinguishable states $\omega_i, \omega_j$ when the receiver acts strategically. 
%For the ratio of the signaling scheme updated the $k-$th time, denoted by $\frac{\pi^{(k)}(s_0|\omega_j)}{\pi^{(k)}(s_0|\omega_i)}$, %when it is the $k-$th update of the signaling scheme, 
%recording the receiver's actions in all repetition of $\frac{\pi^{(k)}(s_0|\omega_j)}{\pi^{(k)}(s_0|\omega_i)}$  would pollute the efficient learning of the binary search, since the receiver may still deviates from the best response, possibly because he is unable to do the above tradeoff calculation or he just does so on purpose. Thus, recording only the first action during the repetition uses of a signaling scheme ratio protects the efficient learning of the binary search to the largest extent.
%The first action in the $k-$th loop to the largest extent reflects the receiver’s trade-off: he must choose between acting truthfully to maximize his immediate payoff or trying to manipulate the designer’s learning. Since the repeated rounds make lying unattractive, the first response provides the most reliable information about his belief. By using only this initial action, our algorithm keeps the estimate robust against manipulation.

%Our main intuition for learning the receiver's belief is as follows. The algorithm collects the receiver's action that contains information about his belief only \textit{once} when signal $s_0$ is sent to the receiver for the first time in the current loop.(haven't finished)

%{\color{red} TODO: Change a lot of $\eps$ to $\terminate$}

\begin{algorithm}[H]
\caption{Estimating Prior Ratio for Distinguishable States by Binary Search}
\label{alg:binary-search}

\SetKwInOut{Input}{Input}
\SetKwInOut{Output}{Output}
\SetKwInOut{Parameter}{Parameter}

\Input {two states $\omega_i, \omega_j$ whose receiver-optimal actions are different.} 
\Parameter {a termination criterion $\tau > 0$, and an integer $m\ge 1$.}
\Output {an estimation $\hat \rho$ of the ratio $\frac{\rprior(\omega_i)}{\rprior(\omega_j)}$.}

\DontPrintSemicolon
\LinesNumbered

Let $k = 0$, $\ell^{(0)} = 0$, $r^{(0)} = \frac{1}{G p_0}$. \;
Let $a^*_{\omega_i} = \argmax_{a\in A} v(a, \omega_i)$.  Let $\tilde a$ be the first indifferent action $\tilde a_{\omega_j, \omega_i}$. \; % $\tilde a = \argmax_{a \in A} v(a, \omega_2)$. ($a_1 \ne \tilde a$ by assumption) \;
\While{$r^{(k)} - \ell^{(k)} > \terminate$} {
    Let $q = \frac{\ell^{(k)} + r^{(k)}}{2}$. \;
    Let $s_0$ be an arbitrary signal in $S$; let $\pi^{(k)}$ be a signaling scheme that satisfies $\frac{\pi^{(k)}(s_0 | \omega_j)}{\pi^{(k)}(s_0 | \omega_i)} = q$: \newline 
    {
    - if $q \le 1$, let $\pi^{(k)}(s_0 |\omega_j) = q$, $\pi^{(k)}(s_0 |\omega_i) = 1$; \newline
    - if $q > 1$, let $\pi^{(k)}(s_0 |\omega_j) = 1$, $\pi^{(k)}(s_0 |\omega_i) = 1/q$. \newline
    }
    For $\omega \in \Omega\setminus\{ \omega_i, \omega_j\}$, $\pi^{(k)}(s_0|\omega) = 0$. For $s\in S\setminus\{s_0\}$, $\pi^{(k)}(s|\omega)$ can be arbitrary. \; 
    Repeat $\pi^{(k)}$ for multiple periods until signal $s_0$ is realized.  Let $a^{(k)}$ be the action taken by the receiver when $s_0$ is realized. \;
    Repeat $\pi^{(k)}$ for additional $m$ periods, while ignoring the receiver's actions. \;
    \eIf {$a^{(k)} = a^*_{\omega_i}$} {
        Let $\ell^{(k+1)} = q, r^{(k+1)} = r^{(k)}$.
    } {
        Let $r^{(k+1)} = q, \ell^{(k+1)} = \ell^{(k)}$. 
    }
    $k = k+1$. \;
}
Output $\hat \rho = \ell^{(k)} \cdot \frac{v(\tilde a, \omega_j) - v(a_1^*, \omega_j)}{v(a_{\omega_i}^*, \omega_i) - v(\tilde a, \omega_i)}$. 
\end{algorithm}
\vspace{0.2em}

The performance of Algorithm \ref{alg:binary-search} is characterized by Lemma \ref{lem:ratio}, which shows that a good estimate of $\frac{\rprior(\omega_i)}{\rprior(\omega_j)}$ can be obtained within a very short amount of time:  Algorithm \ref{alg:binary-search} requires at most $O(\log T)$ periods to reach a precision $\tau=O(\frac{1}{T})$.

\begin{lemma} 
\label{lem:ratio}
%For a strategic receiver, 
The output of Algorithm \ref{alg:binary-search}, $\hat \rho$, satisfies $\big| \hat \rho - \frac{\rprior(\omega_i)}{\rprior(\omega_j)}\big| \le \frac{1}{G} \big(\terminate + \frac{\gamma^m}{(1-\gamma) G^3 p_0^3}\big) \le \frac{1}{G p_0} \big(\terminate + \frac{\gamma^m}{(1-\gamma) G^3 p_0^3}\big) \frac{\rprior(\omega_i)}{\rprior(\omega_j)}$, and Algorithm~\ref{alg:binary-search} terminates in at most $(\frac{1}{p_0} + m ) \log_2 \frac{1}{Gp_0 \terminate}$ periods in expectation. 
\end{lemma}
\begin{proof}
    See Appendix \ref{proof:ratio}. 
 \end{proof}

%The derivation of Lemma \ref{lem:ratio} is supported by our finding (Lemma \ref{lem:binary-search-strategic-receiver}) that we are able to incentivize the strategic receiver to perform consistently with single-period best responses by controlling the distance of the signaling scheme ratio estimate $\frac{\pi^{(k)}(s_0|\omega_j)}{\pi^{(k)}(s_0|\omega_i)}$ from the targeted ratio $\frac{\pi^{*}(s_0|\omega_j)}{\pi^{*}(s_0|\omega_i)}$.
A key part in the proof of Lemma \ref{lem:ratio} is the following Lemma \ref{lem:binary-search-strategic-receiver}, which shows that the strategic receiver is incentivized to perform consistently with single-period best responses if the signaling scheme ratio estimate $\frac{\pi^{(k)}(s_0|\omega_j)}{\pi^{(k)}(s_0|\omega_i)}$ differs from the targeted ratio $\frac{\pi^{*}(s_0|\omega_j)}{\pi^{*}(s_0|\omega_i)}$ by a positive distance $\diss = O(\gamma^m)$, {where the targeted ratio $\frac{\pi^{*}(s_0|\omega_j)}{\pi^{*}(s_0|\omega_i)}$ is the solution to 
$$\frac{\rprior(\omega_j)}{\rprior(\omega_i)} = \frac{\pi^*(s_0 | \omega_i)}{\pi^*(s_0|\omega_j)} \cdot \frac{v(a^*_{\omega_i}, \omega_i) - v(\tilde a_{\omega_j, \omega_i}, \omega_i)}{v(\tilde a_{\omega_j, \omega_i}, \omega_j) - v(a^*_{\omega_i}, \omega_j)}.$$
}
%In other words, we have already obtained an accurate estimate of the target ratio $\frac{\pi^{*}(s_0|\omega_j)}{\pi^{*}(s_0|\omega_i)}$ when the strategic receiver starts to manipulate. 

\begin{lemma}\label{lem:binary-search-strategic-receiver}
Choose $\diss = \frac{\gamma^m}{(1-\gamma)G^2 p_0^2}{}$. 
In Line 6 of Algorithm \ref{alg:binary-search}, when signal $s_0$ is sent, 
\begin{itemize}
\item if $\frac{\pi^{(k)}(s_0|\omega_j)}{\pi^{(k)}(s_0|\omega_i)} < (1-\diss) \frac{\pi^*(s_0|\omega_j)}{\pi^*(s_0|\omega_i)}$, then the receiver will take action $a_{\omega_i}^*$;
\item if $\frac{\pi^{(k)}(s_0|\omega_j)}{\pi^{(k)}(s_0|\omega_i)} > (1+\diss) \frac{\pi^*(s_0|\omega_j)}{\pi^*(s_0|\omega_i)}$, then the receiver will not take action $a_{\omega_i}^*$. 
\end{itemize} 
\end{lemma}

To see the intuition of Lemma \ref{lem:binary-search-strategic-receiver}, note that when recommended by a signal $s_0$ according to Algorithm \ref{alg:binary-search}, the strategic receiver is incentivized to take actions truthfully because his loss from deviation is larger than his gain from deviation. Regarding the gain, the receiver's accumulated gain from deviating is upper bounded by a positive number due to his discounted factor. Regarding the loss, the receiver's loss is lower bounded by a positive number that depends on how close our ratio estimate $\frac{\pi^{(k)}(s_0|\omega_j)}{\pi^{(k)}(s_0|\omega_i)}$ is from the targeted ratio $\frac{\pi^{*}(s_0|\omega_j)}{\pi^{*}(s_0|\omega_i)}$, as well as the positive margin $G$ by the unique optimal action assumption (Assumption \ref{ass:receiver-optimal-action}). Then, at a controlled distance of ratio estimate $\frac{\pi^{(k)}(s_0|\omega_j)}{\pi^{(k)}(s_0|\omega_i)}$ from ratio $\frac{\pi^{*}(s_0|\omega_j)}{\pi^{*}(s_0|\omega_i)}$, the receiver's loss lower bound is strictly larger than his gain upper bound, making him incentivized not to deviate from the recommendation. 

As an illustration, Figure \ref{fig:ratio_v_1} shows the case of $\frac{\pi^{(k)}(s_0|\omega_j)}{\pi^{(k)}(s_0|\omega_i)} < (1-\diss) \frac{\pi^*(s_0|\omega_j)}{\pi^*(s_0|\omega_i)}$, in which the strategic receiver is incentivized to take action $a_{\omega_i}^*$. The payoff brought by action $a_{\omega_i}^*$ is the highest for that case. Figure \ref{fig:ratio_v_2} shows the case of $\frac{\pi^{(k)}(s_0|\omega_j)}{\pi^{(k)}(s_0|\omega_i)} > (1-\diss) \frac{\pi^*(s_0|\omega_j)}{\pi^*(s_0|\omega_i)}$, in which the strategic receiver is incentivized not to take action $a_{\omega_i}^*$. The payoff brought by action $a_{\omega_i}^*$ is no longer the highest.

\begin{figure}[htbp]
\centering
\begin{subfigure}[b]{0.48\textwidth}
\includegraphics[width=\linewidth]{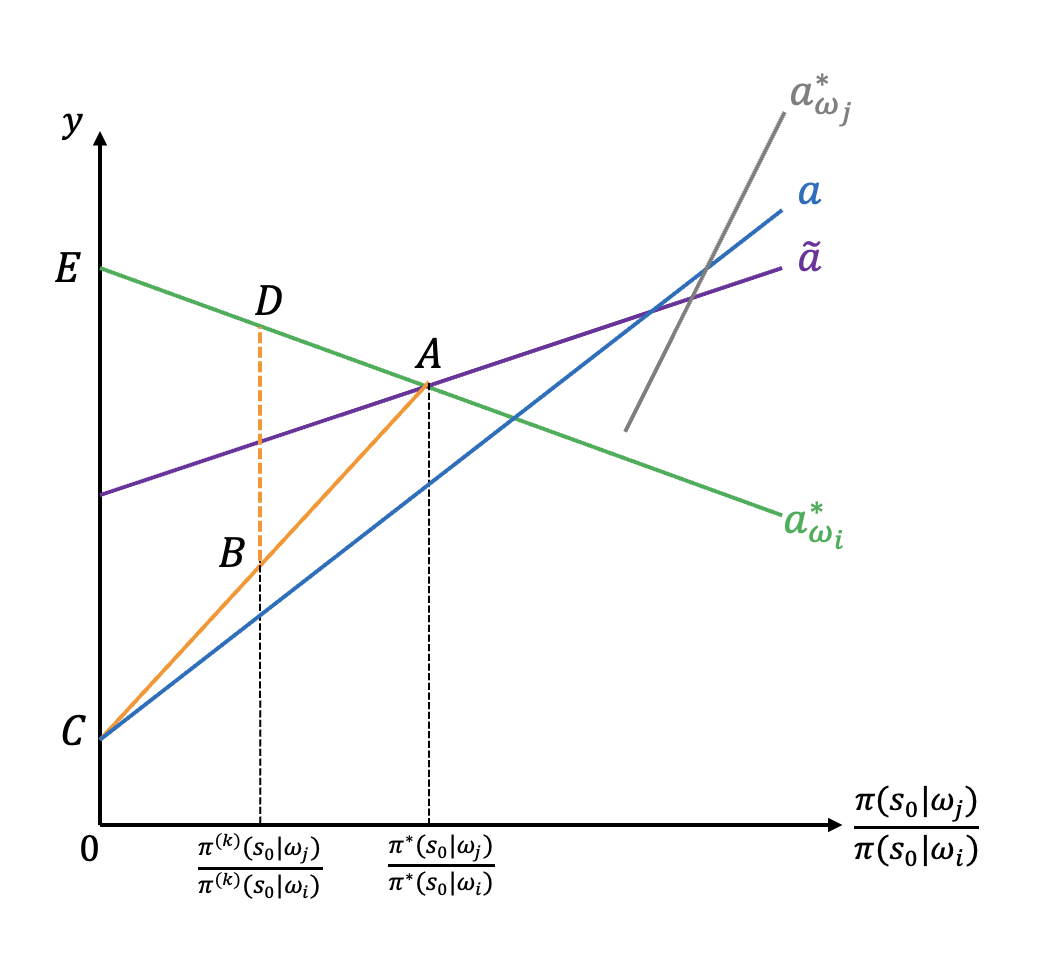}
\caption{The case of $\frac{\pi^{(k)}(s_0|\omega_j)}{\pi^{(k)}(s_0|\omega_i)} < (1-\diss) \frac{\pi^*(s_0|\omega_j)}{\pi^*(s_0|\omega_i)}$}
\label{fig:ratio_v_1}
\end{subfigure}
\begin{subfigure}[b]{0.48\textwidth}
\includegraphics[width=\linewidth]{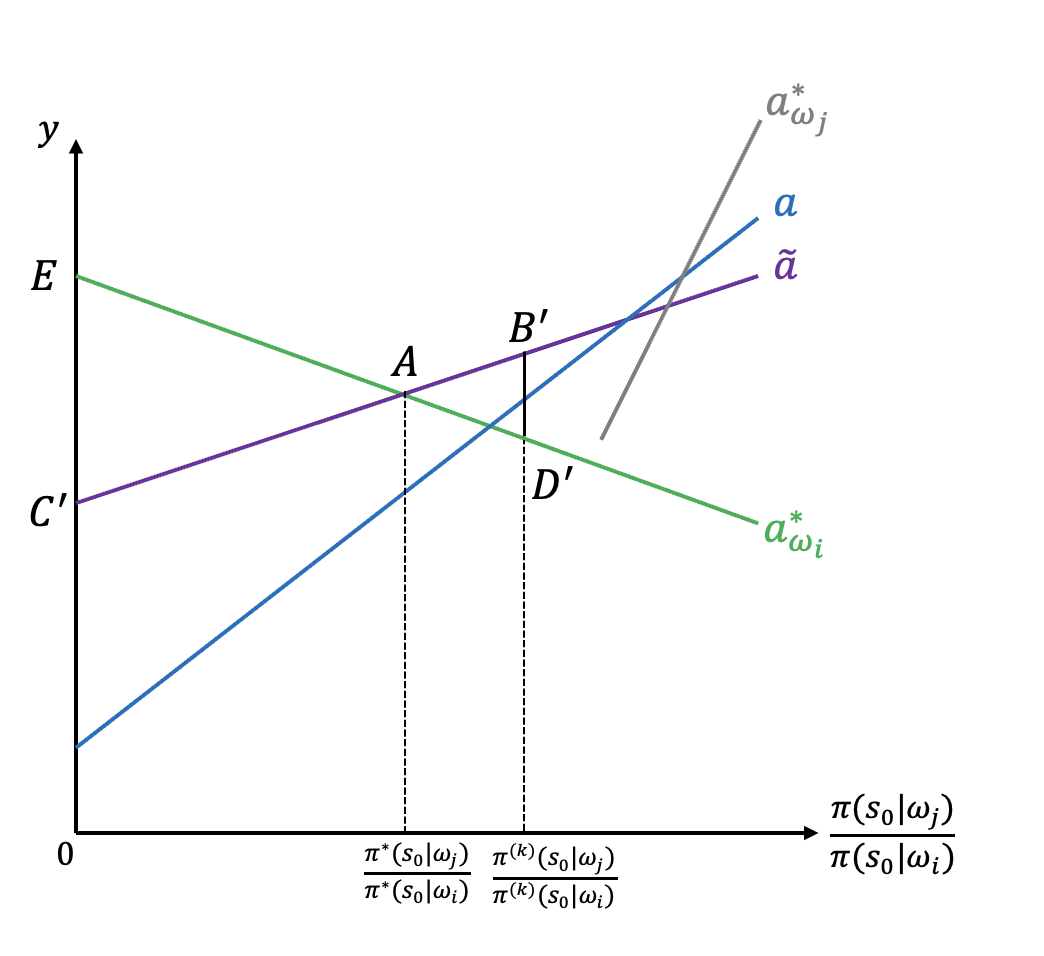}
\caption{The case of $\frac{\pi^{(k)}(s_0|\omega_j)}{\pi^{(k)}(s_0|\omega_i)} > (1+\diss) \frac{\pi^*(s_0|\omega_j)}{\pi^*(s_0|\omega_i)}$}
\label{fig:ratio_v_2}
\end{subfigure}
\caption{\centering Normalized utility of the receiver for different actions. The $x$-coordinate is the signaling ratio $\frac{\pi(s_0|\omega_j)}{\pi(s_0|\omega_i)}$. Each line corresponds to an action $a$, with equation $y_a(x) = v(a, \omega_i) + \frac{\rprior(\omega_j)}{\rprior(\omega_i)} \cdot x \cdot v(a, \omega_j)$. }
\label{ratio_v}
\end{figure}

% the following crucial lemma, Lemma \ref{lem:binary-search-strategic-receiver}, % which shows that even when the estimated ratio $\frac{\pi^{(k)}(s_0|\omega_2)}{\pi^{(k)}(s_0|\omega_1)}$ is far from the true ratio $\frac{\pi^*(s_0|\omega_2)}{\pi^*(s_0|\omega_1)}$, the receiver cannot take suboptimal actions., which is presented as follows.
% which shows that the strategic receiver's behavior can be consistent with best responses in each period even when the ratio estimate $\frac{\pi^{(k)}(s_0|\omega_j)}{\pi^{(k)}(s_0|\omega_i)}$ in the $k-$th update is far from the ratio $\frac{\pi^*(s_0|\omega_j)}{\pi^*(s_0|\omega_i)}$ for the receiver's belief $\rprior$.

\begin{proof}
See Appendix \ref{proof: binary-search-strategic-receiver}.
\end{proof}

% Receiver wants to match $v(a_1, \omega_1) = 1$ and $v(a_0, \omega_0) = 0$. 

% Sender wants $u(a_1, \omega_1) = 1$.  $u(a_1, \omega_0) = 1000$. $u(a_0, \omega_0) = 1$, $u(a_0, \omega_1) = 0$. 

% Prior 

What if the receiver-optimal actions of a pair of states are the same? Algorithm \ref{alg:any-pair-of-states} deals with such a case, allowing the information designer to estimate the prior ratio between any pair of states.
The idea of Algorithm \ref{alg:any-pair-of-states} is that, if states $\omega_i$ and $\omega_j$ have the same receiver-optimal actions, then both of them must be distinguishable from one state in a distinguishable pair of states, say $\omega_k \in \Omega$, by Assumption \ref{ass:two-states-different-actions}.  %$\omega_k \in \{\omega_1, \omega_2\}$. 
Thus, we obtain the estimate of $\frac{\rprior(\omega_i)}{\rprior(\omega_j)}$ by estimating the ratios $\frac{\rprior(\omega_i)}{\rprior(\omega_k)}$ and $\frac{\rprior(\omega_j)}{\rprior(\omega_k)}$ separately. 

\begin{algorithm}[H]
\caption{Estimating Belief Ratio for any Pair of States $(\omega_i, \omega_j)$}
\label{alg:any-pair-of-states}

\SetKwInOut{Input}{Input}
\SetKwInOut{Output}{Output}
\SetKwInOut{Parameter}{Parameter}

% \Input {any two states $\omega_i, \omega_j \in \Omega$} 
% \Parameter {a termination criterion $\eps > 0$, and an integer $m\ge 1$.}
\Parameter {$\terminate > 0$ and $m\ge 1$}
% \Output {an estimation $\hat \rho_{ij}$ of the ratio $\frac{\rprior(\omega_i)}{\rprior(\omega_j)}$}

\DontPrintSemicolon
\LinesNumbered

If $\omega_i$ and $\omega_j$ are distinguishable, i.e., $a^*_{\omega_i} \ne a^*_{\omega_j}$, then run Algorithm~\ref{alg:binary-search} on $\omega_i$ and $\omega_j$. % with parameters $\eps$ and $m$. \;

Otherwise, i.e., $a^*_{\omega_i} = a^*_{\omega_j}$, find $\omega_k \in \Omega$ such that $a_{\omega_k}^* \ne a^*_{\omega_i} = a^*_{\omega_j}$.  Run Algorithm~\ref{alg:binary-search} %with parameter $\eps$
to obtain an estimate $\hat \rho_{ik}$ for $\frac{\rprior(\omega_i)}{\rprior(\omega_k)}$ and an estimate $\hat \rho_{jk}$ for $\frac{\rprior(\omega_j)}{\rprior(\omega_k)}$.  Return $\hat\rho_{ij} = \frac{\hat \rho_{ik}}{\hat \rho_{jk}}$. 
\end{algorithm}

% \ce{I have put the proof of Lemma \ref{lem:relax-ratio} in the appendix.}

% The estimate $\hat{\mu}$ of the true prior $\mu^*$ will be easily recovered once we have obtained estimated ratios $\hat \rho_{ij}$ for any pair of states $\omega_i, \omega_j$ by Algorithm~\ref{alg:any-pair-of-states}. We show that the estimated prior $\hat \mu$ must be close to the true prior $\mu^*$ with a distance at most $\frac{6|\Omega|\eps}{p_0^3}$: 

\paragraph{Full algorithm for learning the belief $\rprior$.}
% After showing how to estimate the prior ratio for any pair of states,  we then
% use Algorithm \ref{alg:any-pair-of-states} as a subroutine to design Algorithm \ref{alg:known-prior} for the information designer to achieve $O\big( \log T)$ regret.  The main idea of  Algorithm \ref{alg:known-prior} is as follows. First, we estimate the ratio $\frac{\mu^*(\omega_i)}{\mu^*(\omega_j)}$ for any pair of states $\omega_i, \omega_j \in \Omega$.  In that way, we reconstruct the unknown prior $\mu^*$  with a high precision, i.e., getting an estimated prior $\hat \mu$ such that $\| \hat \mu - \mu^* \|_1 \le O(\eps)$.  Then, we use the robustification technique (Lemma~\ref{lem:robust-pi}) to obtain a signaling scheme $\hat \pi$ that is persuasive and $O(\eps)$-approximately optimal for $\mu^*$. 
% Using $\hat \pi$ for the remaining periods thus incurs a small regret. The regret of Algorithm \ref{alg:known-prior} is formally characterized in Theorem \ref{thm:prior-aware}.

After showing how to learn the belief ratio for any pair of states, we now present the full algorithm in Algorithm \ref{alg:learn-prior} for the designer to learn the receiver's belief $\rprior$. Algorithm \ref{alg:learn-prior} first uses Algorithm~\ref{alg:any-pair-of-states} to learn the ratio $\frac{\rprior(\omega_i)}{\rprior(\omega_1)}$ for each state $\omega_i$, $i=2, \ldots, |\Omega|$ and state $\omega_1$, and then reconstructs the estimate of the unknown belief from those ratio estimates. 
%That allows us to reconstruct the unknown prior $\rprior$ with a high precision, i.e.,

\vspace{0.2em}
\begin{algorithm}[H]
\caption{Learning the Belief of the Strategic Receiver}
\label{alg:learn-prior}

\SetKwInOut{Input}{Input}
\SetKwInOut{Output}{Output}
\SetKwInOut{Parameter}{Parameter}

% \Input {Utility functions $u, v$. Total number of periods $T$.}
\Parameter {desired estimation accuracy $\eps > 0$}
% \Parameter {$\terminate > 0$ and $m\ge 1$}

\DontPrintSemicolon

Let $\terminate = \frac{G p_0^3 \eps}{24|\Omega|}, m = \Big\lceil \frac{1}{\ln(1/\gamma)}\ln(\frac{24|\Omega|}{(1-\gamma) G^4 p_0^6 \eps}) \Big\rceil$. 

For every state $\omega_i\in \Omega$, $i=2, \ldots, |\Omega|$, apply Algorithm \ref{alg:any-pair-of-states} to state pair $(\omega_i, \omega_1)$ with parameters $\terminate$ and $m$ to learn the ratio $\frac{\rprior(\omega_i)}{\rprior(\omega_1)}$, obtaining ratio estimate $\hat \rho_{i1}$. %  Let $T_0$ be the number of periods taken in this process.  \;

Return belief estimate $\hat \mu \in \Delta(\Omega)$ by  
$\begin{cases}
\hat\mu(\omega_1) = 1/\big(1 + \sum_{i=2}^{|\Omega|} \hat \rho_{i1}\big); \\
\hat \mu(\omega_i) = \hat \rho_{i1} \hat \mu(\omega_1), \text{ for } i=2, \ldots, |\Omega|.
\end{cases} 
$  \; 
\end{algorithm}

Algorithm \ref{alg:learn-prior} gives a belief estimate $\hat \mu$ with an $\ell_1$ distance precision $\eps$ at $\| \hat \mu - \rprior \|_1 \le \eps$ within a short running time of $O(\log^2\frac{1}{\eps})$ periods. 

\begin{lemma} \label{lemma:good-prior-estimation}
Given $0 < \eps \le 1$,
\begin{itemize}
    \item the belief estimate $\hat \mu$ from Algorithm \ref{alg:learn-prior} satisfies $\| \hat \mu - \rprior \|_1 \le \eps$;
    \item the expected running time of Algorithm \ref{alg:learn-prior} is at most $2|\Omega|\big(\tfrac{1}{p_0} + \frac{1}{1-\gamma}\ln(\frac{24|\Omega|}{(1-\gamma) G^4 p_0^6 \eps}) \big) \log_2 \tfrac{24|\Omega|}{G^2 p_0^4 \eps}$ periods. 
\end{itemize}
\end{lemma}

\begin{proof}
    See Appendix~\ref{proof:good-prior-estimation}. 
 \end{proof}

\subsection{Static Robustification of Signaling Schemes}
\label{sec:robustification}

\paragraph{\bf Computing the signaling scheme optimal for a belief.} In general, given any prior belief $\mu_0 \in \Delta(\Omega)$ of the receiver, assumed to be known to the designer, the optimal signaling scheme in the static Bayesian persuasion problem can be computed easily.
%recognized by the designer as the receiver's belief, the signaling scheme optimal for the designer under her knowledge of the receiver's belief $\mu_0$ can be computed tractably.

In particular, given any prior belief $\mu_0 \in \Delta(\Omega)$ {known to be} the receiver's belief, there exists a signaling scheme $\pi_0$ that is \emph{direct} and \emph{persuasive} \citep{kamenica_bayesian_2011} under $\mu_0$: a signaling scheme $\pi: \Omega \rightarrow \Delta(A)$ is \emph{direct} if it maps each state to a probability distribution over \textit{actions}; a direct signaling scheme $\pi_0$ is \textit{persuasive} if all actions recommended by $\pi_0$ are optimal for the receiver under prior $\mu_0$: $\sum_{\omega} \mu_0(\omega) \pi_0(a|\omega) \big[ v(a,\omega)-v(a',\omega) \big] \ge 0, \forall a, a' \in A$. Denote by $\mathrm{Pers}(\mu_0)$ the set of persuasive signaling schemes {given} the receiver's belief $\mu_0 \in \Delta(\Omega)$:
\begin{equation}
    \mathrm{Pers}(\mu_0) \coloneqq \Big\{ \pi: \Omega \to \Delta(A) ~ \big | ~  \sum_{\omega \in \Omega} \mu_0(\omega) \pi(a|\omega) \big[ v(a,\omega)-v(a',\omega) \big] \ge 0, ~ \forall a,a' \in A \Big\},
\end{equation}
where $$\sum_{\omega \in \Omega} \mu_0(\omega) \pi_0(a|\omega) \big[ v(a,\omega)-v(a',\omega) \big] ~ \forall a,a' \in A$$ is the \textit{persuasiveness} of the signaling scheme $\pi$ under the receiver's belief $\mu_0$.

The signaling scheme optimal for the designer with $\dprior$ under her knowledge of the receiver's belief $\mu_0$ is solved tractably from the following linear program
\begin{align}
\label{optimal-pi}
    \pi^*(\mu_0) = \argmax_{\pi \in \mathrm{Pers(\mu_0)}} \sum_{\omega \in \Omega} \dprior(\omega) \sum_{a\in A} \pi(s|\omega) u(a, \omega). 
\end{align}

Thus, if the designer knows the receiver's belief $\rprior$, then the designer will obtain her optimal signaling scheme $\pi^*$ for $\rprior$ exactly by the linear program \ref{optimal-pi}. However, after the learning phase, the designer has a learned belief $\hat \mu$, which is only an approximation for the receiver's prior belief $\rprior$.

\paragraph{\bf Static robustification.}
%One technique that we will use to design signaling schemes without knowing the receiver's prior belief is \emph{robustification}. 
After the first phase of learning, we have obtained an estimate $\hat \mu$ of the receiver's prior belief $\rprior$. However, even when the estimate $\hat \mu$ is very close to the prior belief $\rprior$, the signaling scheme $\hat \pi$ optimal for the designer under the estimate $\hat \mu$ 
%Then, a signaling scheme $\hat \pi$ optimal for the estimate $\hat \mu$ can be obtained from solving $\hat \pi = \argmax_{\pi \in \mathrm{Pers(\hat \mu)}} \sum_{\omega \in \Omega} \dprior(\omega) \sum_{a\in A} \pi(s|\omega) u(a, \omega)$. However, $\hat \pi$ 
may not be persuasive or even perform arbitrarily badly under $\rprior$. The reason is that even though a signal may induce two close posterior beliefs from two close prior beliefs $\hat \mu$ and $\rprior$, respectively, the optimal actions for the receiver under those two posterior beliefs may be different, which may lead to an arbitrarily low payoff for the designer under the receiver's prior belief $\rprior$.

{We propose designing a signaling scheme $\pi$ using \emph{static robustification}.}
%Our idea here is to exert a process, \emph{static robustification}, on a direct signaling scheme $\pi$ under the belief estimate $\hat \mu$, which may or may not be optimal under $\hat \mu$. 
{In general, we convert any \textit{direct} signaling scheme $\pi$ under $\hat{\mu}$ to another signaling scheme $\tilde \pi$, such that $\tilde \pi$, when applied to a targeted prior belief (e.g., the receiver's prior $\rprior$), has a weakly higher persuasiveness compared to the case when that typical signaling scheme $\pi$ is applied to the belief estimate $\hat \mu$. Recall that the persuasiveness of a direct signaling scheme under a prior belief measures how any recommended action from that signaling scheme is weakly better than any other unrecommended action. That is, for any $a, a' \in A$,
$$\sum_{\omega}\rprior (\omega) \tilde \pi(a|\omega)[v(a,\omega)-v(a',\omega)] \ge \sum_{\omega}\hat \mu (\omega) \pi(a|\omega)[v(a,\omega)-v(a',\omega)]. $$
%Moreover, the static robustification does not have much influence on the designer's payoff.
%the typical signaling scheme $\pi$ and 
%\textcolor{red}{Page 54. The last line on the page is confusingly phrased.}
%In particular, if what is used for the static robustification is not a typical signaling scheme $\pi$ but the signaling scheme $\hat \pi=\pi^*(\hat \pi)$ that is optimal for the designer under the belief estimate $\hat{\mu}$, then $\hat \pi$ will be converted to another signaling scheme that guarantees approximate optimality for the designer.
\textcolor{black}{Moreover, static robustification causes only a small loss in the designer's payoff. In particular, applying static robustification to $\hat\pi$
%, which is the designer's optimal signaling scheme under the estimated belief $\hat\mu$, 
yields a signaling scheme that is persuasive under the receiver's actual belief $\rprior$ and approximately optimal for the designer.}}
% Our static robustification is applied to any direct signaling scheme, which may not be the optimal one for the belief estimate \citep{zu_learning_2021} or even persuasive \citep{lin_information_2025}.
{By being static, the robustification process is applied to the signaling scheme $\hat \pi$ once, and the same statically robustified signaling scheme $\tilde \pi$ is used repeatedly in each period for the robustification phase till the end. 
%The following Lemma \ref{lem:universal-robustification} formalizes our idea of the static robustification.

\begingroup
\color{black}

Lemma~\ref{lem:universal-robustification} summarizes the effects of static robustification. We briefly introduce the idea of static robustification. 
We first construct a posterior whose persuasiveness is strengthened compared to any direct signaling scheme $\pi$ under belief estimate $\hat\mu$. To construct that posterior, we first consider the posterior $\hat\mu_a$ induced by $\pi$ under belief estimate $\hat\mu$ for each action $a$. Then we consider a \emph{buffer belief} $\eta_a$ for each action $a$, under which action $a$ gives the receiver an expected payoff at least $D$ higher than every other action by Assumption~\ref{ass:D}. We construct posterior $\xi_a$ for each action $a$ by
%Let $q_a=\sum_{\omega\in \Omega} \hat{\mu}(\omega)\pi(a|\omega)$ be the probability that $\pi$ recommends action $a$ under belief estimate $\hat\mu$, and let $\hat\mu_a \in \Delta(\Omega)$ be the corresponding posterior whenever $q_a>0$.  
\begin{align}
\label{eq:staticxi}
    \xi_a=(1-\delta)\hat\mu_a+\delta\eta_a,
\end{align}
for a robustification strength $\delta\in(0,\frac{p_0}{2}]$.
%that determines how far the constructed posterior $\xi_a$ moves toward the buffer belief. 
Thus, if $\pi$ is persuasive under belief $\hat\mu$, then the receiver prefers action $a$ to every alternative by at least $\delta D$ under the constructed posterior $\xi_a$. This positive margin allows persuasiveness to be remained when the receiver's prior $\rprior$ is close to belief $\hat\mu$. %These modified posteriors, weighted by their original probabilities, need not average to $\hat\mu$. 
Then, we restore Bayesian plausibility by choosing a residual belief $\chi$ on the boundary of the belief simplex \(\Delta(\Omega)\) and a weight $y \in [0, \frac{\delta}{p_0}]$ such that
\[
    \hat\mu
    =(1-y)\sum_{a:q_a>0}q_a\xi_a+y\chi,
\]
where $q_a=\sum_{\omega\in \Omega} \hat{\mu}(\omega)\pi(a|\omega)$ is the probability that $\pi$ recommends action $a$ under belief $\hat\mu$.
%Such a choice exists under the assumptions of the lemma, as shown in Appendix~\ref{app:proof:universal-robustification}. 
Thus, we obtain a signaling scheme $\tilde{\pi}$ that induces posterior $\xi_a$ with probability $(1-y)q_a$ and fully reveals state $\omega$ with probability $y\chi(\omega)$ under belief $\hat\mu$. 
%The full revelation signals supply the residual probability needed for the posteriors to average to the estimated prior. 
Finally, we replace each state-revealing signal for full revelation with a receiver-optimal action recommendation to ensure static robustification gives a direct signaling scheme $\tilde\pi$. 
%Expanding $\xi_a$, we have weights $(1-y)(1-\delta)$ on posterior $\hat \mu_a$, $(1-y)\delta$ on buffer belief $\eta_a$ , and $y$ on the full-revelation signaling schemes. %The weight $y$ is chosen to satisfy Bayesian plausibility and need not equal $\delta$.

\par
\endgroup

\begin{lemma}[Static Robustification] \label{lem:universal-robustification}
Assume Assumptions \ref{ass:prior-p0} and \ref{ass:D}.
% Suppose the receiver's priors $\hat \mu$ and $\mu$ satisfy $\| \hat \mu - \mu \|_1 \le \eps \le \frac{p_0^2 D}{4}$.
Choose $\eps \le \frac{p_0^2 D}{4}$. 
Given the belief estimate $\hat \mu$ and the robustification strength $\delta \in [\frac{2\eps}{p_0D}, \frac{p_0}{2}]$, any direct signaling scheme $\pi$ can be converted into another direct signaling scheme $\tilde \pi$ such that
\begin{itemize}
    \item (improved persuasiveness for the receiver) for any prior beliefs $\mu$ and $\hat{\mu}$ with $\| \mu - \hat \mu \|_1 \le \eps$, for any recommendation $a \in A$ from $\tilde \pi$, 
    %which is the robustified direct signaling scheme $\pi$, 
    the persuasiveness of $\tilde \pi$ and $\pi$ under the resulted posterior beliefs $\mu (\cdot | a, \tilde \pi)$, $\hat \mu (\cdot | a, \tilde \pi)$ are close: 
    \begin{equation*}
        \E_{\omega \sim  \mu (\cdot | a, \tilde \pi)} \big[ v(a, \omega) - v(a', \omega) \big]
        \ge \min \big\{\E_{\omega \sim \hat  \mu (\cdot | a,  \pi)}\big[ v(a, \omega) - v(a', \omega) \big] + \delta D - \tfrac{2\eps}{p_0}, ~ 0 \big\}, ~\forall a' \in A\setminus\{a\}, 
    \end{equation*}
    {and, in particular, if the direct signaling scheme $\pi$ is persuasive for $\hat \mu$, then its robustification result $\tilde \pi$ is persuasive for $\mu$;}  
    \item (tiny influence for the designer) when the receiver obeys the recommendation, the designer's payoff from the robustified signaling scheme $\tilde \pi$, compared to that from the direct signaling scheme $\pi$, satisfies $$U(\tilde \pi, \rho_\ob; \dprior) \ge U(\pi, \rho_\ob; \dprior) - \frac{3\delta}{p_0},$$
    where $\rho_\ob$ is the receiver's obedience strategy that puts weight one on the recommended action;
    \item (close signaling schemes) the robustification result $\tilde \pi$ and the direction signaling scheme $\pi$ are close: $$\| \tilde \pi(\cdot | \omega) - \pi(\cdot |\omega) \|_1 \le \frac{4\delta}{p_0^2}, ~ \forall \omega \in \Omega.$$
\end{itemize} 
\end{lemma}

\begingroup
\color{black}

To see the intuition, recall that a signaling scheme obtained from static robustification is constructed based on the posterior $\xi_a$ for each action $a$ as in Eq.~\ref{eq:staticxi}. First, $\hat \mu_a$ is one component of $\xi_a$, but that's under the belief estimated $\hat \mu$, so the receiver under his prior $\rprior$ may still deviate from any recommendation and obtain a gain at most $2\eps/p_0$. However, the buffer belief $\eta_a$ with weight $\delta$, as another component of $\xi_a$, guarantees the receiver a payoff at least $\delta D$ under $\hat\mu$. Thus, before the signaling scheme is finally made direct, the receiver gains at least 
\[
    \delta D-\frac{2\eps}{p_0}\ge 0
\]
at $\delta=2\eps/(p_0D)$. The final step of static robustification, i.e., turning the signaling scheme into a direct one, requires sending the receiver his optimal action in the corresponding state, which also preserves the receiver's obedience incentives. Thus, the direct signaling scheme $\tilde\pi$ from static robustification is persuasive under $\rprior$. 

\par
\endgroup

%To see the high-level idea, \textcolor{red}{This is also very unclear. I think when you say "by the distance...", you mean "because the distance between $\hat \mu$ and $\mu_R$ is less than epsilon".. But the rest needs to be clarified too.}by the distance between $\hat \mu$ and $\rprior$ at $\|\hat \mu - \rprior \|_1 \le \eps$, the designer's optimal signaling scheme $\hat \pi$ under $\hat \mu$ will \textit{not} be persuasive under the receiver's belief $\rprior$ by a margin that is only at most $O(\frac{\eps}{p_0})$. The signaling scheme that induces the belief $\eta_a$ for each $a \in A$ is strictly persuasive for the receiver by a margin of $D$ by Assumption~\ref{ass:D}. The full revelation signaling scheme is weakly persuasive. Thus, the persuasiveness of the mixture, that is, the signaling scheme $\tilde \pi$, is $$D\cdot \delta + 0 \cdot \delta - (1-2\delta)\cdot O(\frac{\eps}{p_0}) \ge 0$$ by a choice of the robustification strength $\delta = O(\frac{\eps}{p_0 D})$.
%As a result, the loss of the designer is at most $$2\delta \cdot O(\frac{1}{p_0}) = O(\frac{\eps}{p_0^2D}).$$

In contrast to previous works that apply a robustifying process to the signaling scheme optimal for the belief estimate \citep{zu_learning_2021} or to signaling schemes persuasive for the belief estimate \citep{li_information_2025}, our static robustification is applied to \emph{any direct} signaling schemes, thus being the most general. If we apply Lemma \ref{lem:universal-robustification} to the designer's optimal signaling scheme $\hat \pi$ under the belief estimate $\hat \mu$ with the robustification strength $\delta = \frac{2\eps}{p_0^2 D}$, then we will obtain an approximately optimal signaling scheme for the actual prior $\rprior$ of the receiver. 
This specific application is summarized in Corollary \ref{cor:robustification}.
%then we will obtain an application of our general static robustification to the specific use for signaling scheme optimal for the belief estimate, which is summerized in Corollary \ref{cor:robustification}.

%Corollary \ref{cor:robustification} is immediately obtained by Lemma \ref{lem:universal-robustification} applied to the designer's optimal signaling scheme $\hat \pi$ under the belief estimate $\hat \mu$ with the robustification strength $\delta = \frac{2\eps}{p_0^2 D}$. 

\begin{corollary}[Robustification for the  optimal signaling scheme]\label{cor:robustification} 
With $\| \hat \mu - \rprior \|_1 \le \eps \le \frac{p_0^2 D}{4}$, the designer's optimal signaling scheme $\hat \pi$ under belief $\hat \mu$ can be converted into another signaling scheme $\tilde \pi$ that is persuasive under belief $\rprior$ and $\frac{6\eps}{p_0^2 D}$-optimal for the designer. 
\end{corollary}

See details of Lemma \ref{lem:universal-robustification} in Appendix~\ref{app:proof:universal-robustification}.

\subsection{Full Learning Algorithm with $O(\log^2 T)$ Regret}
\label{subsec:full}

Finally, we combine the prior-learning algorithm and the static robustification to obtain a signaling scheme $\tilde \pi$ that is $O(\eps)$-approximately optimal for the designer and persuasive for the receiver under belief $\rprior$. 
Using the robustified signaling scheme $\tilde \pi$ in each of the remaining periods thus incurs a small regret. The total regret of Algorithm \ref{alg:known-prior} is formally characterized in Theorem \ref{thm:prior-aware}.

\vspace{0.2em}
\begin{algorithm}[H]
\caption{Learning Algorithm for the Static Benchmark}
\label{alg:known-prior}

\SetKwInOut{Input}{Input}
\SetKwInOut{Output}{Output}
\SetKwInOut{Parameter}{Parameter}

% \Input {Utility functions $u, v$. Total number of periods $T$.}
\Parameter {$\eps > 0$}

\DontPrintSemicolon

% Let $\omega_1 \in \Omega$ be the special state in Assumption~\ref{ass:receiver-optimal-action}. \;
% For each state $\omega_i \in \Omega$, $i = 2, \ldots, |\Omega|$, apply Algorithm~\ref{alg:any-pair-of-states} to $(\omega_i, \omega_1)$ with parameters $\eps$ and $m$ to estimate $\hat \rho_{i1} \approx \frac{\rprior(\omega_i)}{\rprior(\omega_1)}$. Let $T_0$ denote the total number of periods used.\;

Run Algorithm \ref{alg:learn-prior} with parameter $\eps$ to obtain an estimate $\hat \mu$ of receiver's belief. Let $T_0$ be the number of periods taken. \; 
% According to Lemma~\ref{lemma:good-prior-estimation}, we have $\| \hat \mu - \rprior \|_1 \le $.
% Then, use Lemma~\ref{lem:robust-pi} with $\hat \mu$ and $\frac{6|\Omega|\eps}{{p_0^3}}$ to construct a signaling scheme $\pi$ that is persuasive and $\frac{84|\Omega|\eps}{p_0^5 D}$-optimal for $\mu^*$.
Apply Corollary~\ref{cor:robustification} to $\hat \mu$ with parameter $\eps$ to obtain a signaling scheme $\tilde \pi$. 
Use $\tilde \pi$ for the remaining $T - T_0$ periods.
\end{algorithm}

\begin{theorem}
\label{thm:prior-aware}
Assume Assumptions~\ref{ass:prior-p0}, \ref{ass:D}, \ref{ass:receiver-optimal-action}, \ref{ass:two-states-different-actions}. 
Choose $\eps = \frac{p_0^2 D}{4 T}$.  The static regret of Algorithm~\ref{alg:known-prior} is at most
\begin{align}
    % O\Big( |\Omega| \big( \tfrac{1}{p_0} +  \tfrac{1}{\log(1/\gamma)}\log \tfrac{|\Omega|T}{p_0^8 D G^4} \big) \log \tfrac{|\Omega|T}{D G^2 p_0^6} ~ + ~ \tfrac{1}{1-\gamma} \Big) = O\Big( C_{\mathcal I} \cdot \tfrac{\log^2 T}{\log(1/\gamma)} \Big)
    \Reg^{\mathrm{static}}(T; \text{Algorithm \ref{alg:known-prior}}, \mathcal I) ~ \le ~ O\Big( \frac{1}{1-\gamma} \log^2 T \Big). 
     %& ~ \le ~ O\bigg(  \Big( \frac{|\Omega|}{p_0} + \frac{|\Omega|}{1-\gamma}\log\big(\frac{|\Omega| T}{(1-\gamma) p_0 G D}\big)\Big) \log \frac{|\Omega| T}{p_0 G D} \bigg) \\
    %& ~ \le ~ O\Big( \frac{|\Omega|}{1-\gamma} \log^2 T + \frac{|\Omega|}{p_0} \log T \Big) ~ = ~ O\Big( \frac{1}{1-\gamma} \log^2 T \Big). 
\end{align}
% where $C_\mathcal{I} = |\Omega|$ is a constant depending on instance $\mathcal{I}$. 
\end{theorem}

%\ce{Should we move the following proof to the appendix?}
%\tao{Maybe not.  Sometimes keeping a proof in the body is a good idea.  Let's see how the page limit goes. }

\begin{proof}

By Lemma~\ref{lemma:good-prior-estimation}, we have $\| \hat \mu - \rprior \|_1 \le \eps$ and $\E[T_0] \le 2|\Omega|\big(\tfrac{1}{p_0} + \frac{1}{1-\gamma}\ln(\frac{24|\Omega|}{(1-\gamma) G^4 p_0^6 \eps}) \big) \log_2 \tfrac{24|\Omega|}{G^2 p_0^4 \eps}$. 
According to Corollary~\ref{cor:robustification}, the constructed signaling scheme $\tilde \pi$ is persuasive for $\rprior$ and is $\frac{6\eps}{p_0^2 D}$-approximately optimal for the designer.  Thus, the regret of Algorithm \ref{alg:known-prior} satisfies
\begin{align*}
    \Reg^{\mathrm{static}}(T; \Alg, \mathcal I) & ~ \le ~ \underbrace{ \E[ T_0 \cdot 1]}_{\text{regret in the first $T_0$ periods}} ~ + ~ \underbrace{(T - \E[T_0]) \frac{6\eps}{p_0^2 D} }_{\text{regret in the remaining periods}} \\
    & ~ \le ~ 2|\Omega|\Big(\frac{1}{p_0} + \frac{1}{1-\gamma}\ln\big(\frac{24|\Omega|}{(1-\gamma) G^4 p_0^6 \eps}\big)\Big) \log_2 \frac{24|\Omega|}{G^2 p_0^4 \eps} ~ + ~ T \frac{6\eps}{p_0^2 D}
\end{align*}
Choosing $\eps = \frac{p_0^2 D}{4T}$ (which satisfies the condition in Corollary \ref{cor:robustification}), we obtain
\begin{align*}
    \Reg^{\mathrm{static}}(T; \Alg, \mathcal I) & ~ \le ~ 2|\Omega|\Big( \frac{1}{p_0} + \frac{1}{1-\gamma}\ln\big(\frac{96 |\Omega| T}{(1-\gamma) G^4 p_0^8 D}\big)\Big) \log_2 \frac{96|\Omega| T}{G^2 p_0^6 D} ~ + ~ \frac{3}{2} \\
    & ~ = ~ O\Big( \frac{1}{1-\gamma} \log^2 T  \Big). 
    %& ~ = ~ O\Big( \frac{|\Omega|}{1-\gamma} \log^2 T + \frac{|\Omega|}{p_0} \log T \Big). % ~ = ~ O\Big( C_{\mathcal I}' \log^2(T) \cdot \Big). 
\end{align*}

\end{proof}

% To see the intuition of Theorem~\ref{thm:prior-aware}, 
% the algorithm splits the running time into (i) a belief–learning phase and (ii) a robustification-based exploitation phase. 
% To learn the receiver's belief with accuracy $\eps$, we run a binary search algorithm to keep updating the experimented signaling scheme, which takes at most $O(\log(\frac{1}{\eps}))$ distinct experimented signaling schemes. 
% Each experimented signaling scheme is repeated multiple times, and we only use the \emph{first} action taken by the receiver to avoid the receiver's incentive in single-period manipulation, which also helps reduce the learning inaccuracy. We choose $K=\Theta(\log T)$ (or $K=\Theta(\frac{\log T}{1-\gamma})$ with discounting) yields vanishing error probabilities and dynamic incentives. 
% Setting $\varepsilon \asymp 1/T$ keeps exploitation loss at $T\cdot O(\varepsilon)=O(1)$, so the exploration cost dominates. 
% Hence total exploration regret is 
% \[
% \underbrace{\Theta(\log(1/\varepsilon))}_{\#\text{ probes }=\Theta(\log T)} 
% \times 
% \underbrace{\Theta(\log T)}_{\text{length per probe }=K}
% ~=~ O(\log^2 T),
% \]
% and the overall regret is $O(\log^2 T) + O(1) = O(\log^2 T)$.

To see the intuition of Theorem~\ref{thm:prior-aware}, recall the algorithm has two phases: (i) learning the receiver’s belief $\rprior$ for an accuracy $\varepsilon$, and (ii) robustification, which repeatedly uses a robustified signaling scheme that is guaranteed persuasive for the receiver's belief $\rprior$ and approximately optimal for the designer. Phase (i) 
%promising future non-disclosure after a deviation, 
%we make obedience strictly optimal today. 
runs binary search algorithms to keep updating the experimented signaling scheme, which takes at most $O(\log(\frac{1}{\eps}))$ distinct experimented signaling schemes. %Estimating each pairwise odds ratio by the adaptive binary searching yields a $\log(\frac{1}{\eps})$ factor, and 
Further, for each distinct experimented signaling scheme, making the obedience incentive strong enough for a strategic receiver requires an additional repetition sub-phase of length $m = O \! \big(\tfrac{\log(\frac{1}{\eps})}{1-\gamma}\big)$, contributing another $\log(\frac{1}{\eps})$. Thus, we get the $O\!\big(\tfrac{1}{1-\gamma}\log^2(\frac{1}{\eps})\big)$ exploration cost formalized in Lemma~\ref{lemma:good-prior-estimation}. Phase (ii) robustifies the designer's optimal signaling scheme under the estimate $\hat \mu$. % by mixing in tiny, state–contingent slack (robustification strength $\delta$ in the order of $\eps/(p_0^2 D)$). 
That preserves persuasiveness for all beliefs within a distance of $\eps$ of $\hat\mu$, thus also for the receiver's belief $\rprior$, while degrading the designer’s payoff by only $O(\frac{\eps}{p_0^2 D})$ per period (Corollary~\ref{cor:robustification}). Choosing $\eps=\tfrac{p_0^2 D}{4T}$ results in a loss of $T\cdot O(\frac{\eps}{p_0^2 D})=O(1)$ for phase (ii) (exploitation), so the regret is dominated by phase (i) (exploitation), yielding $\Reg^{\mathrm{static}}(T)=O\!\big(\tfrac{1}{1-\gamma}\log^2 T\big)$ as claimed in Theorem~\ref{thm:prior-aware}.

We add a final brief remark on Algorithm \ref{alg:known-prior} and Theorem~\ref{thm:prior-aware}. In the first phase, the designed signaling scheme used in binary searches may be \textit{not} persuasive. The designer’s primary goal is to gather informative responses from the receiver’s actions to accurately estimate the prior through binary search during that phase. Thus, achieving persuasiveness and optimality at each step is secondary to this learning objective.

\section{Characterization of the Dynamic Benchmark}
\label{sec:characterization-dynamic-benchmark}
% We will give an upper bound on $\mathcal U^*_\gamma$. For $\gamma = 0$, $\mathcal U^*_0 = T U^*_{\mathcal BP}$. For $\gamma = 1$, $\mathcal U^*_1 \le U^*_{\mathrm{punishment}}$.

% ===================================================

This section characterizes the dynamic benchmark $\mathcal U^{**}_T(\mathcal I)$, which is the optimal payoff obtainable by the designer if she knows the receiver's belief $\rprior$. {That is, it is} the designer's optimal payoff in a $T$-period repeated Bayesian persuasion problem with full commitment power.  In Section \ref{subsec:opt-ddp}, we show that a certain class of grim-trigger-style strategies are optimal for the designer, that the dynamic benchmark is weakly larger than the static benchmark, and that the designer's optimal strategy can be obtained through dynamic programming. 
In Section \ref{subsec:example}, we present examples to illustrate the dynamic benchmark and the optimal strategy.

\subsection{Optimal Dynamically Direct + Punishment (DDP) Strategy}
\label{subsec:opt-ddp}

\paragraph{Optimal DDP strategy: dynamic benchmark characterization}
We show that the designer's dynamic benchmark $\mathcal U^{**}_T(\mathcal I)$ can be achieved by a class of $T$-period strategies with a \emph{grim-trigger} property: {the designer recommends} actions to the receiver {until} the receiver deviates from the recommendation, and {provides} no information {if} the receiver deviates. 
%Each period's recommendation strategy can depend on history. 
We call such a class of $T$-period strategies ``dynamically direct + punishment (DDP)'' strategies. 
\begin{definition}[DDP strategy]
\label{def:DDP}
A $T$-period strategy $\sigma$ %$\sigma=(\pi_{\sigma}^{(t)})_{t=1}^T$
of the designer is a \emph{dynamically direct + punishment (DDP) strategy} if $\sigma$ 
\begin{itemize}
    \item (dynamically direct) {uses a} direct signaling scheme {if} the receiver {has} never {deviated} from {past} recommendations,
    \item (punishment) and reveals no information (i.e., always sending one signal regardless of the state) \textcolor{black}{if the receiver did not follow the recommendation at some point in the past.}
    %{if} the receiver {has} {deviated in the past}.
\end{itemize}
\end{definition}

\textcolor{black}{The punishment depends only on whether the receiver's observed
action matches the recommendation he received. The designer
does not observe how he would have responded to other
recommendations.
}

%We then introduce two properties that make a DDP strategy $\sigma$ optimal. The first property is \textit{history independence}. 
We {now} introduce two important properties of DDP strategies. The first property is \textit{history independence}, which means that the signaling scheme in each period does not depend on what {happened} in the past, as long as the receiver has been obedient since the beginning. 
%For a DDP strategy $\sigma$, the signaling scheme in each period may depend on history. By the definition of being a DDP strategy, we may neglect $\sigma$'s dependence on history due to the receiver's obedience.

\begin{definition}[History {Independence}]
\label{def:hi}
A DDP strategy $\sigma$ %$\sigma=(\pi_{\sigma}^{(t)})_{t=1}^T$
is \emph{history-independent} if the direct signaling scheme $\pi_{\sigma}^{(t)}$ for each period $t$ depends on the period number $t$ but not on history, as long as the receiver \textcolor{black}{followed all past recommendations.}
%{has} never {deviated} from {past} recommendations. 
\end{definition}

% The second property is \textit{dynamic persuasiveness}.\textcolor{red}{why is this written as an equation? You're stating in words that the receiver's strategy is to obey. Also, this is not really correct. The designer doesn't observe what the receiver's strategy is, only what actions the receiver takes. In other words, perhaps the receiver would obey one recommendation but not another and the designer happens to make the recommendation the receiver obeys.}
\begin{definition}[Dynamic Persuasiveness]
\label{def:dp}
A DDP strategy $\sigma=(\pi_{\sigma}^{(t)})_{t=1}^T$ is \emph{dynamically persuasive} for the receiver if the receiver's $T$-period obedient strategy
\begin{equation*}
    \phi_{\ob} =  \text{obey the action recommendation in every period $t \in \{1, \ldots, T\}$ given any history}
\end{equation*}
{is a best response to $\sigma$.}
%\textcolor{black}{For a DDP strategy $\sigma$, let $\phi_{\ob}$ denote the receiver's strategy that follows each recommendation as long as all past recommendations were followed and, during punishment, chooses an action that maximizes his expected single-period payoff under his prior $\rprior$. The strategy $\sigma$ is \emph{dynamically persuasive} for the receiver if $\phi_{\ob}$ is a best response to $\sigma$.}
\end{definition}

If the direct signaling scheme in each period of a DDP strategy {is} all persuasive in the per-period Bayesian persuasion game, then obedience is clearly a best response for the receiver, and the DDP strategy is dynamically persuasive. However, a dynamically persuasive DDP strategy may include some direct signaling schemes that are \emph{not} persuasive in the static game. We will provide such examples in Section \ref{subsec:example}. 

The first result of this subsection shows that the designer's dynamic benchmark $\mathcal U^{**}_T(\mathcal I)$ can be achieved by a history-independent and dynamically persuasive DDP strategy $\sigma^*$: 
\begin{proposition}[Optimality of DDP Strategy] \label{prop:dynamic-benchmark-DDP}
The dynamic benchmark $\mathcal U^{**}_T(\mathcal I)$ can be achieved
by a pair of strategies $(\sigma^{*}, \phi_{\ob})$, where $\sigma^{*} = (\pi_{\sigma^{*}}^{(t)})_{t=1}^T$ is a history-independent and dynamically persuasive DDP strategy of the designer, and $\phi_\ob$ is the $T$-period obedient strategy of the receiver. 
\end{proposition}

\begin{proof}
\noindent

The proof of Proposition \ref{prop:dynamic-benchmark-DDP} has two steps. 
First, we provide a \emph{dynamic revelation principle} (in a similar spirit to \cite{myerson1986}) in the following Lemma \ref{lemma:dynamic-revelation-principle}, which shows that the designer's $T-$period (global) optimal utility $\mathcal U^{**}_T(\mathcal{I})$ can always be achieved by a history-dependent dynamically persuasive DDP strategy. See Appendix \ref{proof:dynamic-revelation-principle} for the proof of Lemma \ref{lemma:dynamic-revelation-principle}.  

\begin{lemma}[Dynamic revelation principle]
\label{lemma:dynamic-revelation-principle}
There exists an optimal strategy $\sigma^*=(\pi_{\sigma^*}^{(t)})_{t=1}^T$ for the designer that is a history-dependent dynamically persuasive DDP strategy, that is, $\mathcal U^{**}_T(\mathcal{I}) = \mathcal U_T(\sigma^*, \phi_\ob; \dprior)$ and $\phi_\ob = \phi^*(\sigma^*, \rprior)$.   
\end{lemma}

{Next}, we show in Lemma \ref{lemma:exist-ddp-independent} that {history-dependence} is not needed. At a high level, this is because the Bayesian persuasion problems at different periods are independent, as the states are i.i.d.~sampled. As long as the receiver has not deviated in the past, his actions will not contain information that is valuable for the designer.
Thus, the designer does not need to adjust signaling schemes based on the irrelevant history. The formal proof of Lemma \ref{lemma:exist-ddp-independent} is in Appendix \ref{proof:exist-ddp-independent}. 
\begin{lemma}
\label{lemma:exist-ddp-independent}
There exists an optimal dynamically persuasive DDP strategy that is history-independent. 
\end{lemma}

The above two lemmas together prove Proposition \ref{prop:dynamic-benchmark-DDP}.
\end{proof}

% \begin{table}[h!]
% \centering
% \setlength{\tabcolsep}{8pt}
% \renewcommand{\arraystretch}{1.3}
% \begin{tabular}{@{} l cc @{}}
% \toprule
%  & \textbf{history-dependent} & \textbf{history-independent} \\
% \midrule
% \textbf{dynamically persuasive DDP} &
% \cellcolor{pink!25}\textbf{lemma 5} &
% \cellcolor{pink!25}{\begin{tabular}{@{}c@{}}\textbf{lemma 5}\\ \textcolor{blue}{\textbf{lemma 6}}\end{tabular}}\\
% \textbf{not dynamically persuasive DDP} &
%  & \\
% \bottomrule

% \tao{Do we really need this table? }
% \end{tabular}
% \end{table}

\paragraph{Computing the optimal DDP strategy (dynamic benchmark)}
Proposition \ref{prop:dynamic-benchmark-DDP} significantly simplifies the designer's optimization problem. We can write the dynamic benchmark $\mathcal U^{**}_T(\mathcal I)$ as an optimization problem over history-independent DDP strategies subject to the dynamic persuasiveness constraint. 

We now study the dynamic persuasiveness constraint in detail.
%a history-independent DDP strategy consists of a sequence of direct signaling schemes $\sigma = (\pi^{(1)}, \ldots, \pi^{(T)})$. 
For a history-independent DDP strategy $\sigma=(\pi_{\sigma}^{(t)})_{t=1}^T$, \textit{dynamic persuasiveness} requires that for each period $t$, the sequence of ongoing (period $t$ included) direct signaling schemes $(\pi_{\sigma}^{(t)}, \ldots, \pi_{\sigma}^{(T)})$ in $\sigma=(\pi_{\sigma}^{(t)})_{t=1}^T$ should satisfy the following: for any action $a$ recommended by $\pi_{\sigma}^{(t)}$ with positive probability $P_{\rprior, \pi_{\sigma}^{(t)}}(a) = \sum_{\omega \in \Omega} \rprior(\omega)\,\pi_{\sigma}^{(t)}\!\left(a \mid \omega\right) > 0$, for any $a' \in A$,   
\begin{equation}  \label{eq:dynamically-persuasive-definition}
\begin{aligned}
    & \gamma^t \sum_{\omega \in \Omega} \frac{\rprior(\omega) \pi_{\sigma}^{(t)}(a|\omega)}{P_{\rprior, \pi_{\sigma}^{(t)}}(a)} v(a, \omega) + \sum_{t' = t+1}^{T} \gamma^{t'} V(\pi_{\sigma}^{(t')}, \rho_{\ob}; \rprior) & \\
    & \ge ~ \gamma^t \sum_{\omega \in \Omega} \frac{\rprior(\omega) \pi_{\sigma}^{(t)}(a|\omega)}{P_{\rprior, \pi_{\sigma}^{(t)}}(a)} v(a', \omega) + \sum_{t' = t+1}^{T} \gamma^{t'} V_\uninformed(\rprior),
\end{aligned}
\end{equation}
where $V_{\mathrm{uninformed}}(\rprior) = \max_{a\in A} \sum_{\omega \in \Omega} \rprior(\omega) v(a, \omega)$ is the receiver's payoff from being punished.
The left-hand side of Eq. \eqref{eq:dynamically-persuasive-definition} is the receiver's payoff of obedience from period $t$ to $T$ given the action $a$ recommended at period $t$, while the second line is the receiver's payoff from period $t$ to $T$ by deviating to $a'$ at period $t$ and then being punished in the future.
% , with
% \begin{equation}
% V_{\mathrm{uninformed}}(\rprior) = \max_{a\in A} \sum_{\omega \in \Omega} \rprior(\omega) v(a, \omega)
% \end{equation}
% being the receiver's single-period payoff without receiving any information.  
%Here, $a'$ could be the same as $a$, which implies that the receiver could obey the recommendation at period $t$ but deviate at period $t+1$. However, our dynamic persuasiveness precludes that situation. 

%where $\E_{\sigma}$ denotes the expectation over the random sample of $\pi^{(t')}$, which depends on $\sigma$ and all the randomness (in states, signals, actions) before period $t'$. 
An equivalent way to write the dynamic persuasiveness constraint uses the receiver's posterior belief. In particular, denote by $\rprior(\omega | a, \pi_{\sigma}^{(t)}) = \frac{\rprior(\omega) \pi_{\sigma}^{(t)}(a|\omega)}{P_{\rprior, \pi_{\sigma}^{(t)}}(a)}$ the posterior belief of state $\omega$ given action $a$ sent by signaling scheme $\pi^{(t)}$ under belief $\rprior$, Then, Eq. \eqref{eq:dynamically-persuasive-definition} {can be rewritten} as follows. For each period $t$, $\forall a, a' \in A$, 
\begin{equation} \label{eq:dynamically-persuasive-definition-2}
    \gamma^t \E_{\omega \sim \rprior(\cdot | a, \pi_{\sigma}^{(t)})} \big[ v(a, \omega) - v(a', \omega) \big] + \sum_{t' = t+1}^{T} \gamma^{t'} \big( V(\pi_{\sigma}^{(t')}, \rho_{\ob}; \rprior) - V_\uninformed(\rprior) \big) \ge 0. 
\end{equation}

Formally, the \textit{optimization problem} for the dynamic benchmark $\mathcal U^{**}_T(\mathcal{I})$ is 
\begin{align}
    \mathcal U^{**}_T(\mathcal{I}) ~ = ~ \max_{\text{history-independent DDP strategy $\sigma = (\pi_{\sigma}^{(t)})_{t=1}^T$}} ~~ &  \mathcal{U}_T(\sigma;\phi_{\text{ob}}, \dprior) ~ = ~ \sum_{t=1}^T U(\pi_{\sigma}^{(t)}, \rho_\ob; \dprior) \label{eq:global-benchmark-optimization} \\
    \text{subject to} \hspace{5.5em} & \sigma \text{ satisfies \eqref{eq:dynamically-persuasive-definition} or \eqref{eq:dynamically-persuasive-definition-2}}.  \nonumber 
\end{align}

Lemma \ref{lem:dynamic-larger-than-static} shows that the dynamic benchmark is always weakly larger than the static benchmark. Moreover, the dynamic benchmark is strictly larger than the static benchmark in some instances, as we will show in Section \ref{subsec:example}.  
\begin{lemma}
\label{lem:dynamic-larger-than-static}
\textcolor{black}{The dynamic benchmark is weakly larger than the} static benchmark: $\mathcal U^{**}_T(\mathcal I) \ge  \mathcal U^{*}_T(\mathcal I)$. 
\end{lemma}

% \begin{proof}
% Let $\pi^*_{\BP}$ be an optimal signaling scheme for the static Bayesian persuasion benchmark, achieving
% \[
% U^*_{\BP}(\mathcal I) = U(\pi^*_{\BP}, \rho^{\ob}; \mu_D),
% \]
% where $\rho_\ob$ is the receiver's obedience strategy that puts weight 1 on the recommended action.
% Construct a $T$-period dynamic strategy $\sigma^{\mathrm{rep}}$ that repeats $\pi^*_{\BP}$ in every period, independently of history. Because $\sigma^{\mathrm{rep}}$ does not react to deviations, the receiver’s intertemporal optimization problem reduces to the static one in each period. Thus, the receiver’s global best response is to play obediently each period, inducing the same outcome as in the static benchmark.
% The designer’s payoff under $\sigma^{\mathrm{rep}}$ is therefore
% \[
% \mathcal U_T(\sigma^{\mathrm{rep}}, \phi^{\mathrm{ob}}; \mu_D) = \sum_{t=1}^T U(\pi^*_{\BP}, \rho^{\mathrm{ob}}; \mu_D) = T \cdot U^*_{\BP}(\mathcal I).
% \]
% Since the dynamic benchmark takes the maximum over all feasible dynamic strategies, it is at least as large as this payoff. Hence
% \[
% U^{**}_T(\mathcal I) \ge T \cdot U^*_{\BP}(\mathcal I).
% \]
% \end{proof}

\begin{proof}
Let $\pi^*_{\BP}$ be an optimal signaling scheme in the static Bayesian persuasion problem. From Section \ref{sec:robustification}, $\pi^*_\BP$ is persuasive, where the obedient strategy $\rho_\ob$ of the receiver is a single-period best response. Thus, a $T$ times repetition of $\pi^*_\BP$, denoted by $\sigma^{\mathrm{rep}} = (\pi^{*(1)}_\BP, ..., \pi^{*(T)}_\BP)$, is a feasible strategy for the $T$-period game. The receiver's global best response to $\sigma^{\mathrm{rep}}$ consists of obedience in each period. 
%Since this strategy is history-independent, the receiver’s global best response coincides with his period-by-period best response.
Thus, the designer’s $T$-period payoff is $T \cdot U^*_{\BP}(\mathcal I)$. By the optimality of the dynamic benchmark, we immediately have 
\begin{equation*}
\mathcal U^{**}_T(\mathcal I) ~ \ge ~ \mathcal U_T(\sigma^{\mathrm{rep}}, \phi_\ob; \mu_D) ~ = ~ T \cdot U^*_{\mathrm{BP}}(\mathcal I)=\mathcal U^{*}_T(\mathcal I). %\qedhere 
\end{equation*}
\end{proof}

\paragraph{\bf A dynamic program to solve for $\mathcal U^{**}_T(\mathcal I)$.} %\textcolor{red}{you should cut or at least greatly shorten this material.}
Having shown that the designer's dynamic benchmark $\mathcal U^{**}_T(\mathcal I)$ can be achieved by a history-independent dynamically persuasive DDP strategy, we present a \emph{dynamic {programming}} algorithm for obtaining such an optimal DDP strategy. 

Define a value function $F(t, g)$ as the designer's optimal payoff \textit{from period $t$} to period $T$ under a history-independent dynamically persuasive DDP strategy, such that the receiver's continuation payoff from being obedient (from period $t$ to $T$) is weakly larger than the continuation payoff from being punished by receiving no information plus a non-negative margin of $g \ge 0$. Then the designer's optimization problem is, for each period $t = \{T, T-1, ..., 1\}$, 
\begin{align}
\label{eq:opt-general}
    F(t, g)  ~ = ~ \max_{\pi^{(t)}, \ldots, \pi^{(T)}} ~ & \sum_{t'=t}^T U(\pi^{(t')}, \rho_\ob; \dprior) \\
    \text{s.t.} ~~~~ & \sum_{t'=t}^T \gamma^{t'-t} \cdot V(\pi^{(t')}, \rho_\ob; \rprior) ~ \ge ~ \sum_{t'=t}^T \gamma^{t'-t} \cdot V_\uninformed(\rprior) ~ + ~ g \nonumber \\
    & (\pi^{(t)}, \ldots, \pi^{(T)}) \text{ is dynamically persuasive}.  \nonumber 
\end{align}
Essentially, the designer promises the receiver that he will obtain a payoff of at least $\sum_{t'=t}^T \gamma^{t'-t} \cdot V_\uninformed(\rprior) + g$ if he keeps being obedient from period $t$ to $T$.
Set $F(t, g) = -\infty$ if the optimization problem \eqref{eq:opt-general} is infeasible, which leads to 
the dynamic benchmark obtained at $\mathcal{U}^{**}_T(\mathcal{I})  = F(1, 0)$.
% \begin{equation}
%     \label{eq:DP-final-objective}
%     \mathcal{U}^{**}_T(\mathcal{I})  = F(1, 0). 
% \end{equation}

%If the optimization problem \eqref{eq:opt-general} is feasible, then the value function $F(t, g)$ will be computed recursively from the last period. 
We compute the value function $F(t, g)$ recursively from the last period.
That is, we first solve the optimization \eqref{eq:opt-final} for the last period $t = T$ by solving

% \begin{align}\label{eq:opt-final}
%     F(T, g) ~ = ~ \max_{\pi^{(T)}} ~~ & U(\pi^{(T)}, \rho_\ob; \dprior) \label{eq:DP-last-period} \\
%     \text{s.t.} ~~ &  V(\pi^{(T)}, \rho_\ob; \rprior) - V_\uninformed(\rprior) ~ \ge ~ g, \nonumber \\ 
%     & \pi^{(T)} \text{ is persuasive}. \nonumber 
% \end{align}

% Then, for $t \in \{T-1, T-2, \ldots, 1\}$, $g \ge 0$, solve the optimization \eqref{eq:opt-rest}
% \begin{align}\label{eq:opt-rest}
%     F(t, g) ~ = ~ \max_{\pi^{(t)}, g' \ge 0} ~~  & U(\pi^{(t)}, \rho_\ob; \dprior) + F(t+1, g')  \label{eq:DP-period-t-form-1} \\
%     \text{s.t.} ~~~~  &  V(\pi^{(t)}, \rho_\ob; \rprior) - V_\uninformed(\rprior) + \gamma \cdot g' ~ \ge ~ g  \nonumber \\
%     & \E_{\omega \sim \rprior(\cdot | a, \pi^{(t)})} \big[ v(a, \omega) - v(a', \omega) \big] + \gamma \cdot g' ~ \ge ~ 0,  \label{eq:DP-dynamic-persuasive-constraint} \\
%     & \hspace{11em} \forall a \text{ with $P_{\rprior, \pi^{(t)}}(a) > 0$}, ~~ \forall a' \in A.  \nonumber 
% \end{align}

% Last period
\begin{subequations}\label{eq:opt-final}
\begin{align}
F(T,g) ~ &= ~ \max_{\pi^{(T)}} ~U\!\left(\pi^{(T)}, \rho_\ob; \dprior\right) \label{eq:DP-last-period}\\
~~ ~ \text{s.t.} \quad 
& V\!\left(\pi^{(T)}, \rho_\ob; \rprior\right) - V_\uninformed(\rprior) \ge g, \nonumber\\
& \pi^{(T)} \text{ is persuasive}. \nonumber
\end{align}
\end{subequations}
Then, for each $t \in \{T-1, T-2, \ldots, 1\}$, $g \ge 0$, solve the following optimization problem, which makes use of the value function $F(t+1, g')$ for the next period $t+1$:
% Period t
% \begin{subequations}\label{eq:opt-rest}
% \begin{align}
% F(t,g) &= \max_{\pi^{(t)},\, g'\ge 0}\; U\!\left(\pi^{(t)}, \rho_\ob; \dprior\right) + F(t+1,g') \label{eq:DP-period-t-form-1}\\
% \text{s.t.}\quad 
% & V\!\left(\pi^{(t)}, \rho_\ob; \rprior\right) - V_\uninformed(\rprior) + \gamma \cdot g' \ge g, \nonumber\\
% & \E_{\omega \sim \rprior(\cdot \mid a, \pi^{(t)})}
%    \big[ v(a,\omega) - v(a',\omega) \big] + \gamma \cdot g' \ge 0,\quad
%    \forall a \text{ with } P_{\rprior,\pi^{(t)}}(a)>0,\ \forall a' \in A. \label{eq:DP-dynamic-persuasive-constraint}
% \end{align}
% \end{subequations}
%Or, solve the following optimization problem \eqref{eq:DP-period-t-form-2} that is equivalent to \eqref{eq:opt-rest}:   
 \begin{align}
     F(t, g) ~ = ~ \max_{\pi^{(t)}, g' \ge 0} ~~  & U(\pi^{(t)}, \rho_\ob; \dprior) + F(t+1, g')  \label{eq:DP-period-t-form-2} \\
     \text{s.t.} ~~~~  &  V(\pi^{(t)}, \rho_\ob; \rprior) - V_\uninformed(\rprior) + \gamma \cdot g' ~ \ge ~ g  \nonumber \\
     & \sum_{\omega\in \Omega} \rprior(\omega) \pi^{(t)}(a|\omega) \Big( v(a, \omega) - v(a', \omega) + \gamma \cdot g' \Big) ~ \ge ~ 0 \hspace{1em} \forall a, a' \in A.  \nonumber  
 \end{align}

% The following Lemma \ref{lem:DP-characterization} shows we can solve the designer’s dynamic problem by keeping one ``state” variable: the amount of future utility promised to the receiver starting now.
To see the intuition, %the goal of the dynamic program is to optimize 
%we define a value function $F(t, g)$ for each period $t$ to be the designer’s best total payoff from period $t$ onward, where $g$ is the current promised payoff \textit{starting from the current period $t$} made by the designer to the receiver, such that the receiver's payoff from being obedient is larger his payoff from deviation by at least a positive value of $g$. 
we note that $F(t, g)$ is the designer’s best total payoff from period $t$ onward, while promising the receiver that his obedient payoff will be larger than his deviating payoff by at least a margin of $g \ge 0$, \textit{starting from the current period $t$} to the end.
At the current period $t$, the designer chooses the current signaling scheme $\pi^{(t)}$ and the next promised payoff $g'$ that the designer must deliver starting from the next period $t+1$. 
%The current signaling scheme $\pi^{(t)}$ and the next promised payoff $g'$ should satisfy the first constraints in the above optimization problems to incentivize the receiver to obey the recommendation. That is, the receiver is incentivized to be obedient \textit{ex ante}. 
The current signaling scheme $\pi^{(t)}$ and the next promised payoff $g'$ should satisfy the first constraints in the above optimization problems, which ensure that the receiver's obedient payoff is larger than the deviating payoff by a margin of $g$ starting from the \emph{beginning} of period $t$. 
%The choices of $\pi^{(t)}$ and $g'$ should also satisfy the second constraints above, such that the receiver, \textit{after} receiving the current recommended action, will indeed find the current recommendation persuasive, which is the current period $t$'s contribution to the dynamic persuasiveness of the designer's $T-$period strategy. 
The choices of $\pi^{(t)}$ and $g'$ should also satisfy the second constraints above, such that the receiver, \textit{after} receiving the current recommendation $a$, will indeed find the recommendation persuasive, considering the gain from deviating from $a$ to $a'$ and the loss of promised future utility $g'$ after deviation. 
%if we must promise the receiver at least $g$ units of future payoff above the uninformed baseline. 
%At period $t$, the designer chooses both the current signaling scheme $\pi^{(t)}$ and the next promise $g′$ that must be made to the receiver from the next period $t+1$ onward. The first constraint in the above optimization problems is to incentivize the receiver to obey the recommendation by setting a promised payoff that is at least $g$, where the promised payoff of $g$ is calculated from the current period $t$. The second constraint is the dynamic persuasiveness, 

%The key idea of the dynamic program is to keep one ``state” variable: the amount of future utility $g$ promised to the receiver starting from the current period $t$. In each period, we choose today’s recommendation and tomorrow’s promise so that, after seeing today’s recommendation, the receiver’s incentive to deviate today is outweighed by the continuation utility promised in the future, while deviation will trigger the punishment of getting no information. Expressing the dynamic persuasiveness constraint as \eqref{eq:DP-dynamic-persuasive-constraint} separates the current posterior deviation gain from the future continuation, and our normalization simply subtracts the uninformed continuation from both sides. The dynamic program then rolls these promises forward: from the last period back to the first, it picks the best plan that keeps promises feasible and obedience attractive at every step. 

The following Lemma \ref{lem:DP-characterization} shows that the dynamic program above computes the dynamic benchmark correctly, with its full proof in Appendix \ref{proof:lem:DP-characterization}.

\begin{lemma}[Dynamic program for dynamic benchmark]
\label{lem:DP-characterization}
The dynamic program \eqref{eq:opt-final}-\eqref{eq:DP-period-t-form-2} computes an optimal solution $(\sigma^{*}, \phi_{\ob})$ that reaches the dynamic benchmark $\mathcal U^{**}_T(\mathcal I)$, where $\sigma^{*} = (\pi_{\sigma^{*}}^{(t)})_{t=1}^T$ is an optimal DDP strategy that is history-independent and dynamically persuasive. 
\end{lemma}

\subsection{Examples of Optimal DDP Strategies}
\label{subsec:example}
%\textcolor{red}{I think the example will have more impact if you do an example where $\mu_D=\mu_R$.}
Here we illustrate the structure of the optimal history-independent dynamically persuasive DDP strategy that attains the dynamic benchmark $\mathcal U^{**}_T$. 
We consider an example with three states $\Omega =\{\omega_1, \omega_2, \omega_3\}$ and three actions $A = \{a_1, a_2, a_3\}$.
The payoffs of the designer and receiver are 
%as follows (rows for actions, columns for states):
\begin{align*}
\begin{tabular}{c|ccc}
\text{designer's $u$} & $\omega_1$ & $\omega_2$ & $\omega_3$ \\ \hline
$a_1$       & $1$        & $1$        & $1$        \\
$a_2$       & $1/2$        & $1/2$        & $1/2$        \\
$a_3$       & $0$        & $0$        & $0$       
\end{tabular}
\qquad 
\begin{tabular}{c|ccc}
\text{receiver's $v$} & $\omega_1$ & $\omega_2$ & $\omega_3$ \\ \hline
$a_1$       & $1$        & $0$        & $0$        \\
$a_2$       & $0$        & $1$        & $0$        \\
$a_3$       & $0$        & $0$        & $1$       
\end{tabular}.
\end{align*}
In words, the designer prefers $a_1$ the most, $a_2$ the second, $a_3$ the least, regardless of the state, while the receiver wants to match the state.
The designer and the receiver have a common prior $\dprior = \rprior = (0.2, 0.3, 0.5)$. 
% The designer's prior belief is $\dprior = (0.2, 0.3, 0.5)$, 
%is the uniform distribution over the states. The 
% and the receiver's prior belief is $\rprior = (0.2, 0.3, 0.5)$.
%, where he believes $\omega_3$ is most likely, so he will take $a_3$ by default. 
We will compare the optimal DDP strategy against the optimal signaling scheme $\pi^*_\BP$ in the static Bayesian persuasion problem
\begin{align*}
\pi^*_\BP =
\begin{tabular}{c|ccc}
 & $\omega_1$ & $\omega_2$ & $\omega_3$ \\ \hline
$a_1$       & 1 & 0 & 0.4        \\
$a_2$       & 0 & 1 & 0.6        \\
$a_3$       & 0 & 0 & 0       
\end{tabular},
\end{align*}
which provides the designer and the receiver with expected payoffs 
\begin{equation}
    \label{eq:example-static-BP-benchmark}
    U^*_\BP = U(\pi^*_\BP, \rho_\ob; \dprior) = 0.7, \qquad V(\pi^*_\BP, \rho_\ob; \rprior) = 0.5 = V_\uninformed(\rprior). 
\end{equation}
% respectively.
For different choices of $T$ and $\gamma$, we compute an optimal DDP strategy $\sigma^* = (\pi_{\sigma^*}^{(t)})_{t=1}^T$ for the designer using the dynamic programming in Lemma \ref{lem:DP-characterization} and present the results below.

\subsubsection{Optimal DDP strategy for $T=2$ periods}
We start with the simple case where the total number of periods is $T = 2$ and the receiver is patient (having discount factor $\gamma = 1$).
In this case, the optimal DDP strategy $\sigma^* = (\pi_{\sigma^*}^{(1)}, \pi_{\sigma^*}^{(2)})$ for the designer is the following: 
\begin{align*}
\sigma^* = \left( \quad
\pi^{(1)}_{\sigma^*} : 
\begin{tabular}{c|ccc}
 & $\omega_1$ & $\omega_2$ & $\omega_3$ \\ \hline
$a_1$       & $1$        & $1$        & $1$        \\
$a_2$       & $0$        & $0$        & $0$        \\
$a_3$       & $0$        & $0$        & $0$      
\end{tabular}, 
\qquad 
\pi^{(2)}_{\sigma^*} : 
\begin{tabular}{c|ccc}
 & $\omega_1$ & $\omega_2$ & $\omega_3$ \\ \hline
$a_1$       & 1       & 0        & 2/5        \\
$a_2$       & 0        & 1        & 0        \\
$a_3$       & 0        & 0       & 3/5      
\end{tabular} 
\quad \right).
\end{align*}
With an obedient receiver, the designer's expected payoffs % under $\pi_{\sigma^*}^{(1)}$ and $\pi_{\sigma^*}^{(2)}$
%the two signaling schemes
in the two periods 
are
$U(\pi_{\sigma^*}^{(1)}, \rho_\ob; \dprior) = 1$ and $U(\pi_{\sigma^*}^{(2)}, \rho_\ob; \dprior) = 0.55$. 
The (average) dynamic benchmark is % given by the average payoff 
 \begin{equation*}
 \quad \frac{1}{T} \mathcal U^{**}_T = \frac{1}{T} \mathcal U_T(\sigma^*, \phi_\ob; \dprior) = \frac{1}{2}\Big( U(\pi_{\sigma^*}^{(1)}, \rho_\ob; \dprior) + U(\pi_{\sigma^*}^{(2)}, \rho_\ob; \dprior) \Big) = 0.775,
 \end{equation*}
which is larger than the static BP benchmark \eqref{eq:example-static-BP-benchmark}.
The obedient receiver's expected payoffs are $V(\pi_{\sigma^*}^{(1)}, \rho_\ob; \rprior) = 0.2$ and $V(\pi_{\sigma^*}^{(2)}, \rho_\ob; \dprior) = 0.8$,
% \begin{align*}
%     V(\pi_{\sigma^*}^{(1)}, \rho_\ob; \rprior) = 0.2, \qquad  V(\pi_{\sigma^*}^{(2)}, \rho_\ob; \dprior) = 0.8, 
% \end{align*}
with the undiscounted average 
\begin{align*}
    \qquad \frac{1}{T} \mathcal V_T(\sigma^*, \phi_\ob; \dprior) = 0.5 = V_\uninformed(\rprior). 
\end{align*}

Here are some interesting observations about the optimal DDP strategy:
\begin{itemize}
\item \emph{The signaling scheme $\pi_{\sigma^*}^{(1)}$ in the first period is not persuasive}: it always recommends action $a_1$, which is in favor of the designer but not the receiver. %, as the receiver believes that state $\omega_3$ is most likely.
An obedient receiver's payoff under $\pi_{\sigma^*}^{(1)}$, 0.2, is even lower than the uninformed payoff $V_\uninformed(\rprior) = 0.5$. 
\item The signaling scheme $\pi_{\sigma^*}^{(2)}$ in the second period is \textit{persuasive} and provides the receiver with a \textit{high ex-ante payoff} of $V(\pi_{\sigma^*}^{(2)}, \rho_\ob; \rprior) = 0.8$, which is significantly larger than the receiver's uninformed payoff $V_\uninformed(\rprior) = 0.5$. 
\item Due to the high ex-ante payoff promised by $\pi_{\sigma^*}^{(2)}$, the receiver is willing to follow the non-persuasive recommendation from $\pi_{\sigma^*}^{(1)}$ in the first period, because otherwise the receiver will suffer a large loss in the second period. \emph{Therefore, the designer's two-period strategy $\sigma^* = (\pi_{\sigma^*}^{(1)}, \pi_{\sigma^*}^{(2)})$ satisfies dynamic persuasiveness.}  
%\item The \emph{average} direct signaling scheme
 % \begin{align*}
 % \overline \pi_{\sigma^*} = \frac{1}{2}\Big( \pi_{\sigma^*}^{(1)} + \pi_{\sigma^*}^{(2)} \Big) = 
 % \begin{tabular}{c|ccc}
 %  & $\omega_1$ & $\omega_2$ & $\omega_3$ \\ \hline
 % $a_1$       & 1        & 0.5        & 0.7        \\
 % $a_2$       & 0        & 0.5       & 0        \\
 % $a_3$       & 0         & 0         & 0.3      
 % \end{tabular}
 % \end{align*}
 % is \emph{not persuasive} in the static Bayesian persuasion game.  In particular, when $\overline \pi_{\sigma^*}$ recommends action $a_1$, the receiver's posterior belief is $(0.2857, 0.2143, 0.5)$, under which the receiver's best action should be $a_3$.
\end{itemize}

In summary, the optimal two-period DDP strategy $\sigma^* = (\pi_{\sigma^*}^{(1)}, \pi_{\sigma^*}^{(2)})$ exhibits a \emph{first-exploit-then-compensate} structure.
By exploiting the receiver early and compensating him later, the designer attains a higher average payoff than the static benchmark. This strategy requires a sufficiently forward-looking receiver: if the discount factor $\gamma$ is too small, the receiver will not accept later compensation for early exploitation.
The DDP strategy $\sigma^*$ is dynamically persuasive even though the individual signaling schemes are not necessarily persuasive in the static game.

\subsubsection{Optimal DDP strategy for $T=20$ periods}
\label{sec:optimal-DDP-example-T=20}

\begin{figure}
    \centering
    \begin{subfigure}[t]{0.49\textwidth}
    \resizebox{\linewidth}{!}{
        \input{plots/designer-payoff-T=20}
    }
    % \caption{Designer's payoff}
    \end{subfigure}
    \begin{subfigure}[t]{0.49\textwidth}
    \resizebox{\linewidth}{!}{
        \input{plots/receiver-payoff-T=20}
    }
    % \caption{Receiver's payoff}
    \end{subfigure}
     \caption{Per-period payoffs of designer and receiver under the optimal DDP strategy in Section \ref{sec:optimal-DDP-example-T=20}. }
    \label{fig:optimal-DDP-example}
\end{figure}
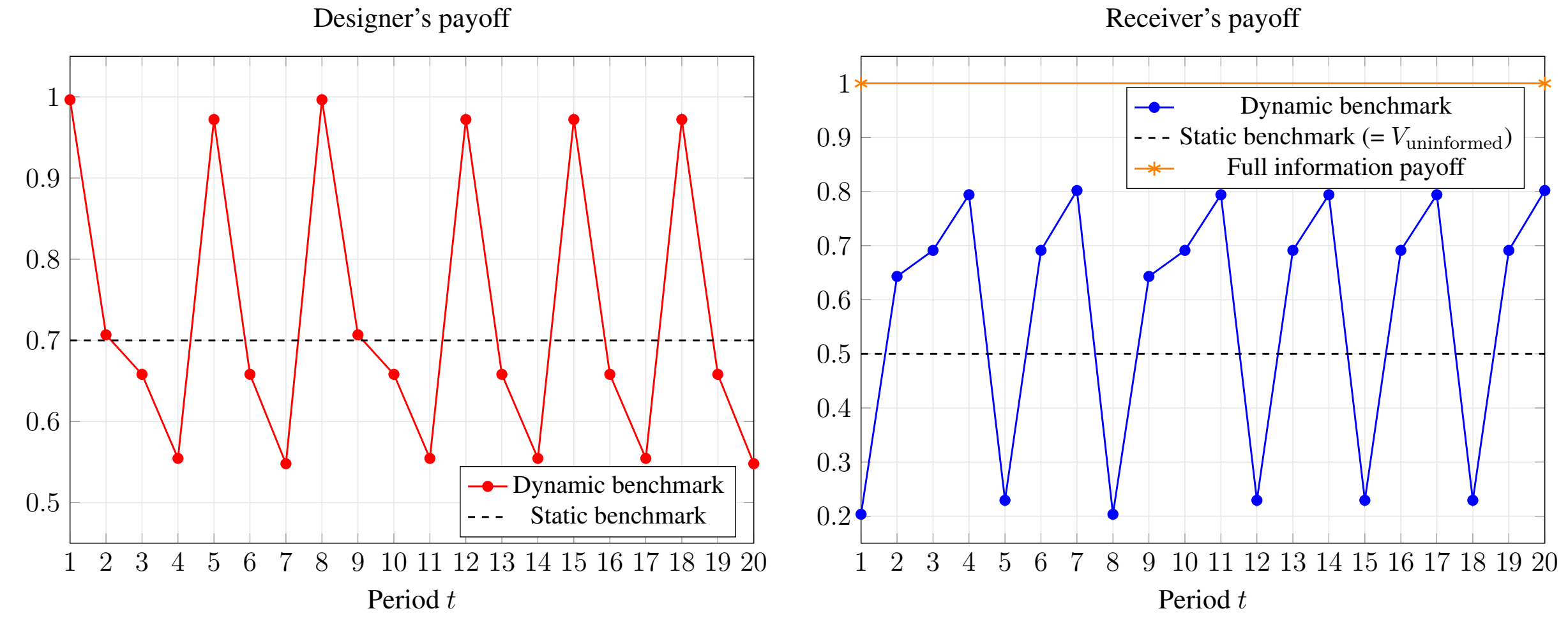

We next consider the $(T = 20)$-period case with the receiver's discount factor $\gamma = 0.7$. 
In Figure \ref{fig:optimal-DDP-example}, we plot the designer's and receiver's per-period payoffs $U(\pi_{\sigma^*}^{(t)}, \rho_\ob; \dprior)$ and $V(\pi_{\sigma^*}^{(t)}, \rho_\ob; \rprior)$ under the direct signaling schemes of the optimal DDP strategy ${\sigma^*} = (\pi_{\sigma^*}^{(t)})_{t=1}^T$, %which are
depicted in {\color{red} red} and {\color{blue} blue} solid curves, respectively. %, in Figure \ref{fig:optimal-DDP-example}.
The average payoff of the designer under the optimal DDP strategy is
% \begin{align}
    $\frac{1}{T} \mathcal U^{**}_T = \frac{1}{T} \sum_{t=1}^T U(\pi_{\sigma^*}^{(t)}, \rho_\ob; \dprior) \approx 0.7279$, 
% \end{align}
which is larger than the static BP benchmark $U^*_\BP = 0.7$. 

We observe \textit{oscillations} in both players' payoffs. % {\color{blue} blue} and {\color{red} red} curves. %, so the optimal DDP strategy's influences on the designer's payoff and the receiver's payoff are not monotone.
The oscillations can be explained by the first-exploit-then-compensate structure %of the optimal $2$-period strategy 
{and by the fact that the designer is more patient than the receiver}. To see one oscillation, as in the $2$-period case, the designer's optimal strategy is to exploit first to obtain a desired payoff and then to compensate the receiver. % to ``pay for the debt.''
The repeated oscillations arise because the receiver is %relatively
impatient while the designer is patient. If compensation is postponed by $T'$ periods, the designer must promise a payoff %with face value 
in the order of $\gamma^{-T'}$ for it to be persuasive for the receiver, which becomes prohibitively costly for the designer. Thus, %to satisfy dynamic persuasiveness,
the designer alternates short exploitation with timely compensation, keeping the receiver’s continuation payoff near the participation threshold ($V_\uninformed$).
%while avoiding a large back-loaded “debt.” 
This generates the periodic pattern in both players' payoffs.

Finally, we remark that the oscillation occurs only for an impatient but sufficiently forward-looking receiver.  If the receiver is impatient with a very small $\gamma$, then the designer's optimal strategy will be to repeat her optimal strategy in the single-period Bayesian persuasion game, without causing oscillations, as demonstrated in the next subsection.

\subsubsection{Effect of $\gamma$ on  Dynamic Benchmark}
\label{subsec:gamma}
We illustrate the effect of the receiver's discount factor $\gamma$ on the designer's dynamic benchmark $\mathcal U^{**}_T$. 
%We computed the optimal DDP strategy $\sigma^* = (\pi_{\sigma^*}^{(t)})_{t=1}^T$ for the designer in the example at the beginning of this subsection with different $\gamma \in [0, 1]$, and plotted in 
Figure \ref{fig:gamma} {plots} the designer's average payoff $\frac{1}{T} \mathcal U^{**}_T = \frac{1}{T} \sum_{t=1}^T U(\pi_{\sigma^*}^{(t)}, \rho_\ob; \dprior)$ and the receiver's normalized payoff $\frac{1}{\sum_{t=1}^T \gamma^t} \sum_{t=1}^T \gamma^t V(\pi_{\sigma^*}^{(t)}, \rho_\ob; \rprior)$ under the optimal DDP strategy {as a function of $\gamma$}.
We observe that the designer's payoff is increasing with $\gamma$, which implies that a more patient receiver leads to a higher dynamic benchmark for the designer. The receiver's normalized payoff, on the other hand, is not monotone with respect to $\gamma$.  When $\gamma$ is small (smaller than $0.5$ in this example), the dynamic benchmark $\frac{1}{T} \mathcal U^{**}_T$ coincides with the static Bayesian persuasion benchmark $U^*_\BP$.

\begin{figure}[htbp]
    \centering
    \begin{subfigure}{0.485\textwidth}
        \centering
        \includegraphics[width=\linewidth]{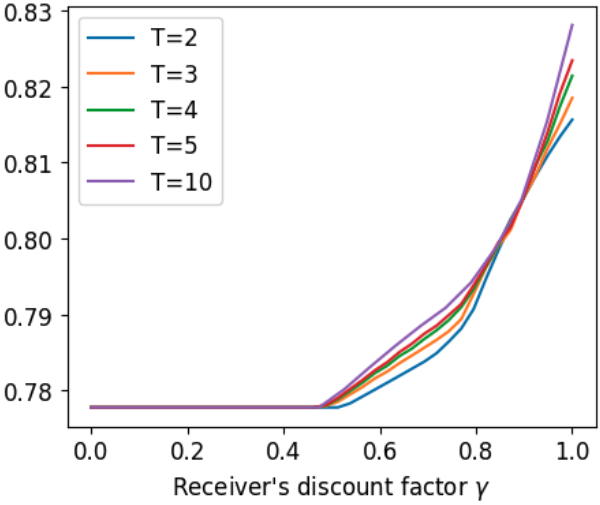}
        \caption{Designer's average payoff}
        \label{fig:gamma-designer}
    \end{subfigure}
    \hfill
    \begin{subfigure}{0.495\textwidth}
        \centering
        \includegraphics[width=\linewidth]{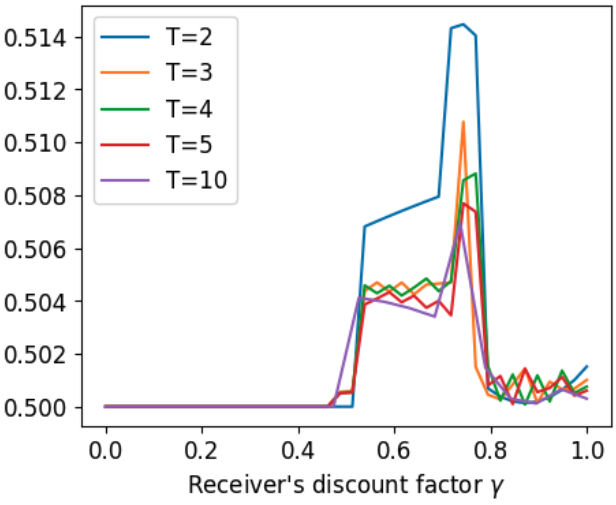}
        \caption{Receiver's normalized payoff}
        \label{fig:gamma-receiver}
    \end{subfigure}

    \caption{Designer's and receiver's payoffs under the optimal DDP strategy, as a function of receiver's discount factor $\gamma$. }
    \label{fig:gamma}
\end{figure}

\textcolor{black}{
%The numerical example illustrates that a higher receiver discount factor can strictly increase the designer's dynamic benchmark. 
More generally, the designer's dynamic benchmark is \textit{weakly} increasing in the
receiver's discount factor $\gamma$
for any fixed instance $\mathcal I$ and horizon $T$, as shown in Proposition~\ref{prop:gamma-dynamic-benchmark-increasing}.
To see why,
consider an optimal history-independent, dynamically persuasive
DDP strategy $\sigma^*_\gamma$. A deviation gives the receiver
a possible current payoff gain but causes him to lose access
to information in all subsequent periods. The current payoff
gain does not depend on his discount factor. To compare the
future loss, note that dynamic persuasiveness ensures that,
at the beginning of each period before punishment, continued
obedience gives the receiver an expected discounted payoff
at least as high as receiving no further information.
Backward induction shows that increasing
the discount factor weakly increases the discounted payoff
loss from punishment. Thus, any deviation that was not attractive under $\gamma$ remains unattractive under $\gamma'>\gamma$,
so $\sigma^*_\gamma$ remains dynamically persuasive.}
\textcolor{black}{The designer can therefore use
$\sigma^*_\gamma$ with the more patient receiver and obtain
the same payoff as before. An optimal strategy for $\gamma'$ gives the designer at least
as much payoff as continuing to use $\sigma^*_\gamma$. Thus, the dynamic
benchmark is weakly increasing in $\gamma$.}

%\textcolor{red}{I don't understand the paragraph that starts at the end of this page and finishes at the top of the next.}
% The observation that the designer's dynamic benchmark $\mathcal U^{**}_T$ is increasing with the receiver's discount factor $\gamma$ is not limited to this example. 
% %We can prove this property for any instance, as formalized in Proposition \ref{prop:gamma-dynamic-benchmark-increasing}.  The intuition is the following. 
% {Note that the} optimal DDP strategy $\sigma^*_\gamma$ for $\gamma$, which is dynamically persuasive for a receiver with discount factor $\gamma$, is also dynamically persuasive for a receiver with a larger discount factor $\gamma' > \gamma$. 
% %for the following reason: 
% With a more patient receiver, the better-than-uninformed total utility promised by the DDP strategy in the future becomes more attractive to the receiver, so the receiver is more willing to be obedient.  Since the DDP strategy $\sigma^*_\gamma$ is dynamically persuasive for $\gamma' > \gamma$, $\sigma_\gamma$ is a feasible solution to the designer's dynamic programming problem under $\gamma'$ and thus is weakly worse than the optimal strategy $\sigma^*_{\gamma'}$ that achieves the dynamic benchmark for the designer under $\gamma'$. 

\begin{proposition}
\label{prop:gamma-dynamic-benchmark-increasing}
For any instance $\mathcal I$ and any $T \ge 1$, 
the designer's dynamic benchmark $\mathcal U^{**}_T(\mathcal I)$ is weakly increasing with the receiver's discount factor $\gamma$. 
\end{proposition}
\begin{proof}
    The full proof is given in Appendix \ref{proof:prop:gamma-dynamic-benchmark-increasing}.
\end{proof}
Though the dynamic benchmark $\mathcal U^{**}_T(\mathcal I)$ is increasing {in} $\gamma$, \textit{the designer} is not always better off in the learning problem with an unknown receiver's belief $\rprior$.
The learning cost of $O(\frac{1}{1-\gamma} \log^2\frac{1}{\eps})$ shown in Lemma \ref{lemma:good-prior-estimation} is also increasing {in} $\gamma$, {as} a more patient receiver has a larger incentive to manipulate the learning process of the designer.  In fact, we will show in Section \ref{sec:lower-bound} that the designer has to suffer a linear regret of $\Omega(T)$ when the receiver is fully patient ($\gamma = 1$).

%Figure \ref{fig:gamma-designer} shows the designer's 
The designer benefits in the dynamic benchmark compared to that in the static benchmark as $T$ goes to infinity. To see that, in Figure \ref{fig:gamma-designer}, the x-axis intersects the designer's average payoff at her single-period Bayesian persuasion optimality, while the highest horizontal line intersects the designer's average payoff at her single-period optimality under IR. %For small values of $\gamma$, the designer's average payoff is the same as the single-period Bayesian persuasion optimality. But the designer's average payoff increases as $\gamma$ approaches to 1.
When $T \to \infty$, when the receiver converges to being fully patient $\gamma \to 1$, then the time-averaged dynamic benchmark converges to the designer's single-period optimality under the receiver's individual rationality constraint (for simplicity, it is named as the designer's optimality under IR), while the designer's optimality under IR is weakly larger than the designer's single-period Bayesian persuasion optimality (i.e., the designer's optimality under the constraint of the signaling scheme being persuasive for the receiver). To see why, the solution to the designer's single-period Bayesian persuasion problem is a feasible solution to her single-period optimization under the receiver's individual rationality constraint. When the receiver is not patient with $\gamma \to 0$, the time-averaged dynamic benchmark converges to the designer's single-period Bayesian persuasion optimality. %In Figure \ref{fig:gamma-designer}, the x-axis intersects the designer's average payoff at her single-period Bayesian persuasion optimality, while the highest horizontal line intersects the designer's average payoff at her single-period optimality under IR.
The idea is formally characterized in Proposition \ref{prop: IR convergence}.

\begin{proposition}
\label{prop: IR convergence}
With Assumptions \ref{ass:prior-p0} and \ref{ass:receiver-optimal-action}, the designer's dynamic benchmark satisfies \[\lim_{T\to\infty} \lim_{\gamma\to 1} \frac{1}{T} \mathcal U^{**}_T(\mathcal I; \gamma) = U^*_{\mathrm{IR}},\]
where $U^*_{\mathrm{IR}}$ is the designer's single-period optimality under the constraint of the receiver's individual rationality: 
\begin{align} \label{eq:V-uninformed-optimization}
    & \max_{\pi: \Omega \to \Delta(A)} U(\pi, \rho_\ob; \dprior) ~~ \mathrm{ s.t. } ~~ V(\pi, \rho_\ob; \rprior) \ge V_\uninformed(\rprior).
\end{align}

\end{proposition}

\begin{proof}
See Appendix \ref{proof:prop: IR convergence}.
\end{proof}

\section{Learning to Achieve No Regret Against the Dynamic Benchmark}
\label{sec:learning-dynamic-benchmark}
For the dynamic {benchmark}, the learning algorithm \textcolor{black}{again} consists of two phases. The first is an exploration phase for learning the receiver's prior belief, which is based on the same algorithm as for the static {benchmark} but with different parameters. %Then, we propose a new robustification process to ensure the receiver's participation in the learning when the designer is aiming for the dynamic benchmark. 
For the second phase, we propose a new robustification process, \emph{dynamic robustification}, to robustify a designer-optimal strategy computed from the learned prior, which ensures approximate optimality for the designer under the receiver's belief $\rprior$.
In this section, we first study the dynamic robustification based on the estimated belief. Then, we show the full algorithm to achieve no regret for the dynamic benchmark.
\subsection{Dynamic Robustification for DDP Strategy}
\label{subsec:dynamic-robustification}
Denote by $\hat \sigma = (\pi_{\hat \sigma}^{(1)}, \ldots, \pi_{\hat \sigma}^{(T)})$ the designer's optimal DDP strategy, which achieves the dynamic benchmark $\mathcal U^{**}_T(\Omega, A, u, v, \dprior, \hat \mu)$ under belief $\hat \mu$. Proposition \ref{prop:dynamic-benchmark-DDP} implies that the strategy $\hat \sigma$ is history-independent and dynamically persuasive for $\hat \mu$. We show that $\hat \sigma$ can be \textit{dynamically robustified} into another DDP strategy $\tilde \sigma = (\pi_{\tilde \sigma}^{(1)}, \ldots, \pi_{\tilde \sigma}^{(T)})$ that is history-independent and dynamically persuasive for a strategic receiver with any belief $\rprior$ satisfying $\| \rprior - \hat \mu \|_1 \le \eps$.

We first show how to convert any signaling scheme $\pi_{\hat \sigma}^{(t)}$ for any period $t$ from the designer's optimal DDP strategy $\hat \sigma$ under belief $\hat \mu$ to a corresponding signaling scheme $\pi_{\tilde \sigma}^{(t)}$. 
%that is %persuasive under belief $\rprior$ and 
%approximately optimal for the {designer}. 

A natural idea is to apply static robustification (Lemma \ref{lem:universal-robustification})
to the signaling scheme $\pi_{\hat \sigma}^{(t)}$ in each period $t$ in the DDP strategy $\hat \sigma$. Recall that static robustification mixes over signaling schemes and thereby
increases the \textit{persuasiveness} of the statically robustified signaling scheme under the receiver's prior belief $\rprior$, i.e.,
%, namely, increasing the \emph{payoff gap} between the recommended action and any other action, under the receiver's \textit{posterior} beliefs formed by different priors and different signaling schemes:   
\begin{equation*}
        \E_{\omega \sim  \rprior (\cdot | a, \tilde \pi^{(t)})} \big[ v(a, \omega) - v(a', \omega) \big]
        \ge \min\big\{\E_{\omega \sim \hat  \mu (\cdot | a,  \pi_{\hat \sigma}^{(t)})}\big[ v(a, \omega) - v(a', \omega) \big] + O(\delta), 0 \big\}, ~\forall a' \in A\setminus\{a\},  
\end{equation*}
where $\tilde \pi^{(t)}$ is the signaling scheme statically robustified from $\pi_{\hat \sigma}^{(t)}$. Thus, the persuasiveness of the statically robustified signaling scheme improves only after the receiver receives a recommendation. \textit{Before} the receiver receives a recommendation, however, static robustification may lower the receiver’s \textit{ex ante}  payoff from obedience, i.e., $V(\tilde \pi^{(t')}, \rho_\ob; \rprior) < V(\pi_{\hat \sigma}^{(t')}, \rho_\ob; \rprior)$ for any $t' > t$. In period $t$, after a recommendation is sent by the statically robustified signaling scheme, the receiver's incentive to obey the recommendation in the current period may become stronger. But the receiver's \textit{ex ante} payoff from obeying the recommendations in \textit{future periods onward} may get worse. Thus, on balance, he may refuse to obey the current recommendation and leave the game. In that way, a DDP strategy obtained by statically robustifying the signaling scheme in each period is not guaranteed to be dynamically persuasive.
{Hence}, we propose %\textit{dynamic robustification},
a new type of robustification, \emph{payoff-improving robustification,} which 
%shows its advantage in 
not only {increases} the persuasiveness of an unrobustified signaling scheme but also weakly {increases} the receiver's \textit{ex ante} payoff from being obedient. 

{Given} a typical signaling scheme $\pi_{\hat \sigma}$ in a DDP strategy $\hat \sigma$, the main idea of the payoff-improving robustification is as follows. We first compute the signaling scheme $\pi_{\tilde \sigma}$ that is obtained from static robustification (Lemma \ref{lem:universal-robustification}) on $\pi_{\hat \sigma}$. Then, we mix $\pi_{\tilde \sigma}$ with the full-revelation signaling scheme $\pi_\full$, which always recommends the optimal action for the receiver at each state.  In particular, if $\pi_{\tilde \sigma}$ is already close to $\pi_\full$ (with distance less than $O(\sqrt{\delta})$), we directly change $\pi_{\tilde \sigma}$ to $\pi_\full$. 
If $\pi_{\tilde \sigma}$ is far from $\pi_\full$ (with distance more than $O(\sqrt{\delta})$), we take a mixture $(1-O(\sqrt{\delta})) \pi_{\tilde \sigma} + O(\sqrt{\delta}) \pi_\full$. Such a mixture weakly improves the receiver's utility (compared to $\pi_{\hat \sigma}$), improves the persuasiveness, while hurting the designer's utility by at most $O(\sqrt{\delta})$.

Lemma \ref{lem:utility-improving-robustification} shows the result of any signaling scheme from the dynamic robustification in a DDP strategy $\hat \sigma$. We omit the time index of the signaling scheme for simplicity. See Appendix \ref{proof:lem:utility-improving-robustification} for the full proof of Lemma \ref{lem:utility-improving-robustification}.\footnote{{Lemma \ref{lem:utility-improving-robustification} can also be used to perform robustification to achieve no-regret against the static benchmark, as Lemma \ref{lem:universal-robustification} does. However, Lemma \ref{lem:utility-improving-robustification} is derived from and conceptually more complex than Lemma \ref{lem:universal-robustification}. Lemma \ref{lem:utility-improving-robustification} gives worse (up to constant factors) results for static robustification than those given by Lemma \ref{lem:universal-robustification}.
% , but the results are only worse up to constant terms.
Thus, we keep using Lemma \ref{lem:universal-robustification} for static robustification and use Lemma \ref{lem:utility-improving-robustification} for dynamic robustification.}}   

\begin{lemma}[Payoff-improving  robustification]
\label{lem:utility-improving-robustification}
Assume Assumptions  \ref{ass:prior-p0}, \ref{ass:D}, {and} \ref{ass:receiver-optimal-action}. 
%For any receiver's belief $\mu$  Choose $\eps \le \frac{p_0^4 D G}{16}$. 
For any signaling scheme $\pi_{\hat \sigma}$ of the designer's optimal DDP strategy under belief $\hat \mu$, with any belief $\mu$ satisfying $\| \mu - \hat \mu \|_1 \le \eps \le \frac{p_0^4 D G}{16}$, with a dynamic robustification parameter $\delta \in [\frac{2\eps}{p_0D}, \frac{p_0^3G}{8}]$, the signaling scheme $\pi_{\hat \sigma}$ can be converted into another signaling scheme $\pi_{\tilde \sigma}$ such that: 
\begin{itemize}
    \item (improved persuasiveness) the persuasiveness of $\pi_{\tilde \sigma}$ and $\pi_{\hat \sigma}$ for the receiver under beliefs $\mu$ and $\hat \mu$, respectively, are close:% $\mu (\cdot | a, \tilde \pi)$, $\hat \mu (\cdot | a, \hat\pi)$ satisfy: 
    \begin{equation*}
        \E_{\omega \sim  \mu (\cdot | a, \pi_{\tilde \sigma})} \big[ v(a, \omega) - v(a', \omega) \big]
        \ge \min \Big\{\E_{\omega \sim \hat \mu (\cdot | a, \pi_{\hat \sigma})}\big[ v(a, \omega) - v(a', \omega) \big] + \delta D - \tfrac{2\eps}{p_0}, ~ 0 \Big\}, %~~ \forall a' \in A\setminus\{a\},
    \end{equation*}
    for any action $a \in A$ recommended by $\pi_{\tilde \sigma}$ and $a' \in A\setminus\{a\}$;
    \item (weakly improving the payoff of the receiver) the receiver's payoff from obedience is weakly increased:  $V(\pi_{\tilde \sigma}, \rho_\ob; \mu) \ge V(\pi_{\hat \sigma}, \rho_\ob; \mu)$; 
    \item ({small} influence on the payoff of the designer) with an obedient receiver, the designer's payoff has a {small} decrease: $U(\pi_{\tilde \sigma}, \rho_\ob; \dprior) \ge U(\pi_{\hat \sigma}, \rho_\ob; \dprior) - 4 \sqrt{\tfrac{\delta}{p_0^3 G}}$. 
\end{itemize} 
\end{lemma}

%\begin{proof}[Proof sketch]
We then apply the above payoff-improving robustification procedure to every signaling scheme $\pi_{\hat \sigma}^{(t)}$ in the designer-optimal DDP strategy $\hat \sigma$ for belief $\hat \mu$, a process we call \emph{dynamic robustification}.
The result of the complete dynamic robustification is {given} in Theorem \ref{thm:dynamic-robustification}.

\begin{theorem}[Dynamic robustification]
\label{thm:dynamic-robustification}
Assume Assumptions  \ref{ass:prior-p0}, \ref{ass:D}, {and} \ref{ass:receiver-optimal-action}. 
Given the designer's optimal DDP strategy $\hat \sigma$ under belief $\hat \mu$, with any belief $\rprior$ satisfying $\| \rprior - \hat \mu \|_1 \le \eps$ for $\eps \le \frac{p_0^4}{1+ p_0\gamma/ (1-\gamma)}\frac{DG}{16}$,
there exists another history-independent DDP strategy $\tilde \sigma = (\pi_{\tilde \sigma}^{(1)}, \ldots, \pi_{\tilde \sigma}^{(T)})$ such that
\begin{itemize}
    \item $\tilde \sigma$ is dynamically persuasive under belief $\rprior$; 
    \item $\tilde \sigma$ is approximately optimal for the designer with
    %the designer's $T$-period payoff is approximately optimal with
    \[ \mathcal U_T(\tilde \sigma, \phi_\ob; \dprior) ~ \ge ~ \mathcal U_T(\hat \sigma, \phi_\ob; \dprior) - 4T \sqrt{\big(\tfrac{1}{p_0} + \tfrac{\gamma}{1-\gamma}\big) \tfrac{2\eps}{p_0^3 G D}}.\]
\end{itemize}
\end{theorem}

We now give the intuition for Theorem \ref{thm:dynamic-robustification}.
The high-level idea of our dynamic robustification procedure is to robustify the signaling scheme of each period $t$ in the DDP strategy $\hat \sigma$ backward from $T$. In particular, for each period $t=T, T-1, \ldots, 1$, we convert $\pi_{\hat \sigma}^{(t)}$ to another signaling scheme $\pi_{\tilde \sigma}^{(t)}$ by Lemma~\ref{lem:utility-improving-robustification} while treating the dynamically robustified tail $(\pi_{\tilde \sigma}^{(t+1)},\ldots,\pi_{\tilde \sigma}^{(T)})$ as fixed. % \textcolor{red}{Intuitively, each step buys a \emph{safety buffer} $\delta_t D$ in the one-shot incentive at $t$ under the perturbed belief $\rprior$, while paying at most a small (square-root) utility tax for the designer.}

Given a recommended action $a$ at time $t$ and any deviation $a'$, Lemma~\ref{lem:utility-improving-robustification} ensures the persuasiveness of $\pi_{\tilde \sigma}^{(t)}$ under the belief $\rprior$ after robustification compared to the persuasiveness of $\pi_{\hat \sigma}^{(t)}$ under the estimate $\hat \mu$ by
\[
\E_{\omega\sim \rprior(\cdot\mid a,\pi_{\tilde \sigma}^{(t)})}\!\big[v(a,\omega)-v(a',\omega)\big]
~\ge~
\min\!\Big\{\E_{\omega\sim \hat\mu(\cdot\mid a,\pi_{\hat \sigma}^{(t)})}\!\big[v(a,\omega)-v(a',\omega)\big]
+\delta_t D-\tfrac{2\eps}{p_0},\,0\Big\},
\]
where a \emph{buffer} $\delta_t D$ and a loss from the belief estimate %\emph{belief–mismatch penalty} 
$\frac{2\eps}{p_0}$ are exerted on the persuasiveness of $\pi_{\hat \sigma}^{(t)}$ under the estimate $\hat \mu$. %Lemma~\ref{lem:utility-improving-robustification} \textcolor{red}{also says the receiver’s payoff from obedience weakly increases when we robustify in later periods,} and it bounds the designer’s one-period utility loss by $4\sqrt{\delta_t/(p_0^3 G)}$.

But that is not enough. Then, we consider the dynamic persuasiveness constraint to secure the receiver's obedience in period $t$. The dynamic persuasiveness in period $t$ is

\[
\underbrace{\gamma^t \E_{\omega\sim \rprior(\cdot\mid a,\pi_{\tilde \sigma}^{(t)})}[v(a,\omega)-v(a',\omega)]}_{\text{today’s incentive}}
+\underbrace{\sum_{t'=t+1}^{T}\gamma^{t'}\!\big(V(\pi_{\tilde \sigma}^{(t')},\rho_{\ob};\rprior)-V_{\uninformed}(\rprior)\big)}_{\text{discounted future payoffs from obedience}}
~\ge~0. \tag{DP$_t$}
\]
There are two intuitive cases. First, when the persuasiveness of $\pi_{\tilde \sigma}^{(t)}$ under  $\rprior$ is already non-negative, then the (already robustified) tail is dynamically persuasive, so the future accumulated payoffs from obedience is non-negative as well. Thus, the dynamic persuasiveness constraint (DP$_t$) holds immediately.

The second case is that the persuasiveness of $\pi_{\tilde \sigma}^{(t)}$ under  $\rprior$ is already negative. Then we consider the help from the already robustified tail, which comes from two sources: (i) robustifying the tail cannot hurt the receiver (Lemma~\ref{lem:utility-improving-robustification}), and
(ii) $V(\cdot,\cdot;\cdot)$ is $1$-Lipschitz in the belief (Lemma~\ref{lem:utility-Lipschitz}),
so the gap from $\hat\mu$ to $\rprior$ costs at most $2\eps$ per period. Putting the contribution from those two sources of the tail, the left-hand side of the dynamic persuasiveness constraint (DP$_t$) is at least $\gamma^t\! \big(\delta_t D-\tfrac{2\eps}{p_0}\big)-\tfrac{2\eps\,\gamma^{t+1}}{1-\gamma}.$ Choosing the dynamic robustification parameter $\delta_t \;=\; \big(\tfrac{1}{p_0}+\tfrac{\gamma}{1-\gamma}\big)\,\frac{2\eps}{D}$, we obtain a non-negative right-hand side of (DP$_t$), so the dynamic persuasiveness constraint (DP$_t$) holds.

Starting from $t=T$ and moving backward, the same argument secures (DP$_t$) for each period $t$.
Thus the completely dynamically robustified strategy $\tilde\sigma$ is dynamically persuasive for every $\rprior$ with $\|\rprior-\hat\mu\|_1\le\eps$.

By the upper bound of the cost from each robustification at each period $t$ on the designer,
$4\sqrt{\frac{\delta_t}{p_0^3 G}}$ (Lemma~\ref{lem:utility-improving-robustification}),
as well as our choice of $\delta_t$ above, we have
\[
\mathcal U_T(\tilde\sigma,\phi_{\ob};\dprior)
~\ge~
\mathcal U_T(\hat\sigma,\phi_{\ob};\dprior)
-4T\,\sqrt{\Big(\tfrac{1}{p_0}+\tfrac{\gamma}{1-\gamma}\Big)\,\frac{2\eps}{p_0^3 G D}}.
\]
See the full proof in Appendix \ref{proof:thm:dynamic-robustification}.

\subsection{Full Learning Algorithm with $O(\log^2 T)$ Regret}

The full algorithm for the learning performance {relative to the dynamic benchmark} is shown in Algorithm \ref{alg:strategic-benchmark}. Algorithm \ref{alg:strategic-benchmark} has two phases: a short \emph{learn-the-prior} phase and a long \textcolor{black}{\emph{exploitation}} phase. The first phase is learning the belief $\rprior$, which is the same as in Algorithm \ref{alg:known-prior}, but with different choices of parameters. The second phase is the dynamic robustification.

\vspace{0.2em}
\begin{algorithm}[H]
\caption{Learning Algorithm for the Dynamic Benchmark}
\label{alg:strategic-benchmark}

\SetKwInOut{Input}{Input}
\SetKwInOut{Output}{Output}
\SetKwInOut{Parameter}{Parameter}

% \Input {Utility functions $u, v$. Total number of periods $T$.}
\Parameter {$\eps > 0$}

\DontPrintSemicolon

Run Algorithm \ref{alg:learn-prior} with parameter $\eps$ to obtain an estimate $\hat \mu$ of the receiver's belief. Let $T_0$ be the number of periods taken. \; 

Compute the optimal DDP strategy $\hat \sigma$ for the $(T-T_0)$-period game assuming that the receiver has a prior belief $\hat \mu$ (by using, e.g., the dynamic program in Lemma~\ref{lem:DP-characterization}). \; 

Apply dynamic robustification (Theorem \ref{thm:dynamic-robustification}) to $\hat \sigma$ to obtain DDP strategy $\tilde \sigma$. 
Use $\tilde \sigma$ for the remaining $T-T_0$ periods. 
\end{algorithm}

Theorem \ref{thm:stronger-benchmark} shows the regret incurred by Algorithm \ref{alg:strategic-benchmark} is at most in the order of $O(\log^2 T)$.
We first provide a brief intuition for proof here.
Algorithm \ref{alg:strategic-benchmark} has two phases: a short \emph{learn-the-prior} phase and a long \textcolor{black}{\emph{exploitation}} phase. In the learning phase we run the estimation routine until the receiver’s belief is identified up to $\ell_1$–error $\varepsilon$, which (even against a strategic receiver) takes
\[
T_0 \;=\; O\!\left(\frac{1}{1-\gamma}\,\log^2\!\frac{1}{\varepsilon} \right)
\]
% \[
% T_0 \;=\; O\!\left(\frac{1}{1-\gamma}\,\log^2\!\frac{1}{\varepsilon}
% \;+\; \frac{1}{p_0}\,\log\!\frac{1}{\varepsilon}\right)
% \]
rounds. In the exploitation phase we compute the optimal DDP plan for $\hat\mu$ and then apply \emph{dynamic robustification} so that obedience remains optimal for any belief within $\varepsilon$ of $\hat\mu$. Dynamic robustification preserves persuasiveness while incurring at most a $T\cdot O(\sqrt{\varepsilon})$ loss in the designer’s value. Choosing $\varepsilon$ in the order of $T^{-2}$ makes this second-phase loss negligible relative to $T_0$, so the total dynamic regret is driven by the learning time, yielding
% \[
% O\!\left(\frac{|\Omega|}{1-\gamma}\,\log^2 T \;+\; \frac{|\Omega|}{p_0}\,\log T\right).
% \]
\[
O\!\left(\frac{1}{1-\gamma}\,\log^2 T \right).
\]

\begin{theorem}
\label{thm:stronger-benchmark}
Assume Assumptions~\ref{ass:prior-p0}, \ref{ass:D}, \ref{ass:receiver-optimal-action}, \ref{ass:two-states-different-actions}. 
Choose $\eps = \frac{p_0^4}{1+ p_0\gamma/ (1-\gamma)}\frac{DG}{16 T^2}$ with $\gamma \in (0, 1)$.  The dynamic regret of Algorithm~\ref{alg:strategic-benchmark} is at most
\begin{align}
    \Reg^{\mathrm{dynamic}}(T; \text{Algorithm~\ref{alg:strategic-benchmark}}, \mathcal I) ~ \le ~ O\Big( \frac{1}{1-\gamma} \log^2 T 
    %\frac{|\Omega|}{1-\gamma} \log^2 T 
    %+ \frac{|\Omega|}{p_0} \log T 
    \Big).
\end{align}
%where $C_\mathcal{I}$ is an instance-dependent constant {\color{red}{depending on $\mathcal{I}=\{u,v,\rprior,\dprior\}$.}} 
\end{theorem}

\begin{proof}
The proof will use the following 
% We first construct
Lemma~\ref{lem:larger-than-T0} (with proof given in Appendix \ref{app:larger-than-T0}) to compare the dynamic benchmarks for $T$-period game and $(T-T_0)$-period game.

\begin{lemma}
\label{lem:larger-than-T0}
For any integers $T \ge T_0 \ge 0$ and beliefs $\mu_D,\hat \mu$,
$$\mathcal U^{**}_{T-T_0}(\dprior, \hat \mu) \ge \mathcal U^{**}_{T}(\dprior, \hat \mu) - T_0.$$
\end{lemma}

By the dynamic robustification result (Theorem \ref{thm:dynamic-robustification}), $\tilde \sigma$ is dynamically persuasive for a receiver with prior $\rprior$ and approximately optimal for the designer in the $(T - T_0)$-period game under a prior $\hat \mu$:
\begin{align*}
     \mathcal U_{T-T_0}(\tilde \sigma, \phi_\ob; \dprior) & ~ \ge ~ \mathcal U_{T-T_0}(\hat \sigma, \phi_\ob; \dprior) - 4(T-T_0) \sqrt{\big(\tfrac{1}{p_0} + \tfrac{\gamma}{1-\gamma}\big) \tfrac{2\eps}{p_0^3 G D}} \\
     & ~ = ~ \mathcal U^{**}_{T-T_0}(\dprior, \hat \mu) - 4(T-T_0) \sqrt{\big(\tfrac{1}{p_0} + \tfrac{\gamma}{1-\gamma}\big) \tfrac{2\eps}{p_0^3 G D}} 
\end{align*}
where the equality is because $\hat \sigma$ is optimal for the $(T-T_0)$-period game under receiver prior $\hat \mu$.  Using Lemma \ref{lem:larger-than-T0}, the total utility of the designer in $T$ periods is at least
\begin{align*}
     \mathcal U_T ~ \ge ~ 0 + \mathcal U_{T-T_0}(\tilde \sigma, \phi_\ob; \dprior) & ~ \ge ~ \mathcal U^{**}_{T}(\dprior, \hat \mu) - T_0 - 4(T-T_0) \sqrt{\big(\tfrac{1}{p_0} + \tfrac{\gamma}{1-\gamma}\big) \tfrac{2\eps}{p_0^3 G D}}. 
\end{align*}
Applying dynamic robustification (Theorem \ref{thm:dynamic-robustification}) to the optimal DDP strategy under receiver prior $\rprior$, we obtain an approximately optimal DDP strategy under receiver prior $\hat \mu$, so $\mathcal U_T^{**}(\dprior, \hat \mu) \ge \mathcal U_T^{**}(\dprior, \rprior) - 4T \sqrt{\big(\tfrac{1}{p_0} + \tfrac{\gamma}{1-\gamma}\big) \tfrac{2\eps}{p_0^3 G D}}$. 
Thus, we have
\begin{align*}
     \mathcal U_T & ~ \ge ~ \mathcal U_T^{**}(\dprior, \rprior) - 4T \sqrt{\big(\tfrac{1}{p_0} + \tfrac{\gamma}{1-\gamma}\big) \tfrac{2\eps}{p_0^3 G D}} - T_0 - 4(T-T_0) \sqrt{\big(\tfrac{1}{p_0} + \tfrac{\gamma}{1-\gamma}\big) \tfrac{2\eps}{p_0^3 G D}} \\
     & ~ \ge ~ \mathcal U^{**}_{T}(\dprior, \rprior) - T_0 - 8T \sqrt{\big(\tfrac{1}{p_0} + \tfrac{\gamma}{1-\gamma}\big) \tfrac{2\eps}{p_0^3 G D}}. 
\end{align*}Choose $\eps = \frac{p_0^4}{1+ p_0\gamma/ (1-\gamma)}\frac{DG}{16 T^2}$, which satisfies the requirement for $\eps$ in Theorem \ref{thm:dynamic-robustification}.
By Lemma \ref{lemma:good-prior-estimation}, $T_0 \le  2|\Omega|\Big(\tfrac{1}{p_0} + \tfrac{1}{1-\gamma}\ln\big(\tfrac{24|\Omega|}{(1-\gamma) G^4 p_0^6 \eps} \big) \Big) \log_2 \tfrac{24|\Omega|}{G^2 p_0^4 \eps}$. So the designer's regret is at most
\begin{align*}
     & \Reg^{\mathrm{dynamic}}(T; \text{Algorithm~\ref{alg:strategic-benchmark}}, \mathcal I) ~ = ~ T_0 + 8T \sqrt{\big(\tfrac{1}{p_0} + \tfrac{\gamma}{1-\gamma}\big) \tfrac{2\eps}{p_0^3 G D}} \\
     & ~ \le ~ 2|\Omega|\Big(\tfrac{1}{p_0} + \tfrac{1}{1-\gamma}\ln\big(\tfrac{24|\Omega|}{(1-\gamma) G^4 p_0^6 \eps} \big) \Big) \log_2 \tfrac{24|\Omega|}{G^2 p_0^4 \eps} + 8T \sqrt{\big(\tfrac{1}{p_0} + \tfrac{\gamma}{1-\gamma}\big) \tfrac{2\eps}{p_0^3 G D}} \\
     & ~ = ~ O\Big( \frac{1}{1-\gamma} \log^2 T  \Big).
     %& ~ = ~ O\Big( \frac{|\Omega|}{1-\gamma} \log^2 T + \frac{|\Omega|}{p_0} \log T \Big).
\end{align*}
\end{proof}

%Interestingly, our two learning algorithms, Algorithm \ref{alg:known-prior} and \ref{alg:strategic-benchmark} incur regrets both in the order of $O (\log^2 T)$, against the static benchmark and the dynamic benchmark, respectively. The superiority of both of our learning algorithms is reflected by the logarithmic upper bounds $O (\log^2 T)$ on their regrets. That implies that the designer's average payoff converges to the optimality benchmark at a fast speed, in the order of $O(\frac{\log^2 T}{T})$.

%\textcolor{red}{As previously mentioned, it would be better to say that this is interesting because it's showing that the learning speed for the dynamic is no slower than for the static, something that one might not expect.}
Interestingly, the regret bounds of our two learning algorithms, Algorithms \ref{alg:known-prior} and \ref{alg:strategic-benchmark} (against the static benchmark and the dynamic benchmark, respectively) are both logarithmic in time horizon: $O (\log^2 T)$. {While empirical estimation by observing the frequency distribution of states is not relevant here, as the receiver's belief is subjective, we note that our algorithm gives faster learning. Learning a distribution via empirical estimation leads to regret at most $O(\sqrt{T})$, which is exponentially worse than our regret of $O (\log^2 T)$.}
\section{Necessity of Receiver's Discount: Lower Bound on Regrets}
\label{sec:lower-bound}

% ====
% SP game
% theorem 4: regret defined over the myopic benchmark
% lemma: regret over strategic benchmark, equivalence
% ====

% ====
% SP game

% lemma:  Myopic benchmark = strategic benchmark.

% theorem 4: $\Reg^m(T) = \Reg^s(T) \ge \Omega(1/(1-\gamma)$
% ====

% ====
% theorem 4: there exists an instance where: (1) the myopic benchmark and the strategic benchmark are equal;
% (2) the designer's regret $\Reg^m(T) = \Reg^s(T) \ge \Omega(1/(1-\gamma)$. 

% 6.1 Construction of the instance: Proof of Theorem 4

% SP game

% lemma: Myopic benchmark = strategic benchmark. 

% Lemma: Regret >= $T_\gamma$
% ====

In this section, we prove that the assumption that the receiver has a discount factor of $\gamma < 1$ is necessary for the designer to achieve no regret asymptotically (or, sub-linear regret). 
In particular, if the receiver has a discount factor $\gamma = 1$, then no matter what learning algorithm is used by the designer, there always exists an information design instance such that the designer's regret is at least a constant times $T$. 

{We now introduce the notion of \emph{regret lower bound}.
A regret lower bound means that \textit{for any algorithm {$\Alg$},} 
there exists some instance $\mathcal I$, i.e., for some unknown prior $\rprior$,
such that
\begin{equation*}
    \Reg(T; {\Alg}, \mathcal I) ~ \ge ~ c_{\mathcal I} \cdot g(T) ~ = ~ \Omega( g(T)), \quad \forall T,
\end{equation*}
for some instance-dependent constant $c_{\mathcal I}>0$.
The function $g(T)$ captures the unavoidable cost of not knowing $\rprior$ no matter what learning algorithm is used, % at the beginning of the algorithm's running,
which fundamentally shows how hard the problem of information design with an unknown prior is. }

{Denote by $T_\gamma = \sum_{t=1}^T \gamma^t$ the receiver's \emph{effective horizon}. Theorem \ref{thm:lower-bound} shows that, with respect to both static benchmark and dynamic benchmark, %the worst-case
the regret of any learning algorithm $\Alg$ for the designer is lower bounded by a \textit{linear} function of the receiver's effective horizon. }

%\st{In fact, we will prove the result for general discount factor sequences besides exponential discounting. Let $(\gamma^t)_{t=1}^T$ be any weakly decreasing discount factor sequence of the receiver. Let $T_\gamma = \sum_{t=1}^T \gamma^t$ be the receiver's \emph{effective horizon}. The following theorem shows that the worst-case regret of any learning algorithm $\Alg$ for the designer must be linear in the receiver's effective horizon, with respect to both the static benchmark and the dynamic benchmark.} 

\begin{theorem}[Regret lower bound for both benchmarks]
 \label{thm:lower-bound}
 %\st{Let $(\gamma^t)_{t=1}^T$ be any weakly decreasing sequence in $[0, 1]$, with $T_\gamma = \sum_{t=1}^T \gamma^t$.}
 For any learning algorithm $\Alg$, there exists an instance $\mathcal I$ with equal static and dynamic benchmarks $\mathcal U^*_T(\mathcal I) = \mathcal U^{**}_T(\mathcal I)$ such that the designer's regret with respect to both benchmarks is at least $\Omega(T_\gamma)$.  In particular, 
 \begin{itemize}
    \item for $\gamma \in [0, 1 - \frac{1}{T})$, the designer's regret is at least $\Omega(\frac{1}{1-\gamma})$. 
    \item for $\gamma \in [1 - \frac{1}{T}, 1]$, the designer's regret is at least $\Omega(T)$. 
\end{itemize}

 \end{theorem}

{We remark that the instance above satisfies the regularity assumptions in Theorem \ref{thm:prior-aware}, which implies that the designer's regrets come from the receiver's strategic decision-making instead of irregularities. Theorem \ref{thm:lower-bound} implies that the designer cannot obtain sub-linear regret against a patient receiver ($\gamma = 1$). {This} shows the necessity of an impatient receiver in our model.}

\subsection{Proof of the Regret Lower Bound}

% The rest of this section gives an overview of the proof \textcolor{red}{of Theorem \ref{thm:lower-bound}}. 
% \textcolor{blue}{We} construct a family of single-period information design instances that capture a single-period auction problem, where the receiver's subjective belief corresponds to the buyer's private value of the item in the auction.
% Using the fact that an auctioneer who does not know the buyer's private value suffers a constant revenue loss \citep{amin_learning_2013}, we prove that the designer, not knowing the receiver's belief, must suffer a constant regret in the single-period game.
% \textcolor{blue}{We then \textit{reduce} the $T$-period game to the single-period game to show that the designer's total regret is at least $\Omega(T_\gamma)$.}

{
The rest of this section proves Theorem \ref{thm:lower-bound}.
Our proof uses the idea of \emph{reduction}. %\footnote{This is a technique in theoretical computer science to prove the hardness of a problem.}
We will construct a family of new problems, which is a family of \emph{single-period information design games with menu}, where the designer (not knowing the receiver's prior belief) offers a menu of signaling schemes for the receiver to choose. In this single-period game, the designer's regret is defined to be the difference between the optimal payoffs the designer can obtain with and without knowing the receiver's prior. 
We then show that, \emph{if} there exists a learning algorithm for the \emph{$T$-period} information design game with regret smaller than $c_0 \cdot T_\gamma$, \emph{then} there will be a menu for the single-period game with regret smaller than $c_0$.
However, because the designer does not know the receiver's prior, we can show that no menu has regret smaller than some constant $c_0 > 0$. Therefore, no learning algorithm has regret better than $c_0 \cdot T_\gamma$, thus proving a regret lower bound of $\Omega(T_\gamma)$.  
}

% \textcolor{blue}{
% \begin{definition}
%     Let problem $A$ have a known regret lower bound $L_A(T)$. Problem $A$ \textit{reduces} to problem $B$ for regret lower bounds if there exist
%     \begin{enumerate}
%         \item a mapping that turns any instance of $A$ into an instance of $B$, and
%         \item a way to convert any algorithm $\Alg_B$ for $B$ into an algorithm $\Alg_A$ for $A$, such that for all $T$,
%         \begin{align}
%             \Reg_A(\Alg_A; T) \leq c \cdot \Reg_B(\Alg_B; T) + o(L_A(T))
%         \end{align}
%         for some constant $c>0$. Consequently, if every algorithm on $A$ incurs regret at least $L_A(T)$, then every algorithm on $B$ must incur regret at least $\frac{1}{c}L_A(T)$.
%     \end{enumerate}
% \end{definition}
% }    
% The high-level idea is to reduce a single-period auction design problem to a $T$-period information design problem where the receiver's subjective prior corresponds to the buyer's private value of the item in the auction.  It is known that an auctioneer who does not have access to the buyer's private value suffers some revenue loss \cite{amin_learning_2013}, and such a loss translates to the regret of any learning algorithm in the information design problem.

\paragraph{\bf Single-period information design problem with menu} 
%\textcolor{red}{I think this proof should be moved to the appendix. It is kind of interesting but really breaks up the flow of the paper.}
To prove Theorem \ref{thm:lower-bound}, we {start by constructing a family of single-period information design problems \emph{with menus}.}  {At a high level, these problems} capture single-buyer single-seller auctions.  
Suppose there are two states $\Omega = \{\omega_0, \omega_1\}$ and two actions $A = \{a_0, a_1\}$.
% For example, $\omega_0$ could represent a bad state and $\omega_1$ a good state, while $a_1$ means buying the recommended financial product and $a_0$ means not buying it.
The designer always prefers the receiver to take $a_1$ regardless of the state, i.e.,
\begin{equation}
u(a_1, \omega) = 1, \quad u(a_0, \omega) = 0, \quad \forall \omega \in \Omega.
\end{equation}
% For instance, an online financial platform, acting as the designer, always prefers that the user buys its recommended product.
% The receiver’s utility function reflects that buying in the good state ($\omega_1$) brings the highest utility of 1, while not buying yields 0. In the bad state ($\omega_0$), however, not buying the product is better, giving a utility of 0, whereas buying leads to a loss with utility –1.
% The receiver's utility for action $a_0$ is always $0$ and the utility for action $a_1$ is positive if the action matches the state $\omega_1$ and negative otherwise.
The receiver wants {take an action} to match the state, i.e.,%\footnote{We allow negative utility $v(a_1, \omega_0) = -1$ for convenience. One can define an equivalent instance with positive utility by adding a constant. }
\begin{equation}
    v(a_0, \omega_0) = 0, \quad v(a_0, \omega_1) = 0, \quad v(a_1, \omega_0) = -1, \quad v(a_1, \omega_1) = 1. 
\end{equation}
The designer's {prior belief} $\dprior$ is 
\begin{equation}
    \dprior(\omega_0) = 1 - p_0, \quad \dprior(\omega_1) = p_0, \quad \text{with $0 < p_0 < 1/2$}.
\end{equation}
The receiver's {prior belief} $\rprior$ is parameterized by a number $\tilde \mu_R \in[0, 1]$ that is unknown to the designer. In particular, 
\begin{equation}
    \rprior(\omega_0) = \tfrac{1}{1+\tilde \mu_R}, \quad \rprior(\omega_1) = \tfrac{\tilde \mu_R}{1+\tilde \mu_R}.
\end{equation}
%{\color{red} (REMEMBER to explain those terms, e.g., 1=buy) }

Consider a single-period game where the designer first elicits the receiver's belief and then uses a direct signaling scheme to recommend an action to the receiver. Formally, the designer designs a \emph{menu}, which is a family of direct signaling schemes $\mathcal M = \{ \pi_{\mu} \}$.  If the receiver reports belief $\mu$ and obeys the recommendation from the signaling scheme $\pi_\mu$, then he gets the following single-period payoff 
\begin{align}
    V^{\sp}(\mathcal M, \rprior, \mu) & = \sum_{\omega \in \Omega} \rprior(\omega) \sum_{a\in A} \pi_\mu(a|\omega) v(a, \omega) \nonumber \\
    & = - \rprior(\omega_0) \pi_\mu(a_1|\omega_0) + \rprior(\omega_1) \pi_\mu(a_1|\omega_1) \nonumber \\
    & = \tfrac{1}{1+\tilde \mu_R}\Big( \tilde \mu_R \pi_\mu(a_1|\omega_1) - \pi_\mu(a_1|\omega_0) \Big).  \label{eq:V-sp-definition}
\end{align}
%The intuition for $V^{\sp}(\mathcal M, \rprior, \mu)$ in Eq.~\eqref{eq:V-sp-definition} is in a similar spirit of auction design.
% The intuition for Eq.~\eqref{eq:V-sp-definition} is in a similar spirit of auction design.
Eq.~\eqref{eq:V-sp-definition} captures the payoff of a buyer in an auction.
Specifically, the receiver's {prior belief} $\tilde \mu_R$ corresponds to the buyer's private value of an item; $\pi_\mu(a_1|\omega_1)$ captures the probability that the receiver gets the item; $\pi_\mu(a_1|\omega_0)$ is the receiver's payment.
We say the menu $\mathcal M = \{ \pi_{\mu} \}$ is
\begin{itemize}
    \item \emph{incentive compatible} if the receiver is willing to report his {prior belief $\rprior$} (his private value) truthfully: $\forall \mu_R, \forall \mu$, $V^{\sp}(\mathcal M, \rprior, \rprior) \ge V^{\sp}(\mathcal M, \rprior, \mu)$;
    \item \emph{individually rational} if the receiver is weakly better off with the menu compared to receiving no information: $\forall \mu_R$, $V^{\sp}(\mathcal M, \rprior, \rprior) \ge V_{\mathrm{uninformed}}(\rprior) = \max_{a\in A} \sum_{\omega \in \Omega} \rprior(\omega) v(a, \omega)$.
\end{itemize}

\paragraph{Benchmark for the single-period game with menu}
{If the designer knows the receiver's prior belief $\rprior$, then the designer does not need to offer a menu of multiple signaling schemes; she can just use the following optimal single-period signaling scheme: }
\begin{equation}
    \pi^*_\rprior(a_1|\omega_1) = 1, \quad \pi^*_\rprior(a_0|\omega_1) = 0, \quad \pi^*_\rprior(a_1|\omega_0) = \tilde \mu_R, \quad \pi^*_\rprior(a_0|\omega_0) = 1-\tilde \mu_R,
\end{equation}
% (namely, recommending $a_1$ with the largest possible probability while satisfying individual rationality), 
where $\pi^*_\rprior(a_1|\omega_1) = 1$ means always allocating the item to the buyer, and $\pi^*_\rprior(a_1|\omega_0) = \tilde \mu_R$ means setting the payment equal to the buyer's value $\tilde \mu_R$.
{See Appendix \ref{app:two-benchmark-equal} for a proof of the optimality of $\pi^*_{\rprior}$.} 
The optimal signaling scheme gives the designer an expected payoff of
\begin{equation} \label{eq:single-period-optimal-utility}
    U_{\sp}^*(\rprior) = \dprior(\omega_0) \pi^*_\rprior(a_1|\omega_0) + \dprior(\omega_1) \pi^*_\rprior(a_1|\omega_1) = (1-p_0) \tilde \mu_R + p_0.
\end{equation}
{Here $U_{\sp}^*(\rprior)$ is the benchmark for the single-period game with menu, against which we define the single-period regret.}

\paragraph{Single-period regret}
When the designer does not know the receiver's {prior belief} $\rprior$ and uses a certain incentive compatible and individually rational menu $\mathcal M$, she gets a single-period payoff {of}
\begin{equation} \label{eq:sender-utility}
    U^{\sp}(\mathcal M, \rprior) = (1-p_0) \pi_{\rprior}(a_1|\omega_0) + p_0 \pi_{\rprior}(a_1|\omega_1). 
\end{equation}
Compared to the optimal {payoff} \eqref{eq:single-period-optimal-utility}, the designer's single-period regret is
\begin{equation}
    \Reg^{\sp}(\mathcal M, \rprior) = (1-p_0) \tilde \mu_R + p_0 -  U^{\sp}(\mathcal M, \rprior).
\end{equation}
{We show that the single-period regret of any menu is at least $1/16$ in the worst case.}
\begin{lemma} \label{lemma:single-period-regret}
For any incentive compatible and individually rational menu $\mathcal M$, there exists an instance with $p_0 = \frac{1}{8}$ and $\tilde \mu_R \in [\frac{3}{8}, 1]$, which corresponds to a {prior belief} $\rprior$ satisfying the regularity Assumption \ref{ass:prior-p0}, such that $\Reg^{\sp}(\mathcal M, \rprior) \ge \frac{1}{16}$. 
\end{lemma}
\begin{proof}
    See Appendix \ref{proof:single-period-regret}. 
\end{proof}

\paragraph{\bf Reduction from $T$-period game to single-period game}
% We then reduce the $T$-period information design game to a single-period game. 
{
We then consider the $T$-period information design game that repeats the above single-period instance $\mathcal I$ for $T$ periods but without the menu. We will show that any learning algorithm for the $T$-period game with a small regret can be converted into a single-period menu with a small regret, which will then imply a lower bound for the former regret. 
Before presenting the conversion, we show that the static benchmark and dynamic benchmark for the $T$-period game (without menu) are the same and equal to $T$ times the benchmark $U^*_\sp(\rprior)$ for the single-period game with menu: 
% 
% We note that $U^*_\sp(\rprior)$ is exactly the Bayesian persuasion benchmark $U^*_\BP(\mathcal I)$ where the receiver's prior belief $\rprior$ is known:}
% \begin{equation}
%     U^*_\sp(\rprior) = U^*_\BP(\mathcal I). 
% \end{equation}
% Moreover, the Bayesian persuasion benchmark turns out to be equal to the {\color{blue}dynamic} benchmark in this instance, so the designer's regret with respect to the two benchmarks will be the same. 
% 
\begin{lemma} \label{lem:two-benchmark-equal}
% In the above information design instance, %the designer's myopic benchmark and strategic benchmark are both equal to $U^*_\sp(\rprior)$:
The benchmarks satisfy
$\mathcal U^{*}_T(\mathcal I) = \mathcal U^{**}_T(\mathcal I) = T \cdot U^*_{\sp}(\rprior)$. 
\end{lemma}
\begin{proof}
    See Appendix \ref{app:two-benchmark-equal}. 
\end{proof}
}

{
We now convert any learning algorithm  $\Alg$ of the designer for the $T$-period game (without menu) to a menu $\mathcal M$ for the single-period game.
Given learning algorithm $\Alg$, a strategic receiver with prior $\rprior$ best responds by a $T$-period strategy $\phi^*(\Alg, \rprior) = (\phi^{(t)})_{t=1}^T$.
The pair ($\Alg$, $\phi^*(\Alg, \rprior))$ generates a stochastic process $(\pi^{(t)}, \rho^{(t)})_{t=1}^T$ of the designer's signaling schemes and the receiver's single-period strategies. 
We consider the average probability $\sigma_{\rprior}(a | \omega)$ that the receiver takes action $a$ conditioning on state $\omega$ when the receiver uses $\phi^*(\Alg, \rprior)$ in response to the designer's algorithm $\Alg$, weighted by the receiver's discount factor: 
\begin{equation}
    \sigma_{\rprior} (a|\omega):=\E_{(\pi^{(t)}, \rho^{(t)}) \sim \Alg, \phi^*(\Alg, \rprior)} \bigg[\frac{1}{T_\gamma}\sum_{t=1}^T \gamma^t \sum_{s\in S}\pi^{(t)}(s|\omega)\rho^{(t)}(a|s)\bigg ]. 
\end{equation}
% is the probability for each action $a \in A$ conditional on each state $\omega \in \Omega$, 
%where $T_\gamma := \sum_{t=1}^T \gamma^t$.
%and $\frac{1}{T_\gamma}\sum_{t=1}^T \gamma^t$ is to technically make $s_{\rprior}(\cdot|\omega) \in \Delta(A)$ a well-defined distribution.
%Note that 
Since $\sigma_\rprior(\cdot|\omega)$ is a distribution over the set of actions $A$, $\sigma_{\rprior}$ can be regarded as a direct signaling scheme.
}
Consider the menu $\mathcal M = \{ \sigma_{\mu_R} \}_{\mu_R \in \Delta(\Omega)}$ of the direct signaling scheme $\sigma_{\mu_R}$ associated with {prior belief $\mu_R$}.
Because the receiver is best-responding, the menu $\mathcal M$ constructed in this way must be incentive-compatible and individually rational for the receiver; this is reminiscent to a ``revelation principle'', which we formalize below (with proof given in Appendix \ref{proof:icir}): 
% and we  formally prove this below (in Appendix \ref{proof:lower-bound}). 
\begin{claim}
\label{claim:icir}
The menu $\mathcal M = \{ \sigma_{\mu_R} \}_{\mu_R \in \Delta(\Omega)}$ is incentive compatible and individually rational.  
\end{claim}

{
Then, we prove that the designer's $T$-period regret is at least her single-period regret times the receiver's effective horizon $T_\gamma$, formalized in Lemma \ref{lem:T-regret-single-period-regret}. 
Intuitively, this is because
% the receiver can manipulate for $T_\gamma$ periods 
the designer suffers a single-period regret of $\Reg^{\sp}(\mathcal M, \rprior)$ in every period where she has not fully learned the receiver's prior belief, due to the receiver's manipulation, and the receiver's manipulation lasts for $T_\gamma$ periods. See details in Appendix \ref{proof:T-regret-single-period-regret}. 

\begin{lemma} \label{lem:T-regret-single-period-regret}
The designer's regret satisfies $\Reg(T; \Alg, \mathcal I) \ge T_\gamma \cdot \Reg^{\sp}(\mathcal M, \rprior)$. 
\end{lemma}

According to Lemma \ref{lemma:single-period-regret}, for any incentive compatible and individually rational menu $\mathcal M$, there exists an instance such that the designer's single-period regret $\Reg^{\sp}(\mathcal M, \rprior)$ is at least $\frac{1}{16}$.
That implies $\Reg(T; \Alg, \mathcal I) \ge \frac{1}{16} T_\gamma$, which proves Theorem \ref{thm:lower-bound}. 
}

\section{Discussion}

We study the information design problem when the information designer does not know the belief of {a strategic} receiver. We design learning algorithms so that the designer can learn the receiver's belief through repeated interactions. Our learning algorithms are robust to the receiver’s strategic manipulation of the learning process of the designer.  

We define regret relative to two benchmarks to measure the performance of the learning algorithms. The static benchmark is $T$ times the single-period {optimum} for the designer for the known belief. The dynamic benchmark is the global optimality for the designer for the known belief. Our learning algorithms achieve no regret against both benchmarks at fast speeds of $O(\log^2 T)$. Our work thus provides a robust learning foundation for the information design problems with unknown subjective beliefs of strategic receivers. 

We offer a brief discussion of several natural extensions. First, while we consider the receiver's belief as a convergence outcome of his misspecified learning, it {would be} interesting to {extend the analysis to consider a receiver who is learning about the distribution.} %\textcolor{red}{As we discussed, it would be better not to phrase the question in terms of whether the players' beliefs will converge to the true distribution. Unless their beliefs are misspecified, we know they will. The question is how this learning affects the designer.}
%know what the game looks like where there is a learning designer and a learning strategic receiver who starts his learning with any belief. 
Will the two players' learning converge to the true distribution of the states? Or will their learning converge to a belief that is neither the true distribution of the states nor the receiver's belief ($\rprior$ in the paper) formed by his own misspecified learning? Will any factors in the game, such as their patience comparisons, determine which learning outcome they will reach?

Second, it is interesting to {extend our results to consider a designer with a discount factor close to but strictly less than 1.}
\textcolor{black}{Our results require the designer to be more patient than the receiver.}
%know what will happen when the designer is nearly patient with a discount factor, say, $\gamma^D,$ close to 1.
%As our model shows in the paper, the designer cannot be more impatient than the receiver. 
But what if the designer is close to being patient, i.e., $\gamma_D$ close to 1? And what if the designer is only more patient than the receiver but is still impatient, i.e., $\gamma < \gamma_D < 1$? Such questions may have interesting implications for real-world problems. For example, the online platform may have limited patience with a certain user because interactions require time, but time is valuable.

Third, our learning algorithms indeed enable the designer to achieve no regret at fast speeds. But our learning algorithms rely on using the receiver's actions, which may contain the receiver's privacy and security information. %\textcolor{red}{You seem to say "the receiver's privacy" when you mean "the receiver's private information." I found this discussion unclear and was not sure why privacy issues would be interesting in this context. I'd suggest either clarifying why this is interesting or dropping it.} 
Can the designer still learn the receiver's belief at a fast speed without learning the receiver's privacy? And if learning the privacy cannot be avoided, can the designer use algorithms that unlearn the receiver's privacy to make the learning reliable and trustworthy for the receiver? 

Finally, we are also interested in whether our contributions, including the learning of a strategic receiver's belief as well as the dynamic incentives of a strategic receiver, to the information design setting can be generalized or extended to other possible settings. We leave those as open questions for future work.

\paragraph{Acknowledgment} Declaration of financial support or competing interests: None.

\newpage

\printbibliography
%\bibliography{bib}




 
% LNCS Homepage, \url{http://www.springer.com/lncs}, last accessed 2023/10/25
% \end{thebibliography}

\newpage

\appendix

\section{Useful Lemmas}
\label{section:useful_lemmas}

\begin{lemma}[Lipschitz continuity of utility]
\label{lem:utility-Lipschitz}
The designer's and receiver's expected utility functions, $U(\pi, \rho; \dprior), V(\pi, \rho; \rprior)$, are:
\begin{itemize}
    \item $1$-Lipschitz continuous with respect to priors under the $\ell_1$-norm:
    \[| U(\pi, \rho; \mu_1) - U(\pi, \rho; \mu_2)| ~ \le ~ \| \mu_1 - \mu_2 \|_1.\] 
    \item $1$-Lipschitz continuous with respect to $\pi$ under the matrix $\ell_1$-norm:
    \[| U(\pi_1, \rho; \dprior) - U(\pi_2, \rho; \dprior)| ~ \le ~ \max_{\omega \in \Omega } \sum_{s\in S} \big| \pi_1(s|\omega) - \pi_2(s|\omega) \big|.\]
\end{itemize}
Analogous inequalities hold for $V(\cdot, \rho; \cdot)$. 
\end{lemma}

\begin{proof}[Proof]
Write
\[
U(\pi,\rho;\mu_D)=\sum_{\omega}\mu_D(\omega)\sum_{s}\pi(s\mid\omega)\sum_{a}\rho(a\mid s)u(a,\omega).
\]

\emph{(Lipschitz in the prior).} For fixed $(\pi,\rho)$, the inner expectation lies in $[0,1]$, hence
\[
\big|U(\pi,\rho;\mu_1)-U(\pi,\rho;\mu_2)\big|
=\Big|\sum_{\omega}(\mu_1-\mu_2)(\omega)\cdot \underbrace{\sum_{s,a}\pi(s\mid\omega)\rho(a\mid s)u(a,\omega)}_{\in[0,1]}\Big|
\le \sum_{\omega}|\mu_1-\mu_2|(\omega)=\|\mu_1-\mu_2\|_1.
\]

\emph{(Lipschitz in the signaling scheme).} For fixed $(\rho,\mu_D)$, set $\bar u_s(\omega):=\sum_a\rho(a\mid s)u(a,\omega)\in[0,1]$. Then
\[
U(\pi_1,\rho;\mu_D)-U(\pi_2,\rho;\mu_D)=\sum_{\omega}\mu_D(\omega)\sum_{s}\big(\pi_1-\pi_2\big)(s\mid\omega)\,\bar u_s(\omega).
\]
For each $\omega$, the inner term is the difference of expectations of a $[0,1]$-valued function under two distributions on $S$, so by the dual characterization of total variation,
\[
\Big|\sum_{s}\big(\pi_1-\pi_2\big)(s\mid\omega)\,\bar u_s(\omega)\Big|
\le \mathrm{TV}\!\big(\pi_1(\cdot\mid\omega),\pi_2(\cdot\mid\omega)\big).
\]
Averaging over $\omega$ (since $\mu_D$ is a probability measure) gives the stated bound. Finally,
$\mathrm{TV}(p,q)=\tfrac12\sum_s|p(s)-q(s)|\le \tfrac{|S|}{2}\max_s|p(s)-q(s)|$ yields the matrix $\ell_1$–norm version.
The proof for $V$ is identical with $u$ replaced by $v$ and $\mu_D$ by $\mu_R$.
\end{proof}

\begin{lemma}[Lipschitz continuity of posterior]
\label{lem:continuity-posterior}
Let $\pi : \Omega \to \Delta(S)$ be any signaling scheme.  Let $\mu, \mu' \in \Delta(\Omega)$ be two priors.  Let $\mu_s$, $\mu'_s$ be the posterior belief induced by signal $s$ under $\pi$ and prior $\mu$, $\mu'$ respectively.  Suppose $\min_{\omega \in \Omega} \mu(\omega) \ge p_0 > 0$.  Then $\| \mu_s - \mu'_s \|_1 \le \frac{2}{p_0} \| \mu - \mu' \|_1$. 
\end{lemma}
\begin{proof}
Let $\pi(s) = \sum_{\omega \in \Omega} \mu(\omega) \pi(s|\omega)$ and $\pi'(s) = \sum_{\omega \in \Omega} \mu'(\omega) \pi(s|\omega)$ be the probability of signal $s$ under prior $\mu$ and $\mu'$ respectively.  By the definition of $\mu_s, \mu'_s$ and by triangle inequality, 
\begin{align*}
    \| \mu_s - \mu'_s \|_1 & = \sum_{\omega \in \Omega} \big| \tfrac{\mu(\omega) \pi(s|\omega)}{\pi(s)} - \tfrac{\mu'(\omega) \pi(s|\omega)}{\pi'(s)} \big| \\
    & \le \sum_{\omega \in \Omega} \big| \tfrac{\mu(\omega) \pi(s|\omega)}{\pi(s)} - \tfrac{\mu'(\omega) \pi(s|\omega)}{\pi(s)} \big| + \sum_{\omega \in \Omega} \big| \tfrac{\mu'(\omega) \pi(s|\omega)}{\pi(s)} - \tfrac{\mu'(\omega) \pi(s|\omega)}{\pi'(s)} \big|.
\end{align*}

For the first term above, $\sum_{\omega \in \Omega} \big| \tfrac{\mu(\omega) \pi(s|\omega)}{\pi(s)} - \tfrac{\mu'(\omega) \pi(s|\omega)}{\pi(s)} \big| = \sum_{\omega \in \Omega} \tfrac{\pi(s|\omega)}{\pi(s)} |\mu(\omega) - \mu'(\omega)|$. 
We note that, $\forall \omega \in \Omega$, 
\begin{align}\label{eq:pi-ratio-le-p0}
    \tfrac{\pi(s|\omega)}{\pi(s)} = \tfrac{\pi(s|\omega)}{\sum_{\omega' \in \Omega} \mu(\omega') \pi(s|\omega')} \le \tfrac{\pi(s|\omega)}{p_0 \sum_{\omega' \in \Omega}\pi(s|\omega')} \le \tfrac{1}{p_0}. 
\end{align}
Thus, $
    \sum_{\omega \in \Omega} \big| \tfrac{\mu(\omega) \pi(s|\omega)}{\pi(s)} - \tfrac{\mu'(\omega) \pi(s|\omega)}{\pi(s)} \big| \le \sum_{\omega \in \Omega} \tfrac{1}{p_0} |\mu(\omega) - \mu'(\omega)| = \tfrac{1}{p_0} \| \mu - \mu' \|_1$. 

For the second term, 
\begin{align*}
    \sum_{\omega \in \Omega} \big| \tfrac{\mu'(\omega) \pi(s|\omega)}{\pi(s)} - \tfrac{\mu'(\omega) \pi(s|\omega)}{\pi'(s)} \big| & = \sum_{\omega \in \Omega} \mu'(\omega) \pi(s|\omega) \big| \tfrac{\pi'(s) - \pi(s)}{\pi(s) \pi'(s)} \big| \\
    & = \sum_{\omega \in \Omega} \mu'(\omega) \pi(s|\omega) \big| \tfrac{\sum_{\omega'\in\Omega} (\mu'(\omega') - \mu(\omega')) \pi(s|\omega')}{\pi(s) \pi'(s)} \big| \\
    & \le \sum_{\omega \in \Omega} \mu'(\omega) \pi(s|\omega) \tfrac{\sum_{\omega'\in\Omega} |\mu'(\omega') - \mu(\omega')|  \cdot \max_{\omega'\in\Omega} \pi(s|\omega')}{\pi(s) \pi'(s)} \\
    & = \| \mu' - \mu \|_1 \sum_{\omega \in \Omega} \tfrac{\mu'(\omega) \pi(s|\omega)}{\pi'(s)} \tfrac{\max_{\omega'\in\Omega} \pi(s|\omega')}{\pi(s)} \\
    \text{by \eqref{eq:pi-ratio-le-p0}} ~ & \le \| \mu' - \mu \|_1 \sum_{\omega \in \Omega} \tfrac{\mu'(\omega) \pi(s|\omega)}{\pi'(s)} \tfrac{1}{p_0} \\
    & = \tfrac{1}{p_0} \| \mu' - \mu \|_1.
\end{align*}

Therefore, we obtain $ \| \mu_s - \mu'_s \|_1 \le \tfrac{2}{p_0}\| \mu' - \mu \|_1$.
\end{proof}

\begin{lemma}[coalescing signals]\label{lemma:coalesce-2}
% When the receiver has prior $\hat \mu$,
% \begin{itemize}
%     \item if $\tilde \pi$ recommends action $a \in A$, then for any other action $a' \ne a$, the receiver's posterior belief is $\xi_a$ and $\E_{\omega \sim \xi_a}[v(a, \omega) - v(a', \omega)] \ge (1-\delta)\E_{\omega \sim \hat \mu_a}[v(a, \omega) - v(a', \omega)] + \delta D$. 
%     \item if $\tilde \pi$ reveals state $\omega \in \Omega$, then the receiver's posterior belief is $\omega$ deterministically. 
% \end{itemize}
Let $s_1, s_2 \in S$ be two signals from a signaling scheme $\pi$, to which the receiver responds by the same action $\rho(s_1) = \rho(s_2) \in A$.
Then, after coalescing signals $s_1$ and $s_2$ into $s$ by letting $\pi^c(s|\omega) = \pi(s_1|\omega) + \pi(s_2|\omega)$, we have:
\begin{itemize}
    \item Assuming that the receiver also responds by action $\rho(s) = \rho(s_1) = \rho(s_2)$ for the coalesced signal, the information designer's and the receiver's expected utilities do not change after coalescing. 
    \item For any actions $a, a' \in A$, if the receiver's posterior $\mu_{s_1}$ under $s_1$ satisfies $\E_{\omega \sim \mu_{s_1}}[v(a, \omega) - v(a', \omega) ] \ge X_1$ and the posterior $\mu_{s_2}$ under $s_2$ satisfies $\E_{\omega \sim \mu_{s_2}}[v(a, \omega) - v(a', \omega) ] \ge X_2$, then the receiver's posterior under $s$ satisfies
    \begin{equation*}
        \E_{\omega \sim \mu_s}[v(a, \omega) - v(a', \omega) ] \ge \min\{X_1, X_2\}. 
    \end{equation*}
\end{itemize}
\end{lemma}

\begin{proof}
The expected utilities of the two players clearly do not change after coalescing. We then verify the second claim. 
By definition, the receiver's posterior $\mu_s$ satisfies 
\begin{align*}
    & \E_{\omega \sim \mu_s} \big[ v(a, \omega) - v(a', \omega) \big] \\
    & = \sum_{\omega \in \Omega} \frac{\hat \mu(\omega) \pi^c(s|\omega)}{\sum_{\omega' \in \Omega} \hat \mu(\omega') \pi^c(s|\omega')} \Big[ v(a, \omega) - v(a', \omega) \Big] \\
    & = \sum_{\omega \in \Omega} \frac{\hat \mu(\omega) \pi(s_1|\omega)}{\sum_{\omega' \in \Omega} \hat \mu(\omega') \pi^c(s|\omega')} \Big[ v(a, \omega) - v(a', \omega) \Big] + \sum_{\omega \in \Omega} \frac{\hat \mu(\omega) \pi(s_2|\omega)}{\sum_{\omega' \in \Omega} \hat \mu(\omega') \pi^c(s|\omega')} \Big[ v(a, \omega) - v(a', \omega) \Big] \\
    & = \frac{\sum_{\omega' \in \Omega} \hat \mu(\omega') \pi(s_1|\omega')}{\sum_{\omega' \in \Omega} \hat \mu(\omega') \pi^c(s|\omega')}\sum_{\omega \in \Omega} \frac{\hat \mu(\omega) \pi(s_1|\omega)}{\sum_{\omega' \in \Omega} \hat \mu(\omega') \pi(s_1|\omega')} \Big[ v(a, \omega) - v(a', \omega) \Big] \\
    & \qquad + \frac{\sum_{\omega' \in \Omega} \hat \mu(\omega') \pi(s_2|\omega')}{\sum_{\omega' \in \Omega} \hat \mu(\omega') \pi^c(s|\omega')}\sum_{\omega \in \Omega} \frac{\hat \mu(\omega) \pi(s_2|\omega)}{\sum_{\omega' \in \Omega} \hat \mu(\omega') \pi(s_2|\omega')} \Big[ v(a, \omega) - v(a', \omega) \Big] \\
    & = \lambda \cdot \E_{\omega \sim \mu_{s_1}}\big[ v(a, \omega) - v(a', \omega)\big] + (1-\lambda) \cdot \E_{\omega \sim \mu_{s_2}}\big[ v(a, \omega) - v(a', \omega)] \\
    & \ge \min\{X_1, X_2\}, 
\end{align*}
where $\lambda = \frac{\sum_{\omega' \in \Omega} \hat \mu(\omega') \pi(s_1|\omega')}{\sum_{\omega' \in \Omega} \hat \mu(\omega') \pi^c(s|\omega')} \in [0, 1]$. 
\end{proof}

\section{Omitted Proofs in Section \ref{sec:static-benchmark}}

\subsection{Proof of Lemma \ref{lem:ratio}}
\label{proof:ratio}
% There are two cases for the left endpoints: 
% $\ell^{(k)} < (1 - \delta) \frac{\pi^*(s_0|\omega_2)}{\pi^*(s_0|\omega_2)}$ and $\ell^{(k)} \ge (1 - \delta) \frac{\pi^*(s_0|\omega_2)}{\pi^*(s_0|\omega_2)}$.
% There are two cases for the left endpoints: 
% $r^{(k)} > (1 + \delta) \frac{\pi^*(s_0|\omega_2)}{\pi^*(s_0|\omega_2)}$ and $r^{(k)} \le (1 + \delta) \frac{\pi^*(s_0|\omega_2)}{\pi^*(s_0|\omega_2)}$.

\begin{claim} 
% ==== Suppose the initial $\ell^{(0)} < (1-\delta) \frac{\pi^*(s_0|\omega_2)}{\pi^*(s_0|\omega_2)}$ $r^{(0)}$ ===
During the binary search algorithm (Algorithm \ref{alg:binary-search}), the left endpoint always satisfies $\ell^{(k)} \le (1+\diss) \frac{\pi^*(s_0|\omega_j)}{\pi^*(s_0|\omega_i)}$, and the right endpoint always satisfies $r^{(k)} \ge (1-\diss) \frac{\pi^*(s_0|\omega_j)}{\pi^*(s_0|\omega_i)}$. 
Additionally, $\ell^{(k)} \le r^{(k)}$. (This claim includes the initial step where $k = 0$.) 
\end{claim}
\begin{proof}
First, we verify that the claim holds at the initial step $k = 0$. 
By definition, we have $\ell^{(0)}=0 \le (1+\diss) \frac{\pi^*(s_0|\omega_j)}{\pi^*(s_0|\omega_i)}$, and $r^{(0)}= \frac{1}{Gp_0}$. Because by definition $\tilde a_{\omega_j, \omega_i}$ is the first indifference action between states $\omega_i$ and $\omega_j$, $\tilde a_{\omega_j, \omega_i}$ is weakly better than $a_{\omega_j}^*$ when the signaling ratio is $\frac{\pi^*(s_0|\omega_j)}{\pi^*(s_0|\omega_i)}$ for a myopic receiver:
\begin{align*}
    v(\tilde a_{\omega_j, \omega_i}, \omega_i) - v(a_{\omega_j}^*, \omega_i) + \frac{\rprior(\omega_j)}{\rprior(\omega_i)} \frac{\pi^*(s_0|\omega_j)}{\pi^*(s_0|\omega_i)} \Big(  v(\tilde a_{\omega_j, \omega_i}, \omega_j) - v(a_{\omega_j}^*, \omega_j) \Big) ~ \ge ~ 0. 
\end{align*}
Note that $v(\tilde a_{\omega_j, \omega_i}, \omega_i) - v(a_{\omega_j}^*, \omega_i) \le 1$ and $ v(\tilde a_{\omega_j, \omega_i}, \omega_j) - v(a_{\omega_j}^*, \omega_j) \le -G < 0$, so
\begin{align*}
    1 - \frac{\rprior(\omega_j)}{\rprior(\omega_i)} \frac{\pi^*(s_0|\omega_j)}{\pi^*(s_0|\omega_i)} \cdot G ~ \ge ~ 0, 
\end{align*}
which implies
\begin{equation} \label{eq:pi*/pi*<=1/p_0G}
    \frac{\pi^*(s_0|\omega_j)}{\pi^*(s_0|\omega_i)} ~ \le ~ \frac{\rprior(\omega_i)}{\rprior(\omega_j)G} ~ \le ~ \frac{1}{p_0 G}. 
\end{equation}
Thus, $r^{(0)} = \frac{1}{p_0 G} \ge (1-\diss) \frac{\rprior(\omega_i)}{\rprior(\omega_j)G}$. 

Then, consider step $k \ge 0$. The algorithm chooses the midpoint $q = \frac{\ell^{(k)} + r^{(k)}}{2}$ and uses a signaling scheme satisfying $\frac{\pi^{(k)}(s_0|\omega_j)}{\pi^{(k)}(s_0|\omega_i)} = q$. When signal $s_0$ is realized:
\begin{itemize}
    \item if the receiver takes action $a_{\omega_i}^*$, then by the second item of the Lemma \ref{lem:binary-search-strategic-receiver}, $q \le (1+\diss) \frac{\pi^*(s_0|\omega_j)}{\pi^*(s_0|\omega_i)}$. The algorithm sets $\ell^{(k+1)}$ to be $q$, so $\ell^{(k+1)}$ satisfies the claim; 
    \item if the receiver does not take action $a_{\omega_i}^*$, then by the first item of Lemma \ref{lem:binary-search-strategic-receiver}, $q \ge (1 - \diss) \frac{\pi^*(s_0|\omega_j)}{\pi^*(s_0|\omega_i)}$. The algorithm sets $r^{(k+1)}$ to be $q$, so $r^{(k+1)}$ satisfies the claim.
\end{itemize}
\end{proof}

\begin{proof}[Proof of Lemma \ref{lem:ratio}]
When Algorithm \ref{alg:binary-search} ends, we have $|r^{(k)} - \ell^{(k)}| \le \terminate$. Then, using the above claim, we have
\begin{equation*}
    \Big| \frac{\pi^*(s_0|\omega_j)}{\pi^*(s_0|\omega_i)} - \ell^{(k)} \Big| ~ \le ~ \terminate + \diss \frac{\pi^*(s_0|\omega_j)}{\pi^*(s_0|\omega_i)},
\end{equation*}
which implies
\begin{align*}
    \Big| \frac{\rprior(\omega_i)}{\rprior(\omega_j)} - \hat \rho \Big| & ~ = ~ \Big| \frac{\pi^*(s_0|\omega_j)}{\pi^*(s_0|\omega_i)} - \ell^{(k)} \Big| \cdot \frac{v(\tilde a_{\omega_j, \omega_i}, \omega_j) - v(a_{\omega_i}^*, \omega_2)}{v(a_{\omega_i}^*, \omega_i) - v(\tilde a_{\omega_j, \omega_i}, \omega_i)} \\
    & ~ \le ~ \Big(\terminate + \diss \frac{\pi^*(s_0|\omega_j)}{\pi^*(s_0|\omega_i)}\Big) \cdot \frac{1}{G} \\
    & ~ \le ~  \Big(\terminate + \frac{\delta}{Gp_0}\Big) \cdot \frac{1}{G} && \text{by \eqref{eq:pi*/pi*<=1/p_0G}} \\
    & ~ \le ~ \Big(\terminate + \frac{\gamma^m}{(1-\gamma) G^3 p_0^3}\Big) \cdot \frac{1}{G}  && \text{by Lemma \ref{lem:binary-search-strategic-receiver}} \\
    & ~ \le ~ \Big(\terminate + \frac{\gamma^m}{(1-\gamma) G^3 p_0^3}\Big) \cdot \frac{1}{G p_0} \cdot \frac{\rprior(\omega_i)}{\rprior(\omega_j)}. 
\end{align*}
where the last step is because $\frac{\rprior(\omega_i)}{\rprior(\omega_j)} \ge p_0$. 

Then, we consider the expected running time of Algorithm \ref{alg:binary-search}.
Because the difference $r^{(k)} - \ell^{(k)}$ shrinks by a half after each while loop, the number of while loops, $k$, is at most  
\begin{equation}\label{eq:bound-on-k}
    k ~ \le ~ \log_2 \frac{r^{(0)} - \ell^{(0)}}{\terminate} ~ = ~ \log_2 \frac{1}{G p_0 \terminate}.
\end{equation}
Then, we consider the number of periods needed in each while loop.  By definition, this is equal to the number of periods until signal $s_0$ is realized, plus $m$.
Since one of $\pi^{(k)}(s_0 |\omega_j)$ and $\pi^{(k)}(s_0 |\omega_i)$ is 1 (see Algorithm \ref{alg:binary-search}), the probability (according to the information designer's prior belief) that $s_0$ is realized in each period is at least: 
\begin{equation*}
    \pi^{(k)}(s_0) ~ = ~ \dprior(\omega_i) \pi^{(k)}(s_0 | \omega_i) + \dprior(\omega_j) \pi^{(k)}(s_0 | \omega_j) ~ \ge ~ \min \{ \dprior(\omega_i), \dprior(\omega_j) \} ~ \ge ~ p_0. 
\end{equation*}
So by the property of the geometric random variable, the expected number of periods until a signal $s_0$ is realized is at most
\begin{equation*}
    \frac{1}{\pi^{(k)}(s_0)} ~ \le ~ \frac{1}{p_0}, 
\end{equation*}
% So, the expected number of periods in each while loop is at most $\frac{1}{p_0} + m$,
and the total expected number of periods over $k$ while loops is at most 
\begin{equation*}
    k \cdot \Big(\frac{1}{p_0} + m \Big) ~ \le ~ \Big(\frac{1}{p_0} + m \Big) \log_2 \frac{1}{G p_0 \terminate}. 
\end{equation*}
\end{proof}

\subsection{Proof of Lemma \ref{lem:binary-search-strategic-receiver}}
\label{proof: binary-search-strategic-receiver}
\begin{proof}
    \textit{Case (i): $\frac{\pi^{(k)}(s_0|\omega_j)}{\pi^{(k)}(s_0|\omega_i)} < (1-\diss) \frac{\pi^*(s_0|\omega_j)}{\pi^*(s_0|\omega_i)}$.}  If the receiver best responds in that period, then his optimal action will be $a_{\omega_i}^*$, since he is indifferent between $a_{\omega_i}^*$ and the first indifferent action $\tilde a_{\omega_j, \omega_i}$ at $\frac{\pi^*(s_0|\omega_j)}{\pi^*(s_0|\omega_i)}$ by definition, and also due to the case with $\frac{\pi^{(k)}(s_0|\omega_j)}{\pi^{(k)}(s_0|\omega_i)} < \frac{\pi^*(s_0|\omega_j)}{\pi^*(s_0|\omega_i)}$.  Suppose the strategic receiver deviates to an action $a \ne a_{\omega_i}^*$ instead. Then, he will suffer a loss in the current period, and will potentially obtain some gain in the future. 

Regarding the gain, the signaling schemes used in the next additional $m$ periods are independent of the receiver's current action, and the receiver's payoff is bounded in $[0, 1]$. Thus, the receiver's total gain in the future is at most 
\begin{align*}
    \mathrm{gain} ~ \le ~ 0 + \sum_{t=m+1}^T \gamma^t \cdot 1 ~ \le ~ \frac{\gamma^{m+1}}{1-\gamma}. 
\end{align*}
Then, we compute his loss, denoted by $\mathrm{loss}(a_{\omega_i}^*, a)$, which is the difference between the receiver's payoff of taking action $a_{{\omega_i}}^*$ versus the deviation action $a$ in the current round.
\begin{align*}
    \mathrm{loss}(a_{\omega_i}^*, a) & ~ = ~ \sum_{\omega \in \Omega} \Pr_{\rprior, \pi^{(k)}}[ \omega | s_0] \Big( v(a_{\omega_i}^*, \omega) - v(a, \omega) \Big)  \nonumber \\
    & ~ = ~ \frac{\rprior(\omega_i)\pi^{(k)}(s_0|\omega_i)}{\pi^{(k)}(s_0)}\Big(v(a_{\omega_i}^*,\omega_i)-v(a,\omega_i)\Big) + \frac{\rprior(\omega_j)\pi^{(k)}(s_0|\omega_j)}{\pi^{(k)}(s_0)}\Big(v(a_{\omega_i}^*, \omega_j) - v(a ,\omega_j)\Big).
\end{align*}
% (We omit the superscript and write $\pi^{(k)}$ as $\pi$.)
\begin{claim}
$\mathrm{loss}(a_{\omega_i}^*, a) > \diss  G^2 p_0^2$.
\end{claim}

\begin{figure}
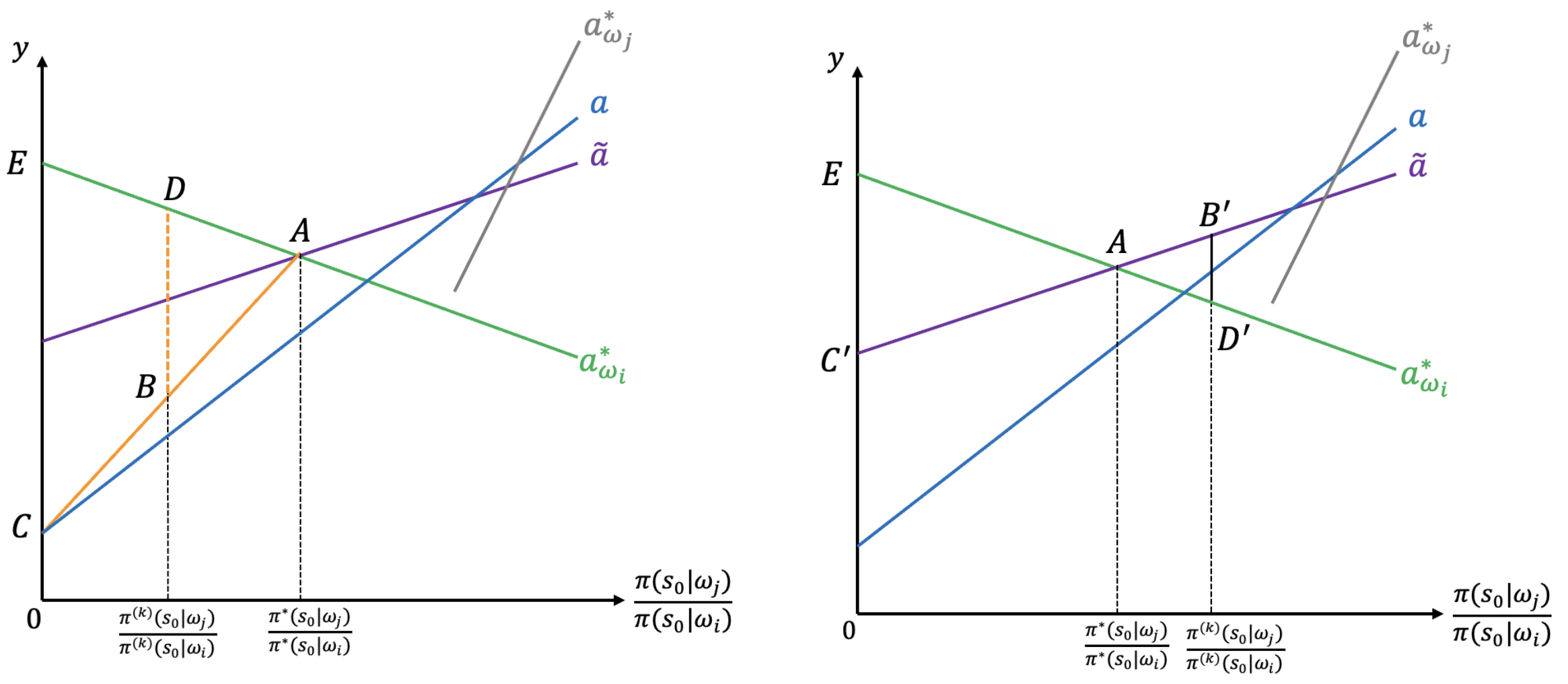

\centering
\begin{subfigure}[b]{0.49\textwidth}
\includegraphics[width=\linewidth]{ratio_v.png}
\caption{The case of $\frac{\pi^{(k)}(s_0|\omega_j)}{\pi^{(k)}(s_0|\omega_i)} < (1-\diss) \frac{\pi^*(s_0|\omega_j)}{\pi^*(s_0|\omega_i)}$}
\label{fig:ratio_v_1}
\end{subfigure}
\begin{subfigure}[b]{0.49\textwidth}
\includegraphics[width=\linewidth]{ratio_v_right_.png}
\caption{The case of $\frac{\pi^{(k)}(s_0|\omega_j)}{\pi^{(k)}(s_0|\omega_i)} > (1+\diss) \frac{\pi^*(s_0|\omega_j)}{\pi^*(s_0|\omega_i)}$}
\label{fig:ratio_v_2}
\end{subfigure}
\caption{\centering Normalized utility of the receiver for different actions. The $x$-coordinate is the signaling ratio $\frac{\pi(s_0|\omega_j)}{\pi(s_0|\omega_i)}$. Each line corresponds to an action $a$, with equation $y_a(x) = v(a, \omega_i) + \frac{\rprior(\omega_j)}{\rprior(\omega_i)} \cdot x \cdot v(a, \omega_j)$. }
\label{ratio_v}
\end{figure}

\begin{proof}
Consider the graph (Figure \ref{fig:ratio_v_1}) where the horizontal axis is the ratio $\frac{\pi(s_0|\omega_j)}{\pi(s_0|\omega_i)}$. For each action $a \in A$, we plot the line representing the receiver's single-period utility of taking action $a$ (after a normalization):
\begin{align*}
    y_a(x) = v(a, \omega_i) + \frac{\rprior(\omega_j)}{\rprior(\omega_i)} \cdot x \cdot v(a, \omega_j), \quad { \text{where} } \quad x = \frac{\pi(s_0|\omega_j)}{\pi(s_0|\omega_i)}. 
\end{align*}

By definition, point $A$ is the indifference point between actions $a_{\omega_i}^*$ and $\tilde a_{\omega_j, \omega_i}$, whose x-coordinate $x_A = \frac{\pi^*(s_0|\omega_j)}{\pi^*(s_0|\omega_i)}$ and y-coordinate $y_A = y_{a_{\omega_i}^*}(x_A)=y_{\tilde a_{\omega_j, \omega_i}}(x_A)$. 
Let $d_x(a_{\omega_i}^*, a) = y_{a_{\omega_i}^*}(x) - y_a(x)$ be the vertical distance between lines $y_{a_{\omega_i}^*}$ and $y_a$ at x-coordinate $x$. 
Because $A$ is the first indifference point, line $y_a$ is below line $y_{a_{\omega_i}^*}$ in the $[0, x_A]$ segment: namely, $d_0(a_{\omega_i}^*, a) > 0$ and $d_{x_A}(a_{\omega_i}^*, a) \ge 0$.  
Consider the line connecting $C = (0, y_a(0) = v(a, \omega_i))$ to $A$, with $B$ being the point on this line at x-coordinate
\begin{align*}
    x_B = \frac{\pi^{(k)}(s_0|\omega_j)}{\pi^{(k)}(s_0|\omega_i)} \le (1-\diss) \frac{\pi^*(s_0|\omega_j)}{\pi^*(s_0|\omega_i)} = (1-\diss) x_A. 
\end{align*}
Since $B$ is between line $y_a$ and line $y_{a_{\omega_i}^*}$, 
%\begin{align*}
    $d_{x_B}(a_{\omega_i}^*, a) \ge BD.$
    % y_{a_1^*}(x_B) - y_B = BD. 
%\end{align*}
Triangle $ABD$ is similar to triangle $ACE$, so
\begin{align*}
    % y_{a_1^*}(x_B) - y_B =
    BD = CE \cdot \frac{AB}{AC} & = CE \cdot \frac{x_A - x_B}{x_A - 0} = \Big( v(a_{\omega_i}^*, \omega_i) - v(a, \omega_i) \Big) \cdot \frac{\frac{\pi^*(s_0|\omega_j)}{\pi^*(s_0|\omega_i)} - \frac{\pi^{(k)}(s_0|\omega_j)}{\pi^{(k)}(s_0|\omega_i)}}{\frac{\pi^*(s_0|\omega_j)}{\pi^*(s_0|\omega_i)}} > G \cdot \diss. 
\end{align*}
Therefore,
%\begin{align*}
    $$G\diss < d_{x_B}(a_{\omega_i}^*, a) = v(a_{\omega_i}^*, \omega_i) + \frac{\rprior(\omega_j)}{\rprior(\omega_i)}\frac{\pi^{(k)}(s_0|\omega_j)}{\pi^{(k)}(s_0|\omega_i)}v(a_{\omega_i}^*, \omega_j) - v(a, \omega_i)-\frac{\rprior(\omega_j)}{\rprior(\omega_i)}\frac{\pi^{(k)}(s_0|\omega_j)}{\pi^{(k)}(s_0|\omega_i)}v(a, \omega_j).$$
%\end{align*}
Multiplying both sides by $\frac{\rprior(\omega_i)\pi^{(k)}(s_0|\omega_i)}{\pi^{(k)}(s_0)}$,
\begin{align*}
    G\diss \frac{\rprior(\omega_i)\pi^{(k)}(s_0|\omega_i)}{\pi^{(k)}(s_0)} & < \frac{\rprior(\omega_i)\pi^{(k)}(s_0|\omega_i)}{\pi^{(k)}(s_0)} ( v(a_{\omega_i}^*, \omega_i) - v(a, \omega_i)  ) + \frac{\rprior(\omega_j)\pi^{(k)}(s_0|\omega_j)}{\pi^{(k)}(s_0)}  (v(a_{\omega_i}^*, \omega_j) - v(a_{\omega_i}^*, \omega_j)) 
    \\
    & = \mathrm{loss}(a_{\omega_i}^*, a).
\end{align*}
We note that $\frac{\rprior(\omega_i)\pi^{(k)}(s_0|\omega_i)}{\pi^{(k)}(s_0)} \ge G p_0^2$,
because $\rprior(\omega_i) \ge p_0$, $\pi^{(k)}(s_0) \leq 1$, $\pi^{(k)}(s_0|\omega_i) = \min \{ 1, \frac{1}{q} \} \ge Gp_0$ from Algorithm \ref{alg:binary-search}.
% Regarding $\pi^{(k)}(s_0|\omega_1) = \min \{ 1, \frac{1}{q} \} \ge Gp_0$, it is from $\frac{\pi^{(k)}(s_0 | \omega_2)}{\pi^{(k)}(s_0 | \omega_1)} = q$: either $\pi^{(k)}(s_0|\omega_1) = 1$ or $\pi^{(k)}(s_0|\omega_1) = \frac{1}{q}$. We take the $\min$ to show the bound. See Lemma {\color{red} to prove.See Lemma 1 for the proof of $\pi^{(k)}$} 
Thus, we obtain $\mathrm{loss}(a_{\omega_i}^*, a) \ge \diss G^2 p_0^2$, which proves the claim. 
% As a tool to simply computation, consider a hypothetical action $a'$ such that $y_{a'}$
 \end{proof}
Comparing gain and loss, given $\diss = \frac{\gamma^m}{(1-\gamma)G^2 p_0^2}$, we have 
\begin{align*}
    \mathrm{loss}(a_{\omega_i}^*, a) ~ > ~ \diss G^2 p_0^2 ~ > ~ \frac{\gamma^{m+1}}{1-\gamma} ~ \ge ~ \mathrm{gain}. 
\end{align*}
Thus, the strategic receiver should not take $a$ for any $a\ne a_{\omega_i}^*$, and thus takes $a_{\omega_i}^*$.

\textit{Case (ii): $\frac{\pi^{(k)}(s_0|\omega_j)}{\pi^{(k)}(s_0|\omega_i)} > (1+\diss) \frac{\pi^*(s_0|\omega_j)}{\pi^*(s_0|\omega_i)}$.}
See Figure \ref{fig:ratio_v_2} for an illustration.  In this case, the optimal action for a myopic receiver is not $a_{\omega_i}^*$, because $\frac{\pi^{(k)}(s_0|\omega_j)}{\pi^{(k)}(s_0|\omega_i)} > (1+\diss) \frac{\pi^*(s_0|\omega_j)}{\pi^*(s_0|\omega_i)}$.  If a strategic receiver takes action $a_{\omega_i}^*$ in the current period, then he suffers a loss in the current period and potentially obtains some gain in the future.  As in the previous case, the gain is at most $\frac{\gamma^{m+1}}{1-\gamma}$. Then, consider the loss. We lower bound the loss by the utility difference between $\tilde a$ and $a_{\omega_i}^*$ in the current period:
\begin{align*}
    \mathrm{loss}(a_{\omega_i}^*) & \ge \sum_{\omega \in \Omega} \Pr_{\rprior, \pi^{(k)}}[ \omega | s_0] \Big( v(\tilde a_{\omega_j,\omega_i}, \omega) - v(a_{\omega_i}^*, \omega) \Big)  \nonumber \\
    & ~ = ~ \frac{\rprior(\omega_i)\pi^{(k)}(s_0|\omega_i)}{\pi^{(k)}(s_0)}\Big(v(\tilde a,\omega_1)-v(a_{\omega_i}^*,\omega_1)\Big) + \frac{\rprior(\omega_2)\pi^{(k)}(s_0|\omega_2)}{\pi^{(k)}(s_0)}\Big(v(\tilde a_{\omega_j,\omega_i}, \omega_j) - v(a_{\omega_i}^*,\omega_j)\Big) \\
    & ~ = ~ \frac{\rprior(\omega_i)\pi^{(k)}(s_0|\omega_i)}{\pi^{(k)}(s_0)} \bigg( v(\tilde a,\omega_i)-v(a_{\omega_i}^*,\omega_i) + \frac{\rprior(\omega_j)}{\rprior(\omega_i)} \frac{\pi^{(k)}(s_0|\omega_j)}{\pi^{(k)}(s_0|\omega_i)} \Big(v(\tilde a_{\omega_j,\omega_i}, \omega_j) - v(a_{\omega_i}^*,\omega_j) \Big) \bigg) \\
    & ~ = ~ \frac{\rprior(\omega_i)\pi^{(k)}(s_0|\omega_i)}{\pi^{(k)}(s_0)} B'D' \qquad \text{(see Figure \ref{fig:ratio_v_2}).} 
\end{align*}
Because triangle $AB'D'$ is similar to triangle $AC'E$, we have
\begin{align*}
    B'D' = C'E \cdot \frac{AB'}{AC'} = \Big( v(a_{\omega_i}^*, \omega_i) - v(a, \omega_i) \Big) \cdot \frac{\frac{\pi^{(k)}(s_0|\omega_j)}{\pi^{(k)}(s_0|\omega_i)} - \frac{\pi^*(s_0|\omega_j)}{\pi^*(s_0|\omega_i)}}{\frac{\pi^*(s_0|\omega_j)}{\pi^*(s_0|\omega_i)}} > G \diss. 
\end{align*}
Same as the previous case, $\frac{\rprior(\omega_i)\pi^{(k)}(s_0|\omega_i)}{\pi^{(k)}(s_0)} \ge G p_0^2$. Therefore, $\mathrm{loss}(a_{\omega_i}^*) > G^2 p_0^2 \diss > \frac{\gamma^{m+1}}{1-\gamma} \ge \mathrm{gain}$, given $\diss = \frac{\gamma^m}{(1-\gamma)G^2 p_0^2}$. So, the strategic receiver should not take $a_{\omega_i}^*$. 
\end{proof}

\label{proof:good-prior-estimation}
\subsection{Proof of Lemma \ref{lemma:good-prior-estimation}}
\label{proof:good-prior-estimation}

\label{proof:relax-ratio}
\begin{lemma}
\label{lem:relax-ratio}
Let $\eps' = \frac{1}{Gp_0} \big(\terminate + \frac{\gamma^m}{(1-\gamma) G^3 p_0^3}\big) \le \frac{1}{2}$. 
For any two states $\omega_i, \omega_j \in \Omega$, the output $\hat \rho_{ij}$ of Algorithm \ref{alg:any-pair-of-states} satisfies $\hat \rho_{ij} \in (1\pm 4\eps') \frac{\rprior(\omega_i)}{\rprior(\omega_j)}$.  Algorithm~\ref{alg:any-pair-of-states} terminates in at most $2(\frac{1}{p_0} + m ) \log_2 \frac{1}{Gp_0 \terminate}$ periods in expectation. 
\end{lemma}

\begin{proof}
The proof uses Lemma~\ref{lem:ratio} twice. See details below. 
Let $\eps' = \frac{1}{Gp_0} \big(\terminate + \frac{\gamma^m}{(1-\gamma) G^3 p_0^3}\big)$. 
If $(\omega_i, \omega_j)$ is a pair of distinguishable states, then Lemma~\ref{lem:ratio} shows that the estimate $\hat \rho_{ij}$ returned by Algorithm \ref{alg:binary-search} satisfies $\hat \rho_{ij} \in (1\pm \eps') \frac{\rprior(\omega_i)}{\rprior(\omega_j)}$. 

If $(\omega_i, \omega_j)$ is not a pair of distinguishable states, then by Lemma~\ref{lem:ratio}, we have the ratio estimates $\hat \rho_{ik} \in (1\pm \eps') \frac{\rprior(\omega_i)}{\rprior(\omega_k)}$ and $\hat \rho_{jk} \in (1\pm \eps') \frac{\rprior(\omega_j)}{\rprior(\omega_k)}$.
So,  
\begin{align*}
    \hat \rho_{ij} ~ = ~ \frac{\hat \rho_{ik}}{\hat \rho_{jk}} ~ \ge ~ \frac{\frac{\rprior(\omega_i)}{\rprior(\omega_k)}(1-\eps')}{\frac{\rprior(\omega_j)}{\rprior(\omega_k)}(1+\eps')} & ~ \ge ~ \frac{\rprior(\omega_i)}{\rprior(\omega_j)}(1-2\eps')  
\end{align*}
and
\begin{align*}
    \hat \rho_{ij} ~ = ~ \frac{\hat \rho_{ik}}{\hat \rho_{jk}} ~ \le ~ \frac{\frac{\rprior(\omega_i)}{\rprior(\omega_k)}(1+\eps')}{\frac{\rprior(\omega_j)}{\rprior(\omega_k)}(1-\eps')} & ~ \le ~ \frac{\rprior(\omega_i)}{\rprior(\omega_j)}(1+4\eps')  
\end{align*}
given $\eps' \le \frac{1}{2}$.  
The multiplicative error of $\hat \rho_{ij}$ is thus at most $4\eps'$. 

Algorithm~\ref{alg:any-pair-of-states} needs twice the running time of Algorithm \ref{alg:binary-search}, which is
%so the expected termination of Algorithm~\ref{alg:any-pair-of-states} takes at most
at most $2(\frac{1}{p_0} + m ) \log_2 \frac{1}{Gp_0 \terminate}$ periods in expectation by Lemma \ref{lem:ratio}.
\end{proof}

\begin{lemma} \label{lemma:good-prior-estimation-old}
Suppose $\frac{1}{Gp_0} \big(\terminate + \frac{\gamma^m}{(1-\gamma) G^3 p_0^3}\big) \le \frac{1}{2}$.
\begin{itemize}
    \item The estimation $\hat \mu$ returned by Algorithm \ref{alg:learn-prior} satisfies 
 $\| \hat \mu - \rprior \|_1 \le \tfrac{12|\Omega|}{Gp_0^3} \big(\terminate + \tfrac{\gamma^m}{(1-\gamma) G^3 p_0^3}\big)$.
 \item The expected running time of Algorithm \ref{alg:learn-prior} is at most $2(|\Omega|-1)(\tfrac{1}{p_0} + m ) \log_2 \tfrac{1}{Gp_0 \terminate}$ periods. 
\end{itemize}
\end{lemma}

\begin{proof}
By Lemma~\ref{lem:relax-ratio}, the estimation $\hat \rho_{i1}$ satisfies $\hat \rho_{i1} \in (1\pm 4 \eps') \frac{\rprior(\omega_i)}{\rprior(\omega_1)}$ with $\eps' = \frac{1}{Gp_0} \big(\terminate + \frac{\gamma^m}{(1-\gamma) G^3 p_0^3}\big)$. Because $\frac{\rprior(\omega_i)}{\rprior(\omega_1)} \le \frac{1}{p_0}$, we have
\begin{align*}
    \Big| \hat \rho_{i1} - \tfrac{\rprior(\omega_i)}{\rprior(\omega_1)} \Big| \le \tfrac{4\eps'}{p_0}. 
\end{align*}
Since $\rprior$ is a probability distribution, we have $1 = \sum_{\omega \in \Omega} \rprior(\omega) = \rprior(\omega_1) + \rprior(\omega_1) \sum_{i=2}^{|\Omega|} \frac{\rprior(\omega_i)}{\rprior(\omega_1)}$, so
\begin{align*}
    \rprior(\omega_1) = \frac{1}{1 + \sum_{i=2}^{|\Omega|} \frac{\rprior(\omega_i)}{\rprior(\omega_1)}}. 
\end{align*}
Because the function $f(x) = \frac{1}{1+x}$ is $1$-Lipschitz ($|f'(x)| = \frac{1}{(1+x)^2} \le 1$), we have 
\begin{align*}
    | \hat \mu(\omega_1) - \rprior(\omega_1) | = \Big| \frac{1}{1 + \sum_{i=2}^{|\Omega|} \hat \rho_{i1}} - \frac{1}{1 + \sum_{i=2}^{|\Omega|} \frac{\rprior(\omega_i)}{\rprior(\omega_1)}}\Big| \le \Big| \sum_{i=2}^{|\Omega|} \hat \rho_{i1} - \sum_{i=2}^{|\Omega|} \tfrac{\rprior(\omega_i)}{\rprior(\omega_1)} \Big| \le |\Omega| \tfrac{4 \eps'}{p_0}. 
\end{align*}
Then, for every $i = 2, \ldots, |\Omega|$. 
\begin{align*}
    | \hat \mu(\omega_i) - \rprior(\omega_i) | & = \big| \hat \rho_{i1} \hat \mu(\omega_1) - \tfrac{\rprior(\omega_i)}{\rprior(\omega_1)} \rprior(\omega_1) \big| \\
    & \le \big| \hat \rho_{i1} - \tfrac{\rprior(\omega_i)}{\rprior(\omega_1)} \big| \hat \mu(\omega_1) ~ + ~ \tfrac{\rprior(\omega_i)}{\rprior(\omega_1)} \big| \hat \mu(\omega_1) -  \rprior(\omega_1) \big| \\
    & \le \tfrac{4\eps'}{p_0} \hat \mu(\omega_1) ~ + ~ \tfrac{\rprior(\omega_i)}{\rprior(\omega_1)} |\Omega| \tfrac{4 \eps'}{p_0}. 
\end{align*}
Therefore,
\begin{align*}
    \| \hat \mu - \rprior \|_1 & = |\hat \mu(\omega_1) - \rprior(\omega_1) |  +  \sum_{i=2}^{|\Omega|} | \hat \mu(\omega_i) - \rprior(\omega_i) | \\
    & \le |\Omega| \tfrac{4 \eps'}{p_0} + |\Omega|\tfrac{4\eps'}{p_0} \hat \mu(\omega_1) + |\Omega|\tfrac{4\eps'}{p_0} \sum_{i=2}^{|\Omega|} \tfrac{\rprior(\omega_i)}{\rprior(\omega_1)} \\
    & \le |\Omega|\tfrac{8\eps'}{p_0} + |\Omega| \tfrac{4\eps'}{p_0} \tfrac{1}{p_0} \\
    & \le \tfrac{12|\Omega|\eps'}{p_0^2} \\
    & = \tfrac{12|\Omega|}{Gp_0^3} \big(\terminate + \tfrac{\gamma^m}{(1-\gamma) G^3 p_0^3}\big), 
\end{align*}
which proves the first claim of Lemma \ref{lemma:good-prior-estimation-old}. 

% Algorithm \ref{alg:learn-prior} terminates in $\le \tfrac{2 |\Omega|}{p_0} \log_2 \tfrac{1}{G^2p_0 \eps}$ periods because it runs Algorithm \ref{alg:any-pair-of-states} for $|\Omega|-1$ times. 

Because Algorithm \ref{alg:learn-prior} runs Algorithm \ref{alg:any-pair-of-states} for $|\Omega|-1$ times and the expected running time of Algorithm \ref{alg:any-pair-of-states} is at most $2(\frac{1}{p_0} + m ) \log_2 \frac{1}{Gp_0 \terminate}$ periods (by Lemma \ref{lem:relax-ratio}), the expected running time of Algorithm~\ref{alg:any-pair-of-states} is at most
\begin{align*} % \label{eq:T_0-small}
   2(|\Omega|-1)(\tfrac{1}{p_0} + m ) \log_2 \tfrac{1}{Gp_0 \terminate}.
\end{align*}
periods. 
\end{proof}

\begin{proof}[\bf Proof of Lemma \ref{lemma:good-prior-estimation}]
Lemma \ref{lemma:good-prior-estimation} follows from Lemma \ref{lemma:good-prior-estimation-old} by plugging in
\[\terminate = \frac{G p_0^3 \eps}{24|\Omega|}, \quad m = \Big\lceil \frac{1}{\ln(1/\gamma)}\ln(\frac{24|\Omega|}{(1-\gamma) G^4 p_0^6 \eps}) \Big\rceil. \]
\end{proof}

\subsection{Proof of Lemma \ref{lem:universal-robustification}}
\label{app:proof:universal-robustification}

Let $\hat \mu \in \Delta(\Omega)$ be a receiver prior satisfying $\min_{\omega \in \Omega} \hat \mu(\omega) \ge p_0 > 0$.   Let $B_1(\hat \mu, \eps) = \{\mu: \|\mu - \hat \mu \|_1 \le \eps \}$ be the set of priors within distance $\eps$ to $\hat \mu$. Suppose $\eps \le \frac{p_0^2D}{4}$. 
Let $\pi$ be a direct signaling scheme.
We want to construct another direct signaling scheme $\tilde \pi$ in two steps: 
(1) First, construct a non-direct signaling scheme $\tilde \pi^\circ$.  
(2) Then, convert $\tilde \pi^\circ$ into a direct signaling scheme $\tilde \pi$ that still satisfies the requirements of the lemma. 

\paragraph{Step (1): Construct a non-direct signaling scheme $\tilde \pi^\circ$.}
Let $\delta$ satisfy 
\begin{align*}
    \tfrac{2\eps}{p_0D} \le \delta \le \tfrac{p_0}{2}.
\end{align*}
Let $P_{\hat \mu, \pi}(a) = \sum_{\omega \in \Omega} \hat \mu(a) \pi(a | \omega)$ be the unconditional probability that $\pi$ sends signal $a$ under prior $\hat \mu$.  Let $\hat \mu_{a, \pi} \in \Delta(\Omega)$ be the posterior belief induced by signal $a$ under signaling scheme $\pi$ and prior $\hat \mu$: 
\begin{align*}
    \hat \mu_{a, \pi}(\omega) = \frac{\hat \mu(\omega) \pi(a|\omega)}{P_{\hat \mu, \pi}(a)}, \quad \forall \omega \in \Omega. 
\end{align*}
According to Assumption~\ref{ass:D}, there exists a belief $\eta_a \in \Delta(\Omega)$ for which $\E_{\omega \sim \eta_a}[v(a, \omega) - v(a', \omega)] \ge D, \forall a'\ne a$.  Consider the convex combination of $\hat \mu_{a, \pi}$ and $\eta_a$ with coefficients $1-\delta, \delta$: 
\begin{equation}\label{eq:delta-combination}
    \xi_a = (1 - \delta) \hat \mu_{a, \pi} + \delta \eta_a. 
\end{equation}
By the linearity of expectation, we have
\begin{align}\label{eq:delta-D-2}
\E_{\omega \sim \xi_a}[v(a, \omega) - v(a', \omega)] \nonumber & = (1-\delta)\E_{\omega \sim \hat \mu_{a, \pi}}[v(a, \omega) - v(a', \omega)] + \delta \E_{\omega \sim \eta_a}[v(a, \omega) - v(a', \omega)] \\
& \ge (1-\delta)\E_{\omega \sim \hat \mu_{a, \pi}}[v(a, \omega) - v(a', \omega)]  + \delta D. 
\end{align}
Let $\xi = \sum_{a\in A} P_{\hat \mu, \pi}(a) \xi_a \in \Delta(\Omega)$, and write $\hat \mu$ as the convex combination of $\xi$ and another belief $\chi \in \Delta(\Omega)$, with some coefficient $y \in [0, 1]$: 
\begin{align} \label{eq:convex-combination-2}
    \hat \mu ~ = ~ (1-y) \xi + y \chi ~ = ~ \sum_{a\in A} (1-y) P_{\hat \mu, \pi}(a) \xi_a ~ + ~ y \chi.
\end{align}

\begin{lemma}[Proposition 1 of \cite{zu_learning_2021}]
\label{lem:small-y}
If $\delta \le \frac{p_0}{2}$, then there exist $\chi$ on the boundary of $\Delta(\Omega)$ and $y \le \frac{\delta}{p_0} \le \frac{1}{2}$ that satisfy \eqref{eq:convex-combination-2}. 
\end{lemma}

Since \eqref{eq:convex-combination-2} is a convex decomposition of the prior $\hat \mu$, according to \cite{kamenica_bayesian_2011}, there exists a signaling scheme $\tilde \pi^\circ$ that induces posterior $\xi_a$ with total probability $(1-y) P_{\hat \mu, \pi}(a)$, $\forall a\in A$, and induces the posterior that puts all probability on $\omega$ with total probability $y \chi(\omega)$, $\forall \omega \in \Omega$.  Namely, $\tilde \pi^\circ$ has signal space $S = A \cup \Omega$ and conditional probability
\begin{align} \label{eq:non-direct-scheme-definition}
    \tilde \pi^\circ(s | \omega) = \begin{cases}
    \frac{(1-y) P_{\hat \mu, \pi}(a) \xi_a(\omega)}{\hat \mu(\omega)} & \text{ for } s = a \in A; \\
    \frac{y\chi(\omega)}{\hat \mu(\omega)} & \text{ for } s = \omega \in \Omega; \\
    0 & \text{ otherwise.}
    \end{cases}
\end{align}
It is not hard to verify that, under prior $\hat \mu$ and signaling scheme $\tilde \pi^\circ$, the posterior induced by signal $a \in A$ is equal to $\xi_a$, and the posterior induced by signal $\omega$ is the deterministic distribution on $\omega$. 

% We show that, whenever $\tilde \pi^\circ$ sends an action recommendation $a \in A$, the recommendation satisfies the requirement of the lemma, under all receiver priors in $B_1(\hat \mu, \eps)$.
\begin{claim}\label{claim:tilde-pi-improvement}
For any prior of the receiver $\mu \in B_1(\hat \mu, \eps)$,
given any action recommendation $a\in A$ from $\tilde \pi^\circ$, the receiver's posterior $\mu_{a, \tilde \pi^\circ}$ satisfies
\[ \E_{\omega \sim \mu_{a, \tilde \pi^\circ}}[v(a, \omega) - v(a', \omega)] \ge (1-\delta)\E_{\omega \sim \hat \mu_{a, \pi}}[v(a, \omega) - v(a', \omega)]  + \delta D - \tfrac{2\eps}{p_0}, ~ \forall a' \in A\setminus\{a\}.\]
And when $\tilde \pi^\circ$ sends signal $\omega \in \Omega$, the receiver will believe that the state is $\omega$ deterministically. 
\end{claim}
\begin{proof}
Under signaling scheme $\tilde \pi^\circ$, let $\mu_{a, \tilde \pi^\circ}$, $\xi_a$ be the posteriors induced by signal $a$ from priors $\mu$, $\hat \mu$.
By the continuity of posterior (Lemma~\ref{lem:continuity-posterior}), we have $\| \mu_{a, \tilde \pi^\circ} - \xi_a \|_1 \le \tfrac{2}{p_0} \| \mu - \hat \mu \|_1 \le \frac{2\eps}{p_0}$.  Then, since the receiver's utility is in $[0, 1]$, for any action $a' \ne a$,
% we have
% \begin{align*}
%     \big| \E_{\omega \sim \mu_a}[v(a, \omega) - v(a', \omega)] - \E_{\omega \sim \xi_a}[v(a, \omega) - v(a', \omega)] \big| \le \| \mu_a - \xi_a \|_1 \le \tfrac{2\eps}{p_0}.
% \end{align*}
% Together with \eqref{eq:delta-D}, we get
we have 
\begin{align*}
    \E_{\omega \sim \mu_{a, \tilde \pi^\circ}}[v(a, \omega) - v(a', \omega)] & ~ \ge ~ \E_{\omega \sim \xi_a}[v(a, \omega) - v(a', \omega)] - \tfrac{2\eps}{p_0} \\
    \text{by \eqref{eq:delta-D-2}} & ~ \ge ~ (1-\delta)\E_{\omega \sim \hat \mu_{a, \pi}}[v(a, \omega) - v(a', \omega)]  + \delta D - \tfrac{2\eps}{p_0}. 
\end{align*}
%with the choice of $\delta = \tfrac{2\eps}{p_0D}$. 
\end{proof}

% We then show that the signaling scheme $\tilde \pi$ reveals the state $\omega$ only with a small probability: 
% \begin{claim}\label{claim:pi-tilde-reveal-state}
% Under prior $\hat \mu$, the probability that $\tilde \pi$ reveals the state is $\sum_{\omega \in \Omega} \hat\mu(\omega) \tilde \pi(\omega | \omega) = y \le \frac{\delta}{p_0}$.
% \end{claim}
% \begin{proof}
% The probability that $\tilde \pi$ reveals the state is 
% \begin{align*}
%     \sum_{\omega \in \Omega} \hat\mu(\omega) \sum_{\omega'\in \Omega}\tilde \pi(\omega' | \omega) =  \sum_{\omega \in \Omega} \hat\mu(\omega) \pi(\omega | \omega) = \sum_{\omega \in \Omega} \hat\mu(\omega) \tfrac{y\chi(\omega)}{\hat \mu(\omega)} = y \le \tfrac{\delta}{p_0}. 
% \end{align*}
% \end{proof}

We show that the recommendation part of signaling scheme $\tilde \pi^\circ$ is close to $\pi$, in the following sense: 
\begin{claim}
\label{claim:signaling_scheme_recommendation_close}
For every $\omega \in \Omega$, the $\ell_1$ distance $\| \tilde \pi^\circ(\cdot|\omega) - \pi(\cdot|\omega) \|_1 = \sum_{a\in A} \big| \tilde \pi^\circ(a|\omega) - \pi(a|\omega) \big| \le \frac{3\delta}{p_0}$.  
\end{claim}
\begin{proof}
By definition,
\begin{align*}
    \sum_{a\in A} \big| \tilde \pi^\circ(a|\omega) - \pi(a|\omega) \big | & ~ = ~ \sum_{a \in A} \Big| \frac{(1-y) P_{\hat \mu, \pi}(a) \xi_a(\omega)}{\hat \mu(\omega)} - \frac{P_{\hat \mu, \pi}(a) \hat \mu_a(\omega)}{\hat \mu(\omega)} \Big| \\
    & ~ \le ~ \sum_{a\in A} \bigg( \frac{P_{\hat \mu, \pi}(a)}{\hat \mu(\omega)} \cdot \big| \xi_a(\omega) - \hat \mu_a(\omega) \big| ~ + ~ y \cdot \frac{P_{\hat \mu, \pi}(a) \xi_a(\omega)}{\hat \mu(\omega)} \bigg) \\
    \text{by \eqref{eq:delta-combination} and $y\le \tfrac{1}{2}$} & ~ \le ~ \sum_{a\in A} \bigg( \frac{P_{\hat \mu, \pi}(a)}{\hat \mu(\omega)} \cdot \delta ~ + ~ 2 y \cdot \frac{(1-y) P_{\hat \mu, \pi}(a) \xi_a(\omega)}{\hat \mu(\omega)} \bigg).
\end{align*}
We have $\sum_{a\in A} P_{\hat \mu, \pi}(a) = 1$ and $\sum_{a\in A} (1-y) P_{\hat \mu, \pi}(a) \xi_a(\omega) \le \hat \mu(\omega)$ by \eqref{eq:convex-combination-2}. So, the above is at most 
\begin{equation*}
    ~ \le ~ \frac{1}{\hat \mu(\omega)} \cdot \delta + 2y \cdot 1 ~ \le ~ \frac{\delta}{p_0} + 2 \frac{\delta}{p_0} ~ = ~ \frac{3\delta}{p_0}, 
\end{equation*}
given $y \le \frac{\delta}{p_0}$. 
% where the third ``$\le$'' is because $\frac{(1-y) P_{\hat \mu, \pi}(a) \xi_a(\omega)}{\hat \mu(\omega)} \le 1$ and $y\le \frac{1}{2}$.
\end{proof}

Then, we show that the information designer's utility under scheme $\tilde \pi^\circ$ is close to her utility under $\pi$:  
\begin{claim}\label{claim:designer-utility-close}
The designer's utility 
$U(\tilde \pi^\circ, \rho_\ob; \dprior) \ge U(\pi, \rho_\ob; \dprior) - \tfrac{3\delta}{p_0}$. 
\end{claim}
\begin{proof}
For the non-direct signaling scheme $\tilde \pi^\circ$, the notation $\rho_\ob$ means: when $\tilde \pi^\circ$ recommends action, the receiver takes the recommended action, and when $\tilde \pi^\circ$ reveals the state, the receiver takes an optimal action for that state. 
The designer's expected utility is thus 
\begin{align*}
    U(\tilde \pi^\circ, \rho_\ob; \dprior) & = \sum_{\omega \in \Omega} \dprior(\omega) \Big( \sum_{a\in A} \tilde \pi^\circ(a | \omega) u(a, \omega) ~ + ~ \tilde \pi^\circ(\omega | \omega) u(a^*_\omega, \omega) \Big) \\
    & \ge \sum_{\omega \in \Omega} \dprior(\omega) \sum_{a\in A} \tilde \pi^\circ(a|\omega) u(a, \omega) \\
    % & \ge \sum_{\omega \in \Omega} \dprior(\omega) \sum_{a\in A} \pi(a|\omega) u(a, \omega) - \tfrac{3\delta}{p_0} \cdot 1 \quad  \\
    & \ge U(\pi, \rho_\ob; \dprior) - \tfrac{3\delta}{p_0} \quad \text{by Claim~\ref{claim:signaling_scheme_recommendation_close} and $U$ is $1$-Lipschitz in $\pi$ (Lemma \ref{lem:utility-Lipschitz})}. 
\end{align*}
\end{proof}

\paragraph{Step (2): Convert $\tilde \pi^\circ$ into a direct signaling scheme $\tilde \pi$.}
Then, we convert $\tilde \pi^\circ$ (whose signal space is $A\cup \Omega$) into a direct signaling scheme $\tilde \pi$ (whose signal space is $A$) by coalescing the signals. Specifically, for each state $\omega \in \Omega$, and action $a \in A$, let $\tilde \pi$ send signal $a$ when $\tilde \pi^\circ$ sends signal $a$ and when $\tilde \pi^\circ$ sends signal $\omega$ if the receiver's optimal action $a^*_\omega \in \argmax_{a\in A} v(a, \omega)$ for state $\omega$ equals $a$:
\begin{align} \label{eq:coalescing-definition}
    \tilde \pi(a | \omega) = \tilde \pi^\circ(a | \omega) + \mathbbm{1}[a = a^*_\omega] \cdot \tilde \pi^\circ(\omega | \omega). 
\end{align}
% Note that $\tilde \pi$ is a valid signaling scheme because $\sum_{a\in A} \tilde \pi(a | \omega) = \sum_{a\in A} \tilde \pi^\circ(a | \omega) + \tilde \pi^\circ(\omega | \omega) = 1$.
By the coalescing lemma \ref{lemma:coalesce-2}, when $\tilde \pi$ recommends action $a$, we have: for a receiver with any prior $\mu \in B_1(\hat \mu, \eps)$, the posterior $\mu_{a, \tilde \pi}$ satisfies 
\begin{align*}
    \E_{\omega \sim \mu_{a, \tilde \pi}}\big[ v(a, \omega) - v(a', \omega) \big] & ~ \ge ~ \min \Big\{ \E_{\omega \sim \mu_{a, \tilde \pi^\circ}}\big[ v(a, \omega) - v(a', \omega) \big], ~ 0 \Big\} \\
    (\text{by Claim \ref{claim:tilde-pi-improvement}}) & ~ \ge ~ \min \Big\{ (1-\delta) \E_{\omega \sim \hat \mu_{a, \pi}}\big[ v(a, \omega) - v(a', \omega) \big] + \delta D - \tfrac{2\eps}{p_0}, ~ 0 \Big\}. 
\end{align*}

We verify that $\tilde \pi$ satisfies the condition for the receiver's posterior in Lemma \ref{lem:universal-robustification}. There are two cases: 
\begin{itemize}
\item If $\E_{\omega \sim \hat \mu_{a, \pi}}\big[ v(a, \omega) - v(a', \omega) \big] \ge 0$, then because $\delta D - \frac{2\eps}{p_0} \ge 0$ by our condition on $\delta$, we have $\min \big\{ (1-\delta) \E_{\omega \sim \hat \mu_{a, \pi}}\big[ v(a, \omega) - v(a', \omega) \big] + \delta D - \tfrac{2\eps}{p_0}, ~ 0 \big\} = 0$, so
\begin{align*}
    \E_{\omega \sim \mu_{a, \tilde \pi}} \big[ v(a, \omega) - v(a', \omega) \big] ~ \ge ~ 0 ~ = ~ \min \Big\{\E_{\omega \sim \hat \mu_{a, \pi}}\big[ v(a, \omega) - v(a', \omega) \big] + \delta D - \tfrac{2\eps}{p_0}, ~ 0 \Big\}. 
\end{align*}

\item If $\E_{\omega \sim \hat \mu_{a, \pi}}\big[ v(a, \omega) - v(a', \omega) \big] < 0$, then we have $ 0 > (1-\delta) \E_{\omega \sim \hat \mu_{a, \pi}}\big[ v(a, \omega) - v(a', \omega) \big] \ge \E_{\omega \sim \hat \mu_{a, \pi}}\big[ v(a, \omega) - v(a', \omega) \big]$. So, 
\begin{align*}
    \E_{\omega \sim \mu_{a, \tilde \pi}} \big[ v(a, \omega) - v(a', \omega) \big] & ~ \ge ~ \min \Big\{(1-\delta) \E_{\omega \sim \hat \mu_{a, \pi}}\big[ v(a, \omega) - v(a', \omega) \big] + \delta D - \tfrac{2\eps}{p_0}, ~ 0 \Big\}\\
    & ~ \ge ~ \min \Big\{\E_{\omega \sim \hat \mu_{a, \pi}}\big[ v(a, \omega) - v(a', \omega) \big] + \delta D - \tfrac{2\eps}{p_0}, ~ 0 \Big\}. 
\end{align*}
\end{itemize}
In both cases, $\tilde \pi$ satisfies the condition. 

We then verify the condition for the designer's utility, which is $U(\tilde \pi, \rho_\ob; \dprior) \ge U(\pi, \rho_\ob; \dprior) - \tfrac{3\delta}{p_0}$. It directly follows from Claim \ref{claim:designer-utility-close}, which shows $U(\tilde \pi^\circ, \rho_\ob; \dprior) \ge U(\pi, \rho_\ob; \dprior) - \tfrac{3\delta}{p_0}$, and Lemma \ref{lemma:coalesce-2}, which shows $U(\tilde \pi^\circ, \rho_\ob; \dprior) = U(\tilde \pi, \rho_\ob; \dprior)$.

Finally, we verify the closeness condition between $\tilde \pi$ and $\pi$. For every $\omega \in \Omega$, 
\begin{align*}
    \sum_{a \in A} \big| \tilde \pi(a | \omega) - \pi(a|\omega) \big| & ~ \le ~ \sum_{a\in A} \Big( \big| \tilde \pi^\circ(a|\omega)  - \pi(a|\omega) \big| +  \mathbbm{1}[a = a^*_\omega] \cdot \tilde \pi^\circ(\omega | \omega) \Big)  \qquad \text{by \eqref{eq:coalescing-definition}} \\
    & ~ \le ~ \tfrac{3\delta}{p_0} + y \tfrac{\chi(\omega)}{\hat \mu(\omega)}  \qquad \quad\qquad \text{by Claim \ref{claim:signaling_scheme_recommendation_close} and Equation \eqref{eq:non-direct-scheme-definition}} \\
    & ~ \le ~ \tfrac{3\delta}{p_0} + \tfrac{\delta}{p_0^2} ~ \qquad \qquad \qquad \text{because $y\le \tfrac{\delta}{p_0}$ and $\hat \mu(\omega) \ge p_0$} \\
    & ~ \le ~ \tfrac{4 \delta}{p_0^2}. 
\end{align*}

\section{Omitted Proofs in Section \ref{sec:characterization-dynamic-benchmark}}

\subsection{Proof of Lemma \ref{lemma:dynamic-revelation-principle}}
\label{proof:dynamic-revelation-principle}

%\begin{proof}[Proof of Lemma~\ref{lemma:dynamic-revelation-principle}]
Fix any feasible designer strategy $\sigma=(\pi_\sigma^{(t)})_{t=1}^T$ and let the receiver play a (history--dependent) best response $\phi^*(\sigma,\rprior)$.  
For each period $t$ and realized public history $h_t$ (signals and actions up to $t\!-\!1$), $\pi_{\sigma}^{(t)}$  induces a posterior $\mu_t(\cdot\mid h_t,s_t)$ after signal $s_t\sim\pi_\sigma^{(t)}(\cdot\mid \omega,h_t)$ is sent under $\rprior$.  
Let $\mathcal{B}_t(h_t,s_t)\subseteq A$ be the nonempty set of myopic best replies at $(h_t,s_t)$:
\[
\mathcal{B}_t(h_t,s_t)\;:=\;\arg\max_{a\in A}~V_t(a;\mu_t(\cdot\mid h_t,s_t)),
\]
and fix a measurable tie--breaking rule $\tau_t$ so that $b_t(h_t,s_t):=\tau_t(\mathcal{B}_t(h_t,s_t))\in\mathcal{B}_t(h_t,s_t)$.

\smallskip
\emph{Direct, dynamically persuasive transformation.}  
Define a \emph{direct} signal in period $t$ by post--processing the original signal via 
\[
g_t:\; s_t\mapsto a_t'=b_t(h_t,s_t)\in A,
\]
and set $\pi_{\sigma^d}^{(t)}(\cdot\mid \omega,h_t):=\big(g_t\circ \pi_\sigma^{(t)}(\cdot\mid \omega,h_t)\big)$, where $\sigma^d=(\pi_{\sigma^d}^{(t)})_{t=1}^T$ denotes the resulting strategy.  
Because $g_t$ is a (history--dependent) garbling of $s_t$, the posterior belief $\mu_t(\cdot\mid h_t,s_t)$ induced by $\pi_\sigma^{(t)}$ is the same posterior belief the receiver holds about states upon receiving the recommendation $a_t'=g_t(s_t)$ under strategy $\sigma^d$ (Bayes plausibility is preserved by post--processing).  
By construction, $a_t'$ maximizes $V_t(\cdot;\mu_t)$, so obeying the recommendation is optimal for the receiver at each $(h_t,a_t')$. Therefore there is a best response $\phi_{\mathrm{ob}}$ under $\sigma^d$ that \emph{obeys} the recommendation every period. That is, $\phi_{\mathrm{ob}}=\phi^*(\sigma^d,\rprior)$ and $\sigma^d$ is dynamically persuasive.

\smallskip
\emph{Outcome equivalence and optimality.}  
Consider the outcome path generated by $(\sigma,\phi^*(\sigma,\rprior))$. At each $(h_t,s_t)$, the action actually taken is some $a_t\in\mathcal{B}_t(h_t,s_t)$.  
Under $(\sigma^d,\phi_{\mathrm{ob}})$, the action taken at the same $(h_t,s_t)$ is exactly $a_t'=b_t(h_t,s_t)\in\mathcal{B}_t(h_t,s_t)$.  
Hence, conditional on $(\omega,h_t,s_t)$, the action under $\sigma^d$ is a selection from the same set of best replies as under $\sigma$; in particular, the receiver payoff is weakly higher and the distribution over next--period histories (and thus all future posteriors) is preserved.  
Therefore the designer's expected payoff is unchanged:
\[
\mathcal U_T(\sigma^d,\phi_{\mathrm{ob}};\dprior)
~=~\mathcal U_T(\sigma,\phi^*(\sigma,\rprior);\dprior).
\]
Applying that transformation to an optimal $\sigma$ yields a history--dependent, dynamically persuasive direct strategy $\sigma^*$ that attains the dynamic optimum:
\[
\mathcal U^{**}_T(\mathcal I)
~=~\mathcal U_T(\sigma^*,\phi_{\mathrm{ob}};\dprior)
\quad\text{with}\quad
\phi_{\mathrm{ob}}=\phi^*(\sigma^*,\rprior).
\]

\subsection{Proof of Lemma \ref{lemma:exist-ddp-independent}}
\label{proof:exist-ddp-independent}
Fix an optimal DDP strategy $\sigma^*$ (possibly history-dependent) with an obedient receiver strategy $\phi_{\ob}$, over a finite horizon $t=1,\dots,T$. Let the state process $\{\omega_t\}_{t=1}^T$ be i.i.d.\ with (true) marginal $\lambda\in\Delta(\Omega)$, independent of past play. For each period $t$ and action $a\in A$, define the on-path joint and conditional probabilities induced by $(\sigma,\phi_{\ob})$:
\[
Q_t(\omega,a)\;:=\;\Pr_{\sigma,\phi_{\ob}}\!\big(\omega_t=\omega,\ a_t=a\big),\qquad
P_t(a\mid\omega)\;:=\;\Pr_{\sigma,\phi_{\ob}}\!\big(a_t=a\mid \omega_t=\omega\big),
\]
where $a_t$ denotes the action recommended (and, on-path, taken) at $t$. By the law of total probability, $Q_t(\omega,a)=\lambda(\omega)\,P_t(a\mid\omega)$.

\medskip
%\noindent\textbf{Construction.}
For each $t$, define a \emph{history-independent} direct scheme $\pi_{\tilde \sigma}^{(t)}:\Omega\to\Delta(A)$ by
\[
\tilde\pi^{(t)}(a\mid\omega)\;:=\;P_t(a\mid\omega)\qquad(\text{choose arbitrarily on }\{\omega:\lambda(\omega)=0\}).
\]
Let $\tilde\sigma=(\pi_{\tilde \sigma}^{(1)},\ldots,\pi_{\tilde \sigma}^{(T)})$, and keep the same off-path continuation rule as under $\sigma^*$: following any deviation, switch to the no-information punishment policy as in $\sigma^*$ for the remainder of the game. % (This preserves the continuation values used in the obedience tests.)

\medskip
%\noindent\textit{(i) Outcome equivalence on the obedience path.}
By construction,
\[
\Pr_{\tilde\sigma,\phi_{\ob}}\!\big(\omega_t=\omega,\ a_t=a\big)
\;=\;\lambda(\omega)\,\pi_{\tilde \sigma}^{(t)}(a\mid\omega)
\;=\;\lambda(\omega)\,P_t(a\mid\omega)
\;=\;Q_t(\omega,a),
\]
so the joint law of $(\omega_t,a_t)$ is identical under $(\sigma^*,\phi_{\ob})$ and $(\tilde\sigma,\phi_{\ob})$ for every $t$. Because payoffs are additively separable across periods, the designer’s total expected payoff is the same under $\tilde\sigma$ as under $\sigma^*$. Thus, $\tilde\sigma$ is optimal if $\sigma$ is.

\medskip
%\noindent\textit{(ii) Dynamic persuasiveness is preserved.}
Fix a period $t$ and a recommended action $a$ with $\Pr(a_t=a)>0$. Let $\mu_R$ be the receiver’s prior. Under $\tilde\sigma$, Bayes’ rule gives the posterior
\[
\mu^{\tilde\sigma}_t(\omega\mid a)\;=\;
\frac{\mu_R(\omega)\,\pi_{\tilde \sigma}^{(t)}(a\mid\omega)}
{\sum_{\omega'}\mu_R(\omega')\,\pi_{\tilde \sigma}^{(t)}(a\mid\omega')}
\;=\;
\frac{\mu_R(\omega)\,P_t(a\mid\omega)}
{\sum_{\omega'}\mu_R(\omega')\,P_t(a\mid\omega')}.
\]
The right-hand side is exactly the \emph{average} of the posteriors induced by $(\sigma^*,\phi_{\ob})$ at all on-path histories that yield recommendation $a$, weighted by their probability conditional on $a$ (law of iterated expectations). So we have
\[
\mu^{\tilde\sigma}_t(\cdot\mid a)\;=\;\overline{\mu}^{\sigma^*}_t(\cdot\mid a)
\]
where $\overline{\mu}^{\sigma^*}_t(\cdot\mid a)$ is the averaged posterior faced by the receiver under $(\sigma^*,\phi_{\ob})$ when recommendation $a$ is observed.

Let $W^{g}_{t+1}$ (resp.\ $W^{p}_{t+1}$) denote the receiver’s continuation utility from $t{+}1$ onward on the obedience (resp.\ punishment) path. By our construction of off-path behavior, these continuations are the same under $\tilde\sigma$ as under $\sigma$:
\[
W^{g,\tilde\sigma}_{t+1}=W^{g,\sigma^*}_{t+1},\qquad
W^{p,\tilde\sigma}_{t+1}=W^{p,\sigma^*}_{t+1}.
\]
Therefore, for every deviation $a'\neq a$, the period-$t$ one-shot deviation inequality under $\tilde\sigma$,
\[
\mathbb{E}_{\omega\sim \mu^{\tilde\sigma}_t(\cdot\mid a)}\!\big[v(a,\omega)-v(a',\omega)\big]
\;+\;\big(W^{g,\tilde\sigma}_{t+1}-W^{p,\tilde\sigma}_{t+1}\big)\;\ge\;0,
\]
coincides with the corresponding inequality under $(\sigma^*,\phi_{\ob})$, which holds because $\sigma$ is dynamically persuasive. By the one-shot deviation principle, obedience is optimal at all on-path information sets under $\tilde\sigma$ as well.

\medskip
Combining the above, $\tilde\sigma$ is an \emph{optimal} DDP strategy and is \emph{history-independent} by construction.

\subsection{Proof of Lemma \ref{lem:DP-characterization}}
\label{proof:lem:DP-characterization}
%\paragraph{Step 0 (Restriction to history-independent DDP).}
To begin, we consider the restriction to history-independent DDP.

By Proposition~\ref{prop:dynamic-benchmark-DDP}, the benchmark
\[
\mathcal U^{**}_T(\mathcal I)
=\max_{\sigma}\,\mathcal U_T(\sigma;\phi_{\ob},\mu_D)
\]
is attained by some \emph{history-independent, dynamically persuasive DDP} strategy $\sigma=(\pi_{\sigma}^{(1)},\ldots,\pi_{\sigma}^{(T)})$. So we may restrict attention to this class and impose the dynamic persuasiveness constraint in its \emph{posterior} form \eqref{eq:dynamically-persuasive-definition-2}.

%\paragraph{Step 1 (A promise state and two primitive constraints).}
We then consider a promise state and two primitive constraints.

For any such $\sigma$, define the \emph{promised payoff above the uninformed utility from period $t$ to $T$} as
\[
g_t(\sigma)\;:=\;\sum_{t'=t}^{T}\gamma^{\,t'-t}\Big(V(\pi_{\sigma}^{(t')},\rho_{\ob};\mu_R)-V_{\uninformed}(\mu_R)\Big)\quad(\ge 0).
\]
Dynamic persuasiveness \eqref{eq:dynamically-persuasive-definition-2} at period $t$ recommendation $a$ is equivalent to
\begin{equation}\label{eq:posterior-IC}
\mathbb E_{\omega\sim \mu_R(\cdot\mid a,\pi_{\sigma}^{(t)})}\!\big[v(a,\omega)-v(a',\omega)\big]\;+\;\gamma \cdot \,g_{t+1}(\sigma)\ \ge\ 0\qquad
\forall a'\in A \text{ and on-path } a,
\end{equation}
i.e., a \emph{posterior} single-period deviation inequality with continuation buffer $\gamma \cdot g_{t+1}$.
Moreover, $g_t(\sigma)$ satisfies the constraint of the receiver's \emph{ex ante} obedience
\begin{equation}\label{eq:promise-keeping}
\big(V(\pi_{\sigma}^{(t)},\rho_{\ob};\mu_R)-V_{\uninformed}(\mu_R)\big)\;+\;\gamma\cdot \,g_{t+1}(\sigma)\ \ge\ g_t(\sigma).
\end{equation}

%\paragraph{Step 2 (DP upper-bounds the benchmark).}
Then we show that the dynamic programming solution upper bounds the benchmark.

We show by backward induction on $t$ that for every feasible history-independent dynamically persuasive $\sigma$,
\begin{equation}\label{eq:DP-upper}
\sum_{t'=t}^{T} U(\pi_{\sigma}^{(t')},\rho_{\ob};\mu_D)\ \le\ F\big(t,\,g_t(\sigma)\big).
\end{equation}
For $t=T$, feasibility requires \eqref{eq:posterior-IC} (which reduces to static persuasiveness) and \eqref{eq:promise-keeping} with $g_T$, which is exactly the constraint of \eqref{eq:DP-last-period}. Thus, we have $U(\pi_{\sigma}^{(T)},\rho_{\ob};\mu_D)\le F(T,g_T(\sigma))$.  

Assume \eqref{eq:DP-upper} holds for $t{+}1$. At period $t$, the pair $\big(\pi_{\sigma}^{(t)},\,g_{t+1}(\sigma)\big)$ satisfies the constraints of the Bellman step \eqref{eq:DP-period-t-form-2} with $g=g_t(\sigma)$ by \eqref{eq:posterior-IC} and \eqref{eq:promise-keeping}. Therefore,
\[
\begin{aligned}
\sum_{t'=t}^{T} U(\pi_{\sigma}^{(t')},\rho_{\ob};\mu_D)
& ~ = ~ U(\pi_{\sigma}^{(t)},\rho_{\ob};\mu_D)\;+\;\sum_{t'=t+1}^{T} U(\pi_{\sigma}^{(t')},\rho_{\ob};\mu_D)\\
& ~ \le ~  U(\pi_{\sigma}^{(t)},\rho_{\ob};\mu_D)\;+\;F\big(t{+}1,\,g_{t+1}(\sigma)\big)\\
& ~ \le ~ F\big(t,\,g_t(\sigma)\big),
\end{aligned}
\]
where the last inequality uses the maximization in \eqref{eq:DP-period-t-form-2}.  
Taking $t=1$ and using that $F(1,g)$ is nonincreasing in $g$ (tightening the promise cannot increase designer value), we get
\[
\mathcal U_T(\sigma;\phi_{\ob},\mu_D)\ \le\ F(1,g_1(\sigma))\ \le\ F(1,0).
\]
Maximizing over feasible $\sigma$ yields $\mathcal U^{**}_T(\mathcal I)\le F(1,0)$.

%\paragraph{Step 3 (The DP is tight: construction and persuasiveness).}
We finally show that the dynamic programming solution is tight.

Let $\{(\pi_{\sigma^*}^{(t)}, g_{t+1}^*)\}_{t=1}^{T}$ be maximizers in \eqref{eq:DP-last-period} and \eqref{eq:DP-period-t-form-2} along an optimal path starting at $F(1,0)$ (set $g_1^*:=0$). By definition of $F(t{+}1,g_{t+1}^*)$, there exists a tail strategy $(\pi_{\sigma^*}^{(t+1)},\ldots,\pi_{\sigma^*}^{(T)})$ delivering the promise $g_{t+1}^*$ and satisfying dynamic persuasiveness from $t{+}1$ onward. Concatenating these choices yields a history-independent DDP strategy $\sigma^{*}=(\pi_{\sigma^*}^{(t)})_{t=1}^T$.

We check the feasibility of $\sigma^*$: the pair $(\pi^{(t)\,*},g_{t+1}^*)$ satisfies \eqref{eq:DP-period-t-form-2}:
\begin{align}
V(\pi_{\sigma^*}^{(t)},\rho_{\ob};\mu_R)-V_{\uninformed}(\mu_R)+\gamma\cdot\,g_{t+1}^*&\ge g_t^*, \label{eq:pk-star}\\
\mathbb E_{\omega\sim \mu_R(\cdot\mid a,\pi_{\sigma^*}^{(t)})}\!\big[v(a,\omega)-v(a',\omega)\big]+\gamma\cdot\,g_{t+1}^*&\ge 0\quad
\forall a'\in A,\ \forall \text{ on-path } a. \label{eq:ic-star}
\end{align}
Thus promised utilities are kept \eqref{eq:pk-star}. For dynamic persuasiveness, fix $t$ and an on-path recommendation $a$.
The deviation at period $t$ changes only the \emph{current} action. Then, the DDP strategy switches to the (uninformative) punishment, whose continuation equals the uninformed payoff, while obedience secures an extra $\gamma \cdot g_{t+1}^*$ above that baseline by construction. Thus, the receiver’s \emph{obedience minus deviation} payoff is
\[
\underbrace{\mathbb E[v(a,\omega)-v(a',\omega)\mid a]}_{\text{current gap}}
\;+\;
\underbrace{\gamma\,g_{t+1}^*}_{\text{continuation gap}}
\ \ge\ 0
\quad(\text{by }\eqref{eq:ic-star}),
\]
so obedience is optimal at every on-path history, and thus $\sigma^{*}$ is dynamically persuasive.

Finally, by the Bellman equalities,
\[
\mathcal U_T(\sigma^{*};\phi_{\ob},\mu_D)
= \sum_{t=1}^T U(\pi_{\sigma^*}^{(t)},\rho_{\ob};\mu_D)
= F(1,0).
\]
Combining with the dynamic programming solution upper bounding the benchmark, we obtain $\mathcal U^{**}_T(\mathcal I)=F(1,0)$, attained by the history-independent, dynamically persuasive DDP strategy $\sigma^{*}$.

\subsection{Proof of Proposition \ref{prop:gamma-dynamic-benchmark-increasing}}
\label{proof:prop:gamma-dynamic-benchmark-increasing}

\begin{proposition}\label{prop:gamma-dynamic-benchmark-increasing}
For any instance $\mathcal I$ and any $T\ge 1$, the designer’s dynamic benchmark $\mathcal U^{**}_T(\mathcal I)$ is weakly increasing in the receiver’s discount factor $\gamma\in[0,1)$.
\end{proposition}

\begin{proof}
Fix $\gamma'<\gamma''$. Consider any DDP strategy $\sigma$ that is dynamically persuasive (i.e., induces obedience) under $\gamma'$, together with the receiver’s obedient strategy $\phi_{\ob}$. The one-step obedience constraints compare current recommended action payoffs with any deviation payoffs \emph{plus} the discounted continuation values. When $\gamma$ increases from $\gamma'$ to $\gamma''$, the continuation part of these constraints is (weakly) upweighted, making obedience \emph{easier} to sustain. Hence any $(\sigma,\phi_{\ob})$ feasible at $\gamma'$ remains feasible at $\gamma''$.

Therefore the feasible set of DDP$+$obedience outcomes at $\gamma''$ contains that at $\gamma'$, so the designer’s optimal value cannot decrease:
\[
\mathcal U^{**}_T(\mathcal I;\gamma') \;\le\; \mathcal U^{**}_T(\mathcal I;\gamma'').
\]
Since $\gamma',\gamma''$ were arbitrary with $\gamma'<\gamma''$, $\mathcal U^{**}_T(\mathcal I)$ is weakly increasing in $\gamma$.
\end{proof}

\subsection{Proof of Proposition \ref{prop: IR convergence}}
\label{proof:prop: IR convergence}

\begin{proof}
Let $\pi^*$ be an optimal signaling scheme that solves problem \eqref{eq:V-uninformed-optimization}, with corresponding designer payoff $U(\pi^*, \rho_\ob; \dprior) = U^*_{\mathrm IR}$. 

\textbf{Proof of upper bound $\frac{1}{T} \mathcal U^{**}_T(\mathcal I; \gamma) \le U^*_{\mathrm IR}$}.
By Proposition \ref{prop:dynamic-benchmark-DDP}, the dynamic benchmark $\mathcal U^{**}_T(\mathcal I; \gamma)$ can be achieved by some history-independent DDP strategies $\sigma = (\pi^{(t)})_{t=1}^T$ that is dynamically persuasiveness.
Define $g_t = \sum_{t'=t}^T \gamma^{t'-t} \big(V(\pi^{(t')}, \rho_\ob; \rprior) - V_\uninformed(\rprior) \big)$.
By dynamic persuasiveness, we have $g_t \ge 0$. 
We also note that  
\begin{align}
\label{EX-t}
    V(\pi^{(t)}, \rho_\ob; \rprior) - V_\uninformed(\rprior) + \gamma g_{t+1} = g_t. 
\end{align}
% Then, the ex-ante obedience satisfies
Let $\bar \pi := \frac{1}{T}\sum_{t=1}^T\pi^{(t)}$ be the average signaling scheme of the DDP strategy $\sigma$. 
We will show $\sum_{t=1}^T \big[ V(\pi^{(t)}, \rho_\ob; \rprior) - V_\uninformed(\rprior) \big] \ge 0$. To see that, first, summing Eq.~\eqref{EX-t} over $t=1, ...,T$, % we get
\[ \sum_{t=1}^T \big[ V(\pi^{(t)}, \rho_\ob; \rprior) - V_\uninformed(\rprior) \big] +  { \gamma } \sum_{t=1}^T g_{t+1} = \sum_{t=1}^Tg_{t}.\]
% Shifting the time index in the way of $\sum_{t=1}^Tg_{t+1}(\sigma)=\sum_{t=2}^{T+1}g_t(\sigma)$ and rearranging the terms, we have
Rearranging, 
\[ \sum_{t=1}^T \big[ V(\pi^{(t)}, \rho_\ob; \rprior) - V_\uninformed(\rprior) \big] = { \sum_{t=1}^T g_t - \gamma \sum_{t=1}^T g_{t+1} =  g_1 + (1-\gamma) \sum_{t=2}^T g_t}.\]
% By $g_{T+1} \ge 0$ and $g_1>0$, we have
{ Because $g_1, g_t \ge 0$ and $\gamma \le 1$, we have }
\[ \sum_{t=1}^T \big[V(\pi^{(t)}, \rho_\ob; \rprior) - V_\uninformed(\rprior)\big] \ge 0. \]
{ This means that the average signaling scheme $\overline \pi$ satisfies the receiver's individual rationality constraint in optimization problem \eqref{eq:V-uninformed-optimization}, so it is a feasible solution, and the designer's payoff is at most 
\begin{align*}
    U(\overline \pi, \rho_\ob; \dprior) \le U^*_{\mathrm IR}. 
\end{align*}
By linearity, $U(\overline \pi, \rho_\ob; \dprior) = \frac{1}{T} \sum_{t=1}^T U(\pi^{(t)}, \rho_\ob; \dprior) = \frac{1}{T} \mathcal U^{**}_T(\mathcal I; \gamma)$. So $\frac{1}{T} \mathcal U^{**}_T(\mathcal I; \gamma) \le U^*_{\mathrm IR}$.  
}

% By linearity, we get 
% \begin{align*}
%    \frac{1}{T} \sum_{t=1}^T U(\pi^{(t)}, \rho_\ob; \dprior) = U(\bar \pi, \rho_\ob; \dprior) \leq U^*_{\text{IR}}.
% \end{align*}
% Thus, 
% \begin{align*}
%     \lim_{T \to \infty} \frac{1}{T} \mathcal{U}^{**}_T(I; \gamma=1) \leq U^*_{\text{IR}}.
% \end{align*}

% For fixed $T$ and any $\gamma \in (0, 1)$, summing \ref{EX-t}, we have 
% \begin{align*}
%     \sum_{t=1}^T[V(\pi^{(t)}, \rho_\ob; \rprior) - V_\uninformed(\rprior)] \ge g_1(\sigma)-\gamma g_2(\sigma) = O(1-\gamma).
% \end{align*}

% By linearity, we get 
% \begin{align*}
%     \lim_{\gamma \to 1} \frac{1}{T}  \sum_{t=1}^T U(\pi^{(t)}, \rho_\ob; \dprior) = U(\bar \pi,\rho_\ob; \dprior) \leq U^*_{\text{IR}}.
% \end{align*}

\textbf{Proof of lower bound $\lim_{\gamma\to 1} \lim_{T\to\infty} \frac{1}{T} \mathcal U^{**}_T(\mathcal I) \ge U^*_{\mathrm IR}$}. 
Let $\pi_\full$ be the ``full-information'' signaling scheme that reveals the optimal action $a_\omega^*$ to the receiver at every state $\omega \in \Omega$, which gives the receiver expected payoff $V(\pi_\full, \rho_\ob; \rprior)$. 
Under Assumptions \ref{ass:prior-p0} and \ref{ass:receiver-optimal-action}, it is easy to show that $V(\pi_\full, \rho_\ob; \rprior) - V_\uninformed(\rprior) \ge p_0 G$.
Let $\pi'$ be a direct signaling scheme obtained by mixing $\pi^*$ with weight $1-\delta'$ and $\pi_\full$ with weight $\delta'$. 
Consider the DDP strategy $\sigma' = (\pi')_{t=1}^T$ that repeats $\pi'$ unless the receiver has deviated. We claim that, under $\sigma'$, the receiver is willing to obey the recommended actions in every period $t$ from $1$ to $T - \sqrt{T}$: to see this, given the recommendation at period $t$, the receiver's obedient payoff from period $t$ to $T$ is at least
\begin{align*}
    \gamma^t \cdot 0 + \sum_{t'=t+1}^T \gamma^{t'} V(\pi', \rho_\ob; \rprior). 
\end{align*}
If the receiver deviates from the recommendation in period $t$, he will obtain payoff at most $1$ in period $t$ and be punished starting from period $t+1$, so his total payoff is at most
\begin{align*}
    \gamma^t \cdot 1 + \sum_{t'=t+1}^T \gamma^{t'} V_\uninformed(\rprior). 
\end{align*}
The difference is
\begin{align*}
     & \gamma^t \cdot 0 + \sum_{t'=t+1}^T \gamma^{t'} V(\pi', \rho_\ob; \rprior) - \gamma^t \cdot 1 - \sum_{t'=t+1}^T \gamma^{t'} V_\uninformed(\rprior) \\
     & = - \gamma^t + \sum_{t'=t+1}^T \gamma^{t'}\Big[V(\pi', \rho_\ob; \rprior) - V_\uninformed(\rprior)\Big] \\
     & = - \gamma^t + \sum_{t'=t+1}^T \gamma^{t'}\Big[ (1-\delta') \big( \underbrace{V(\pi^*, \rho_\ob; \rprior) - V_\uninformed(\rprior)}_{\ge 0}\big) + \delta'\big( \underbrace{V(\pi_\full, \rho_\ob; \rprior) - V_\uninformed(\rprior)}_{\ge p_0 G}\big) \Big ] \\
     & \ge - \gamma^t + \sum_{t'=t+1}^T \gamma^{t'} \delta'p_0G \\
     & = - \gamma^t + \delta'p_0G \cdot  \gamma^{t+1} \cdot \tfrac{1 - \gamma^{T-t}}{1-\gamma} \\
     & \ge 0
\end{align*}
by choosing $\delta' = \frac{1-\gamma}{p_0 G \gamma (1-\gamma^{T-t})}$. % Note that $\delta' \to \frac{1}{p_0G(T-t)} \le \frac{1}{p_0G \sqrt{T}}$ as $\gamma \to 1$. 
Note that $\delta' \to \frac{1-\gamma}{p_0G \gamma} \to 0$ as $T\to\infty$ and $\gamma \to 1$. 
This means that the receiver will obey the recommendations from period $1$ to period $T - \sqrt{T}$. So, the designer's average payoff is at least
\begin{align*}
    \frac{1}{T} \mathcal U(\sigma', \phi^*(\sigma'); \dprior) & \ge \frac{1}{T} \bigg( \sum_{t=1}^{T-\sqrt{T}} U(\pi', \rho_\ob; \dprior) ~~ + ~~ 0 \bigg) \\
    & \ge \frac{1}{T} \bigg( \sum_{t=1}^{T-\sqrt{T}} (1-\delta') U(\pi^*, \rho_\ob; \dprior) ~~ + ~~ 0 \bigg) \\
    & = \frac{T - \sqrt{T}}{T}(1-\delta') \cdot U(\pi^*, \rho_\ob; \dprior) \\
    & \to U(\pi^*, \rho_\ob; \dprior)
\end{align*}
as $T\to\infty$ and $\gamma \to 1$, where the last step is because $\delta' \to 0$. 
\end{proof}

\section{Omitted Proofs in Section \ref{sec:learning-dynamic-benchmark}}

\subsection{Proof of Lemma \ref{lem:utility-improving-robustification}}
\label{proof:lem:utility-improving-robustification}

\begin{proof}
Let $\pi_\full$ be the ``full revelation'' signaling scheme that reveals the optimal action $a^*_\omega = \max_{a\in A} v(a, \omega)$ for the receiver for every state $\omega \in \Omega$.  By Assumption \ref{ass:receiver-optimal-action}, $\pi_\full$ is unique.  Let $\| \pi_{\hat \sigma} - \pi_\full\| = \max_{\omega \in \Omega} \| \pi_{\hat \sigma}(\cdot |\omega) - \pi_\full(\cdot |\omega) \|_1$ be the matrix $\ell_1$-distance between direct signaling schemes $\pi$ and $\pi_\full$. Let $\theta = \sqrt{\tfrac{8\delta}{p_0^3 G}}$. 

\textit{Case (1):  $\| \pi_{\hat \sigma} - \pi_\full \| \le \theta$.} 
In this case, we let $\pi_{\tilde \sigma}$ be $\pi_\full$.
Because $\pi_\full$ is persuasive regardless of the receiver's prior, we immediately have
\[\E_{\omega \sim \mu(\cdot | a, \pi_{\tilde \sigma})}\big[ v(a, \omega) - v(a', \omega) \big] ~ \ge~  0 ~ \ge ~ \min \big\{\E_{\omega \sim \hat \mu(\cdot|a, \pi_{\hat \sigma})}\big[ v(a, \omega) - v(a', \omega) \big] + \delta D - \tfrac{2\eps}{p_0}, ~ 0 \big\},\]
which satisfies the first claim of Lemma \ref{lem:utility-improving-robustification}. 
The receiver's utility satisfies $V(\pi_{\tilde \sigma}, \rho_\ob; \rprior) \ge V(\pi_{\hat \sigma}, \rho_\ob; \rprior)$ because the full revelation signaling scheme maximizes the receiver's utility. 
For the designer's utility, by the Lipschitz continuity with respect to $\pi$ (Lemma \ref{lem:utility-Lipschitz}), we have 
\begin{align*}
    U(\pi_{\tilde \sigma}, \rho_\ob; \dprior) ~ \ge ~ U(\pi_{\hat \sigma}, \rho_\ob; \dprior) - \| \pi_\full - \pi_{\hat \sigma} \| ~ \ge ~ U(\pi_{\hat \sigma}, \rho_\ob; \dprior) - \theta,  
\end{align*}
which satisfies the third claim of Lemma \ref{lem:utility-improving-robustification}. 

\textit{Case (2): $\| \pi_{\hat \sigma} - \pi_\full \| > \theta$.}
In this case, we apply static robustification (Lemma \ref{lem:universal-robustification}) to $\pi$ to obtain direct signaling scheme $\pi_{\tilde \sigma}$. 
According to the third claim of Lemma \ref{lem:universal-robustification}, $\| \pi_{\tilde \sigma} - \pi_{\hat \sigma} \| \le \tfrac{4\delta}{p_0^2}$, so by the Lipschitz continuity of the receiver's utility (Lemma \ref{lem:utility-Lipschitz}),
\begin{align*}
    V(\pi_{\tilde \sigma}, \rho_\ob; \rprior) ~ \ge ~ V(\pi_{\hat \sigma}, \rho_\ob; \rprior) - \tfrac{4\delta}{p_0^2}. 
\end{align*}
Consider the convex combination of $\pi_{\tilde \sigma}$ and $\pi_\full$ with weights $1-\alpha$, $\alpha$, denoted by $\pi_{\tilde \sigma, \alpha}$.
By the coalescing lemma (Lemma \ref{lemma:coalesce-2}), $\tilde \pi_\alpha$ automatically satisfies the first claim of Lemma \ref{lem:utility-improving-robustification}. 
By linearity, the receiver's utility satisfies
\begin{align*}
    & V(\pi_{\tilde \sigma, \alpha}, \rho_\ob; \rprior) - V(\pi_{\hat \sigma}, \rho_\ob; \rprior) \\
    ~ & = ~ (1-\alpha) \Big(  V(\pi_{\tilde \sigma}, \rho_\ob; \rprior) - V(\pi_{\hat \sigma}, \rho_\ob; \rprior) \Big) ~ + ~ \alpha \Big( V(\pi_\full, \rho_\ob; \rprior) - V(\pi_{\hat \sigma}, \rho_\ob; \rprior) \Big) \\
    & \ge ~ - (1-\alpha) \tfrac{4\delta}{p_0^2} ~ + ~ \alpha \cdot \tfrac{p_0 G \theta}{2} \qquad \text{by Claim \ref{claim:difference-from-V-full}} \\
    & \ge ~ 0
\end{align*}
where we choose $\alpha = \frac{8\delta}{p_0^3 G \theta} = \sqrt{\frac{8\delta}{p_0^3 G}}$ ($\alpha \le 1$ given $\delta \le \frac{p_0^3G}{8}$).  So, $\pi_{\tilde \sigma, \alpha}$ weakly improves the receiver's utility compared to $\pi$, satisfying the second claim of Lemma \ref{lem:utility-improving-robustification}. 

The designer's utility under $\pi_{\tilde \sigma, \alpha}$ is
\begin{align*}
    & U(\pi_{\tilde \sigma, \alpha}, \rho_\ob; \dprior) \\
    & ~ = ~ (1-\alpha) U(\tilde \pi, \rho_\ob; \dprior) ~ + ~ \alpha U(\pi_\full, \rho_\ob; \dprior) \\
    & ~ \ge ~ (1-\alpha) \Big( U(\pi_{\hat \sigma}, \rho_\ob; \dprior) - \tfrac{3\delta}{p_0} \Big) ~ + ~ 0 & \text{by Lemma \ref{lem:universal-robustification}}\\
    % & ~ \ge ~ U(\pi, \rho_\ob; \dprior) - \tfrac{3\delta}{p_0} - \tfrac{8\delta}{p_0^3 G \theta} \\
    & ~ \ge ~ U(\pi_{\hat \sigma}, \rho_\ob; \dprior) - \tfrac{3\delta}{p_0} - \sqrt{\tfrac{8\delta}{p_0^3 G}} & \text{given $\alpha = \sqrt{\tfrac{8\delta}{p_0^3 G}}$} \\
    & ~ \ge ~ U(\pi_{\hat \sigma}, \rho_\ob; \dprior) - 3 \sqrt{\tfrac{\delta}{8p_0}} - \sqrt{\tfrac{8\delta}{p_0^3 G}} & \text{as $\delta \le \tfrac{p_0^3 G}{8} \implies \tfrac{\delta}{p_0} \le \tfrac{1}{8}$} \\
    & ~ \ge ~ U(\pi_{\hat \sigma}, \rho_\ob; \dprior) - \sqrt{\tfrac{9\delta}{8 p_0^3 G}} - \sqrt{\tfrac{8\delta}{p_0^3 G}} & \text{$p_0, G \le 1$} \\
    & ~ \ge ~ U(\pi_{\hat \sigma}, \rho_\ob; \dprior) - 4 \sqrt{\tfrac{\delta}{p_0^3 G}},
\end{align*}
which proves the lemma. 
\end{proof}

\begin{claim} \label{claim:difference-from-V-full}
$\| \pi_{\hat \sigma} - \pi_\full \| > \theta \implies V(\pi_\full, \rho_\ob; \rprior) - V(\pi_{\hat \sigma}, \rho_\ob; \rprior) > \frac{p_0G\theta}{2}$, 
\end{claim}
\begin{proof}
For any state $\omega \in \Omega$, define the difference
\begin{align*}
\Delta(\omega):=v(a^*_\omega,\omega)-\sum_{a}\pi_{\hat \sigma}(a\mid\omega)v(a,\omega)
& =\sum_{a\neq a^*_\omega}\pi_{\hat \sigma}(a\mid\omega)\big(v(a^*_\omega,\omega)-v(a,\omega)\big) \\
& \ge G\big(1-\pi_{\hat \sigma}(a^*_\omega \mid \omega)\big) = \tfrac{G}{2} \big\| \pi_{\hat \sigma}(\cdot | \omega)  - \pi_\full(\cdot | \omega) \big\|_1
\end{align*}
The condition
$\|\pi_{\hat \sigma}-\pi_{\full}\| > \theta$ implies $\exists\,\bar\omega \in \Omega$ with
$\| \pi_{\hat \sigma}(\cdot | \omega)  - \pi_\full(\cdot | \omega) \|_1 > \theta$, hence $\Delta(\bar\omega)\ge \frac{G\theta}{2}$.
Therefore,
\[
V(\pi_{\full},\rho_{\ob};\mu_R)-V(\pi_{\hat \sigma},\rho_{\ob};\mu_R)
=\sum_{\omega}\mu_R(\omega)\Delta(\omega)
\ge \mu_R(\bar\omega)\Delta(\bar\omega)\ge \tfrac{p_0G\theta}{2}. 
\]
\end{proof}

\subsection{Proof of Theorem \ref{thm:dynamic-robustification}}
\label{proof:thm:dynamic-robustification}

\begin{proof}[\bf Proof of Theorem \ref{thm:dynamic-robustification}]
We convert $\hat \sigma = (\pi_{\hat \sigma}^{(1)}, \ldots, \pi_{\hat \sigma}^{(T)})$ to $\tilde \sigma = (\pi_{\tilde \sigma}^{(1)}, \ldots, \pi_{\tilde \sigma}^{(T)})$ by applying utility-improving robustification (Lemma \ref{lem:utility-improving-robustification}) to the direct signaling schemes of $\sigma$ in the backward order: namely, from $\pi^{(T)}$ to $\pi^{(1)}$. 

Specifically, for each period $t \le T$, having robustified $\pi_{\hat \sigma}^{(T)}, \ldots, \pi_{\hat \sigma}^{(t+1)}$, we robustify $\pi_{\hat \sigma}^{(t)}$ so that $(\pi_{\tilde \sigma}^{(t)}, \pi_{\tilde \sigma}^{(t+1)}, \ldots, \pi_{\tilde \sigma}^{(T)})$ is dynamically persuasive with respect to $\rprior$ satisfying $\|\rprior - \hat \mu\|_1 \le \eps$. In particular, we want the following to hold: for any action $a$ recommended by $\pi_{\tilde \sigma}^{(t)}$ and any $a' \in A$,  
\begin{align} \label{eq:robustified-period-t-dynamically-persuasive}
    \gamma^t \E_{\omega \sim \rprior(\cdot | a, \pi_{\tilde \sigma}^{(t)})} \Big[ v(a, \omega) - v(a', \omega) \Big] + \sum_{t' = t+1}^{T} \gamma^{t'} \Big( V(\pi_{\tilde \sigma}^{(t')}, \rho_{\ob}; \rprior) - V_\uninformed(\rprior) \Big) \ge 0. 
\end{align}
Applying Lemma \ref{lem:utility-improving-robustification} to $\pi^{(t)}$ with parameter $\delta_t > 0$ (to be chosen later), we obtain another direct signaling scheme $\pi_{\tilde \sigma}^{(t)}$ that satisfies $\E_{\omega \sim \rprior(\cdot | a, \pi_{\tilde \sigma}^{(t)})} \big[ v(a, \omega) - v(a', \omega) \big] \ge \min \big\{\E_{\omega \sim \hat \mu(\cdot | a, \pi_{\hat \sigma}^{(t)})} \big[ v(a, \omega) - v(a', \omega) \big] + \delta_t D - \tfrac{2\eps}{p_0}, 0 \big\}$. There are two cases:

\textit{Case (1): $\E_{\omega \sim \rprior(\cdot | a, \pi_{\tilde \sigma}^{(t)})} \big[ v(a, \omega) - v(a', \omega) \big] \ge 0$:}
In this case, by induction, we have 
\begin{align*}
    \sum_{t' = t+1}^{T} \gamma^{t'} \Big( V(\pi_{\tilde \sigma}^{(t')}, \rho_{\ob}; \rprior) - V_\uninformed(\rprior) \Big) \ge 0
\end{align*}
because the robustified DDP strategy $(\pi_{\tilde \sigma}^{(t')})_{t'=t+1}^T$ is dynamically persuasive. So, we immediately obtain \eqref{eq:robustified-period-t-dynamically-persuasive}.

\textit{Case (2): $\E_{\omega \sim \rprior(\cdot | a, \pi_{\tilde \sigma}^{(t)})} \big[ v(a, \omega) - v(a', \omega) \big] < 0$:}
in this case, by the first claim of Lemma \ref{lem:utility-improving-robustification}, 
\begin{align} \label{eq:term-1}
0 ~ > ~ \E_{\omega \sim \rprior(\cdot | a, \pi_{\tilde \sigma}^{(t)})} \big[ v(a, \omega) - v(a', \omega) \big] ~ \ge ~ \E_{\omega \sim \hat \mu(\cdot | a, \pi_{\hat \sigma}^{(t)})} \big[ v(a, \omega) - v(a', \omega) \big] + \delta_t D - \tfrac{2\eps}{p_0}.
\end{align}
We then consider the receiver's continuation utility from period $t+1$ to $T$. 
\begin{align}
    & \sum_{t' = t+1}^{T} \gamma^{t'} \Big( V(\pi_{\tilde \sigma}^{(t')}, \rho_{\ob}; \rprior) - V_\uninformed(\rprior) \Big) \nonumber \\
    & ~ \ge ~ \sum_{t' = t+1}^{T} \gamma^{t'} \Big( V(\pi_{\hat \sigma}^{(t')}, \rho_{\ob}; \rprior) - V_\uninformed(\rprior) \Big) \nonumber & \text{by the second claim of Lemma \ref{lem:utility-improving-robustification}} \\
    & ~ \ge ~ \sum_{t' = t+1}^{T} \gamma^{t'} \Big( V(\pi_{\hat \sigma}^{(t')}, \rho_{\ob}; \hat \mu) - V_\uninformed(\hat \mu) - 2 \| \rprior - \hat \mu\|_1 \Big) \nonumber & \text{$V$ is $1$-Lipschitz (Lemma \ref{lem:utility-Lipschitz})} \\
    & ~ \ge ~ \sum_{t' = t+1}^{T} \gamma^{t'} \Big( V(\pi_{\hat \sigma}^{(t')}, \rho_{\ob}; \hat \mu) - V_\uninformed(\hat \mu) \Big) -  \tfrac{\gamma^{t+1}}{1-\gamma} 2\eps.  \label{eq:term-2}
\end{align}
Adding \eqref{eq:term-1} and \eqref{eq:term-2}, 
\begin{align}
& \gamma^t \E_{\omega \sim \rprior(\cdot | a, \pi_{\tilde \sigma}^{(t)})} \Big[ v(a, \omega) - v(a', \omega) \Big] + \sum_{t' = t+1}^{T} \gamma^{t'} \Big( V(\pi_{\tilde \sigma}^{(t')}, \rho_{\ob}; \rprior) - V_\uninformed(\rprior) \Big) \nonumber \\
& ~ \ge ~ \underbrace{\gamma^t \E_{\omega \sim \hat \mu(\cdot | a, \pi_{\hat \sigma}^{(t)})} \big[ v(a, \omega) - v(a', \omega) \big]  + \sum_{t' = t+1}^{T} \gamma^{t'} \Big( V(\pi_{\hat \sigma}^{(t')}, \rho_{\ob}; \hat \mu) - V_\uninformed(\hat \mu) \Big)}_{\text{$\ge 0$ because $(\pi_{\hat \sigma}^{(t)}, \ldots, \pi_{\hat \sigma}^{(T)})$ is dynamically persuasive for $\hat \mu$}} \nonumber \\
& \qquad + \gamma^t \delta_t D - \gamma^t \tfrac{2\eps}{p_0} - \tfrac{\gamma^{t+1}}{1-\gamma} 2\eps \nonumber \\
% & ~ \ge ~ \gamma^t\Big( \delta_t D - \big(\tfrac{2}{p_0} + \tfrac{2\gamma}{1-\gamma}\big)\eps \Big) \nonumber \\
& ~ \ge ~ 0  \nonumber
\end{align}
where we choose $\delta_t = \big(\tfrac{1}{p_0} + \tfrac{\gamma}{1-\gamma}\big) \frac{2\eps}{D}$.
Thus, $(\pi_{\tilde \sigma}^{(t)}, \pi_{\tilde \sigma}^{(t+1)}, \ldots, \pi_{\tilde \sigma}^{(T)})$ is dynamically persuasive for a receiver with prior $\rprior$.
Performing the robustification from $t = T$ to $t=1$, we therefore obtain a DDP strategy $\tilde \sigma = (\pi_{\tilde \sigma}^{(1)}, \ldots, \pi_{\tilde \sigma}^{(T)})$ that is dynamically persuasive for a receiver with prior $\rprior$. 

The designer's $T$-period total utility under the original DDP strategy $\sigma = (\pi^{(1)}, \ldots, \pi^{(T)})$ is $\mathcal U_T(\sigma, \phi_\ob; \dprior) = \sum_{t=1}^T U(\pi^{(t)}, \rho_\ob; \dprior)$. 
Under the robustified DDP strategy $\tilde \sigma = (\pi_{\tilde \sigma}^{(1)}, \ldots, \pi_{\tilde \sigma}^{(T)})$, the designer's utility is 
\begin{align*}
    \mathcal U_T(\tilde \sigma, \phi_\ob; \dprior) & ~ = ~ \sum_{t=1}^T U(\pi_{\tilde \sigma}^{(t)}, \rho_\ob; \dprior) \\
    & ~ \ge ~ \sum_{t=1}^T \Big( U(\pi^{(t)}, \rho_\ob; \dprior) - 4\sqrt{\tfrac{\delta_t}{p_0^3 G}} \Big) & \text{by the third claim of Lemma \ref{lem:utility-improving-robustification}} \\
    & ~ = ~ \mathcal U_T(\sigma, \phi_\ob; \dprior) - 4T \sqrt{\big(\tfrac{1}{p_0} + \tfrac{\gamma}{1-\gamma}\big) \tfrac{2\eps}{p_0^3 G D}}, 
\end{align*}
which proves Theorem \ref{thm:dynamic-robustification}.
\end{proof}

\subsection{Proof of Lemma \ref{lem:larger-than-T0}}
\label{app:larger-than-T0}
\begin{proof}
Let $\sigma_T^* = (\pi_{\sigma^*}^{(1)}, \ldots, \pi_{\sigma^*}^{(T)})$ be an optimal history-independent DDP strategy for the $T$-period game, which returns $\mathcal U^{**}_{T}(\dprior, \hat \mu)$. 

Let $\sigma^\mathrm{tail} = (\pi_{\sigma^*}^{(T-T_0+1)}, \ldots, \pi_{\sigma^*}^{(T)})$ be the tail strategy for $\sigma_T^*$. Because the DDP strategy ensures the receiver's obedience through the continuation incentives that depend only on the future, truncating the first $T_0$ periods does not affect incentives from period $T_0\!+\!1$ onward. Thus $\sigma^{\mathrm{tail}}$ is a feasible $(T\!-\!T_0)$-period DDP strategy for the same prior.

On the obedient path, the designer’s per-period payoff is $U(\pi^{(t)},\rho_{\mathrm{ob}};\mu_D) \in [0,1]$. Therefore, the first $T_0$ periods contribute at most $T_0$ to the total:

\begin{align*}
    \mathcal U^{**}_{T}(\mu_D,\hat\mu)
= \sum_{t=1}^{T_0} U(\pi_{\sigma^*}^{(t)},\rho_{\mathrm{ob}};\mu_D)
+ \sum_{t=T_0+1}^{T} U(\pi_{\sigma^*}^{(t)},\rho_{\mathrm{ob}};\mu_D)
\;\le\; T_0 \;+\; U(\sigma^{\mathrm{tail}},\rho_{\mathrm{ob}};\mu_D)
\end{align*}

Maximizing over all feasible $(T\!-\!T_0)$-period strategies gives
\begin{align*}
\mathcal U^{**}_{T-T_0}(\mu_D,\hat\mu)\ \ge\ U(\sigma^{\mathrm{tail}},\rho_{\mathrm{ob}};\mu_D)
\ \ge\ \mathcal U^{**}_{T}(\mu_D,\hat\mu) - T_0.
\end{align*}
\end{proof}

\section{Omitted Proofs in Section \ref{sec:lower-bound}}

\subsection{Proof of Lemma \ref{lemma:single-period-regret}}
\label{proof:single-period-regret}
Given an incentive compatible and individually rational menu $\mathcal M = \{\pi_\mu\}$, the receiver will report his prior $\rprior$ truthfully. 
Consider a weighted welfare of the game, which is the sum of the information designer's utility and $(1+\tilde \mu_R)$ times the receiver's utility:
% \begin{align*}
%     & \frac{1}{1+\theta} U^{\sp}(\mathcal M, \rprior) + V^{\sp}(\rprior, \rprior) \\
%     & = \frac{1}{1+\theta} \bigg((1-p_0) \pi_{\rprior}(a_1|\omega_0) + p_0 \pi_{\rprior}(a_1|\omega_1) + \theta \pi_\rprior(a_1|\omega_1) - \pi(a_1|\omega_0) \bigg) \\
%     & = \frac{1}{1+\theta} \bigg(\theta \pi_\rprior(a_1|\omega_1) + p_0 \Big(\pi_{\rprior}(a_1|\omega_1) - \pi_{\rprior}(a_1|\omega_0) \Big) \bigg). 
% \end{align*}
\begin{align}
    & U^{\sp}(\mathcal M, \rprior) + (1+\tilde \mu_R) V^{\sp}(\rprior, \rprior) \nonumber \\
    & = (1-p_0) \pi_{\rprior}(a_1|\omega_0) + p_0 \pi_{\rprior}(a_1|\omega_1) + \tilde \mu_R \pi_\rprior(a_1|\omega_1) - \pi(a_1|\omega_0) \nonumber \\
    & = (\tilde{\mu}_R + p_0) \pi_\rprior(a_1|\omega_1) - p_0 \pi_{\rprior}(a_1|\omega_0).  \label{eq:welfare}
\end{align}
Replacing $\rprior$ by $\mu$ (and replacing $\tilde \mu_R$ by $\tilde \mu$) in \eqref{eq:welfare} and noting that $V^{\sp}(\mu, \mu) \ge V_{\mathrm{uninformed}} \ge 0$ by individual rationality, we have
\begin{align} \label{eq:whiteboard-right}
    U^{\sp}(\mathcal M, \mu) \le (\tilde \mu + p_0) \pi_\mu(a_1|\omega_1) - p_0 \pi_\mu(a_1|\omega_0).
\end{align}
The information designer's single-period regret
\begin{align*}
    \Reg^{\sp}(\mathcal M, \rprior) & = (1-p_0) \tilde \mu_R + p_0 -  U^{\sp}(\mathcal M, \rprior) \\
    & \ge \pi_\rprior(a_1|\omega_1) \cdot \Big( (1-p_0) \tilde \mu_R + p_0 \Big) -  U^{\sp}(\mathcal M, \rprior) \\
    & =  (\tilde \mu_R +p_0) \pi_\rprior(a_1|\omega_1) -  U^{\sp}(\mathcal M, \rprior) - p_0\tilde \mu_R \pi_\rprior(a_1|\omega_1) \\ 
    \text{by \eqref{eq:welfare}} & = (1+\tilde \mu_R) V^{\sp}(\rprior, \rprior) + p_0 \pi_\rprior(a_1|\omega_0) - p_0 \tilde \mu_R \pi_\rprior(a_1|\omega_1) \\
    \text{by IC} & \ge (1+\tilde \mu_R) V^{\sp}(\rprior, \mu) + p_0 \pi_\rprior(a_1|\omega_0) - p_0 \tilde \mu_R \pi_\rprior(a_1|\omega_1) \\
    & = (1+\tilde \mu_R) \frac{1}{1+\tilde \mu_R}\bigg( \tilde \mu_R \pi_\mu(a_1|\omega_1) - \pi_\mu(a_1|\omega_0) \bigg) + p_0 \pi_\rprior(a_1|\omega_0) - p_0 \tilde \mu_R \pi_\rprior(a_1|\omega_1) \\
    \text{by \eqref{eq:whiteboard-right}} & \ge \bigg( \frac{\tilde \mu_R}{\tilde \mu + p_0} \Big( U^{\sp}(\mathcal M, \mu) + p_0 \pi_\mu(a_1|\omega_0) \Big) - \pi_\mu(a_1|\omega_0) \bigg) + p_0 \pi_\rprior(a_1|\omega_0) - p_0 \tilde \mu_R \pi_\rprior(a_1|\omega_1) \\
    \text{by \eqref{eq:sender-utility}} & = \bigg( \frac{\tilde \mu_R}{\tilde \mu + p_0} \Big( U^{\sp}(\mathcal M, \mu) + p_0 \pi_\mu(a_1|\omega_0) \Big) - U^{\sp}(\mathcal M, \mu) - p_0 \pi_\mu(a_1|\omega_0) + p_0 \pi_\mu(a_1|\omega_1) \bigg) \\
    & \qquad + p_0 \pi_\rprior(a_1|\omega_0) - p_0 \tilde \mu_R \pi_\rprior(a_1|\omega_1) \\
    & = \frac{\tilde \mu_R - \tilde \mu - p_0}{\tilde \mu + p_0}  U^{\sp}(\mathcal M, \mu) \\
    & \qquad + p_0 \bigg( \frac{\tilde \mu_R - \tilde \mu - p_0}{\tilde \mu + p_0} \pi_\mu(a_1|\omega_0) + \pi_\mu(a_1|\omega_1) + \pi_\rprior(a_1|\omega_0) - \tilde \mu_R \pi_\rprior(a_1|\omega_1) \bigg) \\
    & \ge \frac{\tilde \mu_R - \tilde \mu - p_0}{\tilde \mu + p_0}  U^{\sp}(\mathcal M, \mu) + p_0 \bigg( \frac{\tilde \mu_R - \tilde \mu - p_0}{\tilde \mu + p_0} \pi_\mu(a_1|\omega_0) - \tilde \mu_R \bigg). 
\end{align*}
Let $\tilde \mu_R = 1$ and $\tilde \mu = \frac{1}{2} - p_0$. 
If $\Reg^{\sp}(\mathcal M, \rprior) \ge \frac{1}{4}$, then the lemma is proved. If $\Reg^{\sp}(\mathcal M, \rprior) < \frac{1}{4}$, then we have
\begin{align*}
    \frac{1}{4} > \Reg^{\sp}(\mathcal M, \rprior) \ge U^{\sp}(\mathcal M, \mu) + p_0 \Big( \pi_\mu(a_1|\omega_0) - 1 \Big)  \ge U^{\sp}(\mathcal M, \mu) - p_0, 
\end{align*}
which implies that the information designer's regret under receiver prior $\mu$ is
\begin{align*}
    \Reg^{\sp}(\mathcal M, \mu) & = (1-p_0) \tilde \mu + p_0 - U^{\sp}(\mathcal M, \mu) \\
    & \ge (1-p_0)(\frac{1}{2} - p_0) + p_0 - \frac{1}{4} - p_0 \ge \frac{1}{4} - \frac{3}{2} p_0 \ge \frac{1}{16}
\end{align*}
given $p_0 = \frac{1}{8}$.  Using $\mu(\omega_0) = \frac{1}{1+\tilde \mu}$ and $\mu(\omega_1) = \frac{\tilde \mu}{1+\tilde \mu}$, by direct calculation, one can verify that the condition $\mu(\omega) \ge p_0, \forall \omega \in \Omega$, is satisfied. 
So, the lemma holds with $\tilde \mu$.

\subsection{Proof of Lemma \ref{lem:two-benchmark-equal}}
\label{app:two-benchmark-equal}

% Consider the single-period information design instance defined before Lemma \ref{lem:two-benchmark-equal}. 

We summarize the information design instance $\mathcal I$ here: 
\begin{align*}
    \mathcal I =
    \begin{cases}
        \text{designer's utility:} & u(a_1, \omega) = 1, \quad u(a_0, \omega) = 0, \quad \forall \omega \in \Omega = \{\omega_0, \omega_1\};  \\
        \text{receiver's utility:} & v(a_0, \omega_0) = 0, \quad v(a_0, \omega_1) = 0, \quad v(a_1, \omega_0) = -1, \quad v(a_1, \omega_1) = 1; \\
        \text{designer's prior:} & \dprior(\omega_0) = 1 - p_0, \quad \dprior(\omega_1) = p_0, \quad 0 < p_0 < 1/2;   \\
        \text{receiver's prior:} &\rprior(\omega_0) = \tfrac{1}{1+\tilde \mu_R}, \quad \rprior(\omega_1) = \tfrac{\tilde \mu_R}{1+\tilde \mu_R}, \quad 0\le \mu_R \le 1. 
    \end{cases}
\end{align*}
We present some useful facts before proving Lemma \ref{lem:two-benchmark-equal}.
Because $\rprior(\omega_1) \le \rprior(\omega_0)$, the receiver's ex-ante utility when receiving no information is
\begin{align} \label{eq:instance-V-uninformed}
    V_\uninformed(\rprior) ~ = ~ \max_{a\in A} \sum_{\omega \in \Omega}\rprior(\omega) v(a, \omega) ~ = ~ \max\big\{\underbrace{0}_{\text{$a_0$'s utility}}, ~ \underbrace{\rprior(\omega_1) - \rprior(\omega_0)}_{\text{$a_1$'s utility}} \, \big\} ~ = ~ 0. 
\end{align} 
When the designer uses a direct signaling scheme $\pi: \Omega \to \Delta(A)$, the ex-ante utility of an obedient receiver is 
\begin{align}
    V(\pi, \rho_\ob; \rprior) ~ = ~ \sum_{\omega \in \Omega} \rprior(\omega) \sum_{a\in A} \pi(a|\omega) v(a, \omega) ~ = ~ - \rprior(\omega_0) \pi(a_1|\omega_0) + \rprior(\omega_1) \pi(a_1|\omega_1), 
\end{align}
and the ex-ante utility of the designer is
\begin{align}
    U(\pi, \rho_\ob; \dprior) ~ = ~ \sum_{\omega \in \Omega} \rprior(\omega) \sum_{a\in A} \pi(a|\omega) u(a, \omega) ~ = ~ (1-p_0) \pi(a_1|\omega_0)  + p_0 \pi(a_1|\omega_1). 
\end{align}
The persuasiveness constraints for a direct signaling scheme $\pi$ are: 
\begin{align}
    & \text{(for $a_0$)} \quad \E_{\omega \sim \rprior(\cdot | a_0, \pi)}\big[ v(a_0, \omega) - v(a_1, \omega) \big] ~ = ~ \frac{\rprior(\omega_0) \pi(a_0 | \omega_0) - \rprior(\omega_1) \pi(a_0 | \omega_1)}{\rprior(\omega_0) \pi(a_0|\omega_0) + \rprior(\omega_1) \pi(a_0|\omega_1)} ~ \ge ~ 0; \\
    & \text{(for $a_1$)} \quad \E_{\omega \sim \rprior(\cdot | a_1, \pi)}\big[ v(a_1, \omega) - v(a_0, \omega) \big] ~ = ~ \frac{- \rprior(\omega_0) \pi(a_1 | \omega_0) + \rprior(\omega_1) \pi(a_1 | \omega_1)}{\rprior(\omega_0) \pi(a_1|\omega_0) + \rprior(\omega_1) \pi(a_1|\omega_1)} ~ \ge ~ 0; \label{eq:a1-persuasive}
\end{align}
The Bayesian persuasion benchmark $U^*_\BP(\mathcal I)$ is the designer's optimal utility from a persuasive direct signaling scheme, when knowing the receiver's prior and not using menu:   
\begin{align*}
   U^*_\BP(\mathcal I) ~ = ~ \max_{\pi: \Omega \to \Delta(A)} ~ & U(\pi, \rho_\ob; \dprior) \\
    \text{s.t.} \quad~ & 
    \text{(for $a_0$)} \quad \rprior(\omega_0) \pi(a_0 | \omega_0) - \rprior(\omega_1) \pi(a_0 | \omega_1) ~ \ge ~ 0; \\
    & \text{(for $a_1$)} \quad -\rprior(\omega_0) \pi(a_1 | \omega_0) + \rprior(\omega_1) \pi(a_1 | \omega_1) ~ \ge ~ 0.
\end{align*}

\begin{claim} \label{claim:instance-BP-optimal}
$U^*_\BP(\mathcal I) = (1-p_0) \tilde \mu_R + p_0 = U^*_\sp(\rprior)$. 
\end{claim}
\begin{proof}
In the $U^*_\BP(\mathcal I)$ optimization problem, the persuasiveness constraint for $a_0$ is
\begin{align*}
    & \rprior(\omega_0) \pi(a_0 | \omega_0) - \rprior(\omega_1) \pi(a_0 | \omega_1) \\
    & = \rprior(\omega_0)\big(1 - \pi(a_1 | \omega_0)\big) - \rprior(\omega_1) \big(1 - \pi(a_1 | \omega_1) \big) \\
    & = \underbrace{\rprior(\omega_0) - \rprior(\omega_1)}_{\ge 0 \text{ by construction}} ~ \underbrace{- \rprior(\omega_0) \pi(a_1 | \omega_0) + \rprior(\omega_1) \pi(a_1 | \omega_1)}_{\ge 0 \text{ when $a_1$ is persuasive}} ~ \ge ~ 0, 
\end{align*}
which is automatically satisfied when $a_1$ is persuasive.  So, we can drop the persuasiveness constraint for $a_0$ and write the optimization problem as 
\begin{align*}
    U^*_\BP(\mathcal I) ~ = ~ \max_{\pi: \Omega \to \Delta(A)} ~ & (1-p_0) \pi(a_1|\omega_0)  + p_0 \pi(a_1|\omega_1) \nonumber \\
    \text{s.t.} \quad~ & -\rprior(\omega_0) \pi(a_1 | \omega_0) + \rprior(\omega_1) \pi(a_1 | \omega_1) ~ \ge ~ 0. \nonumber 
\end{align*}
The solution to the above optimization problem is clearly $\pi(a_1|\omega_1) = 1$ and $\pi(a_1|\omega_0) = \frac{\rprior(\omega_1)}{\rprior(\omega_0)} = \tilde \mu_R$, with optimal objective value $U^*_\BP(\mathcal I) = (1-p_0) \tilde \mu_R  + p_0$, which equals $U^*_\sp(\rprior)$. 
\end{proof}

To prove Lemma \ref{lem:two-benchmark-equal}, we then analyze the designer's dynamic benchmark $\mathcal U^{**}_T(\mathcal I)$. 
According to Proposition \ref{prop:dynamic-benchmark-DDP}, the dynamic benchmark $\mathcal U^{**}_T(\mathcal I)$ can be achieved by some history-independent dynamically persuasive DDP strategy $\sigma = (\pi^{(1)}, \ldots, \pi^{(T)})$: namely, 
\begin{align*}
    \mathcal U^{**}_T(\mathcal{I}) & ~ = ~ \max_{\text{$\sigma = (\pi^{(1)}, \ldots, \pi^{(T)})$}} ~~ \sum_{t=1}^T U(\pi^{(t)}, \rho_\ob; \dprior)
\end{align*}
subject to $\forall t \in \{1, \ldots, T\}$, $\forall a, a' \in A$, 
\begin{align*}
     \gamma^t \E_{\omega \sim \rprior(\cdot | a, \pi^{(t)})} \Big[ v(a, \omega) - v(a', \omega) \Big] + \sum_{t' = t+1}^{T} \gamma^{t'} \Big( V(\pi^{(t')}, \rho_{\ob}; \rprior) - V_\uninformed(\rprior) \Big) \ge 0. 
\end{align*}
Plugging in Equations \eqref{eq:instance-V-uninformed}-\eqref{eq:a1-persuasive}, we obtain 
\begin{align*}
    \mathcal U^{**}_T(\mathcal{I}) & ~ = ~ \max_{\text{$\sigma = (\pi^{(1)}, \ldots, \pi^{(T)})$}} ~~ \sum_{t=1}^T \Big( (1-p_0) \pi^{(t)}(a_1|\omega_0)  + p_0 \pi^{(t)}(a_1|\omega_1) \Big)
\end{align*}
subject to $\forall t \in \{1, \ldots, T\}$, 
\begin{align*}
    & \gamma^t\frac{\rprior(\omega_0) \pi^{(t)}(a_0 | \omega_0) - \rprior(\omega_1) \pi^{(t)}(a_0 | \omega_1)}{\rprior(\omega_0) \pi^{(t)}(a_0|\omega_0) + \rprior(\omega_1) \pi^{(t)}(a_0|\omega_1)} + \sum_{t' = t+1}^{T} \gamma^{t'} \Big(- \rprior(\omega_0) \pi^{(t')}(a_1|\omega_0) + \rprior(\omega_1) \pi^{(t')}(a_1|\omega_1) \Big) \ge 0, \\
    & \gamma^t\frac{- \rprior(\omega_0) \pi^{(t)}(a_1 | \omega_0) + \rprior(\omega_1) \pi^{(t)}(a_1 | \omega_1)}{\rprior(\omega_0) \pi^{(t)}(a_1|\omega_0) + \rprior(\omega_1) \pi^{(t)}(a_1|\omega_1)} + \sum_{t' = t+1}^{T} \gamma^{t'} \Big(- \rprior(\omega_0) \pi^{(t')}(a_1|\omega_0) + \rprior(\omega_1) \pi^{(t')}(a_1|\omega_1) \Big) \ge 0,  \\
    & \text{and } \sum_{t' = t+1}^{T} \gamma^{t'} \Big(- \rprior(\omega_0) \pi^{(t')}(a_1|\omega_0) + \rprior(\omega_1) \pi^{(t')}(a_1|\omega_1) \Big) \ge 0. 
\end{align*}

We note that the optimal solution to the above problem must have $\pi^{(t)}(a_1|\omega_1) = 1$ for every $t$.
This is because: (1) increasing $\pi^{(t)}(a_1|\omega_1)$ increases the designer's utility; (2) increasing $\pi^{(t)}(a_1|\omega_1)$ will not violate the three constraints because the right-hand-sides of the constraints are all increasing in $\pi^{(t)}(a_1|\omega_1)$.  Note that in first constraint, the right-hand-side is decreasing in $\pi^{(t)}(a_0|\omega_1)$ while $\pi^{(t)}(a_0|\omega_1)= 1 - \pi^{(t)}(a_1|\omega_1)$ is decreasing in $\pi^{(t)}(a_1|\omega_1)$.  So, we can increase $\pi^{(t)}(a_1|\omega_1)$ to the maximal value of $1$.  With $\pi^{(t)}(a_1|\omega_1) = 1$, we can rewrite $\mathcal U^{**}_T(\mathcal{I})$ as
\begin{align*}
    \mathcal U^{**}_T(\mathcal{I}) & ~ = ~ \max_{\text{$\sigma = (\pi^{(1)}, \ldots, \pi^{(T)})$}} ~~ \sum_{t=1}^T \Big( (1-p_0) \pi^{(t)}(a_1|\omega_0)  + p_0 \Big) \\
    \text{s.t.} ~~ & \gamma^t\frac{\rprior(\omega_0) \pi^{(t)}(a_0 | \omega_0) - 0}{\rprior(\omega_0) \pi^{(t)}(a_0|\omega_0) + 0} + \sum_{t' = t+1}^{T} \gamma^{t'} \Big(- \rprior(\omega_0) \pi^{(t')}(a_1|\omega_0) + \rprior(\omega_1) \Big) \ge 0, \\
    & \gamma^t\frac{- \rprior(\omega_0) \pi^{(t)}(a_1 | \omega_0) + \rprior(\omega_1)}{\rprior(\omega_0) \pi^{(t)}(a_1|\omega_0) + \rprior(\omega_1)} + \sum_{t' = t+1}^{T} \gamma^{t'} \Big(- \rprior(\omega_0) \pi^{(t')}(a_1|\omega_0) + \rprior(\omega_1) \Big) \ge 0,  \\
    & \text{and } \sum_{t' = t+1}^{T} \gamma^{t'} \Big(- \rprior(\omega_0) \pi^{(t')}(a_1|\omega_0) + \rprior(\omega_1) \Big) \ge 0. 
\end{align*}
Because $\frac{\rprior(\omega_0) \pi^{(t)}(a_0 | \omega_0) - 0}{\rprior(\omega_0) \pi^{(t)}(a_0|\omega_0) + 0} = 1$, the first constraint is implied by the third constraint. So we can drop the first constraint. Letting $x^{(t)} = \pi^{(t)}(a_1|\omega_0)$, we have 
\begin{align} 
    \mathcal U^{**}_T(\mathcal{I}) & ~ = ~  \max_{0\le x^{(t)} \le 1} \sum_{t=1}^T \Big( (1-p_0) x^{(t)}  + p_0 \Big)  \quad \text{ or equivalently } \quad \max_{0\le x^{(t)} \le 1} \sum_{t=1}^T x^{(t)} \label{eq:instance-global-U-objective}\\
    \text{s.t.} ~~ 
    & \gamma^t\frac{- \rprior(\omega_0) x^{(t)} + \rprior(\omega_1)}{\rprior(\omega_0) x^{(t)} + \rprior(\omega_1)} + \sum_{t' = t+1}^{T} \gamma^{t'} \Big(- \rprior(\omega_0) x^{(t')} + \rprior(\omega_1) \Big) \ge 0,  \label{eq:instance-global-U-constraint-1} \\
    & \text{and } \sum_{t' = t+1}^{T} \gamma^{t'} \Big(- \rprior(\omega_0) x^{(t')} + \rprior(\omega_1) \Big) \ge 0. \label{eq:instance-global-U-constraint-2}
\end{align}

\begin{claim} \label{claim:instance-optimization-solution}
A solution to the optimization problem \eqref{eq:instance-global-U-objective}-\eqref{eq:instance-global-U-constraint-2} is $x^{(t)} = \frac{\rprior(\omega_1)}{\rprior(\omega_0)}$ for every $t \in \{1, \ldots, T\}$. 
\end{claim}
\begin{proof}
We prove this claim by induction on $T$.  When $T = 1$, the optimization problem becomes
\begin{align*} 
    \max_{0\le x^{(1)} \le 1} & ~ x^{(1)} \\
    \text{s.t.} ~~ 
    & ~ \gamma^1\frac{- \rprior(\omega_0) x^{(1)} + \rprior(\omega_1)}{\rprior(\omega_0) x^{(1)} + \rprior(\omega_1)} \ge 0,  
\end{align*}
which is clearly maximized by $x^{(1)} = \frac{\rprior(\omega_1)}{\rprior(\omega_0)}$.

Consider $T \ge 2$. Let $x = (x^{(t)})_{t=1}^T$ be any feasible solution where $x^{(T)} < \frac{\rprior(\omega_1)}{\rprior(\omega_0)}$. We define $\tilde x = (\tilde x^{(t)})_{t=1}^T$ by
\begin{align*}
    \tilde x^{(T)} = \tfrac{\rprior(\omega_1)}{\rprior(\omega_0)}, \quad \tilde x^{(T-1)} = x^{(T-1)} - \tfrac{\gamma^T}{\gamma^{T-1}} \big( \tfrac{\rprior(\omega_1)}{\rprior(\omega_0)} - x^{(T)} \big), \quad \tilde x^{(t)} = x^{(t)}~~ \forall t\le T-2.  
\end{align*}
We show that $\tilde x$ is feasible and has a weakly larger objective value than $x$.  To verify the feasibility of $\tilde x$, we verify the first constraint \eqref{eq:instance-global-U-constraint-1} for every $t$:
\begin{itemize}
    \item For $t \le T-2$, we have
    \begin{align*}
        & \gamma^t\frac{- \rprior(\omega_0) \tilde x^{(t)} + \rprior(\omega_1)}{\rprior(\omega_0) \tilde x^{(t)} + \rprior(\omega_1)} + \sum_{t' = t+1}^{T} \gamma^{t'} \Big(- \rprior(\omega_0) \tilde x^{(t')} + \rprior(\omega_1) \Big) \\
        & = \gamma^t\frac{- \rprior(\omega_0) x^{(t)} + \rprior(\omega_1)}{\rprior(\omega_0) x^{(t)} + \rprior(\omega_1)} + \sum_{t' = t+1}^{T-2} \gamma^{t'} \Big(- \rprior(\omega_0) x^{(t')} + \rprior(\omega_1) \Big) \\
        & \qquad + \gamma^{T-1} \Big(- \rprior(\omega_0) \tilde x^{(T-1)} + \rprior(\omega_1) \Big) + \gamma^T \cdot 0 \\
        & = \gamma^t\frac{- \rprior(\omega_0) x^{(t)} + \rprior(\omega_1)}{\rprior(\omega_0) x^{(t)} + \rprior(\omega_1)} + \sum_{t' = t+1}^{T-2} \gamma^{t'} \Big(- \rprior(\omega_0) x^{(t')} + \rprior(\omega_1) \Big) \\
        & \qquad + \gamma^{T-1} \Big(- \rprior(\omega_0) \Big(x^{(T-1)} - \tfrac{\gamma^T}{\gamma^{T-1}} \big( \tfrac{\rprior(\omega_1)}{\rprior(\omega_0)} - x^{(T)} \big) \Big) + \rprior(\omega_1)  \Big) \\
        & = \gamma^t\frac{- \rprior(\omega_0) x^{(t)} + \rprior(\omega_1)}{\rprior(\omega_0) x^{(t)} + \rprior(\omega_1)} + \sum_{t' = t+1}^{T-2} \gamma^{t'} \Big(- \rprior(\omega_0) x^{(t')} + \rprior(\omega_1) \Big) \\
        & \qquad + \gamma^{T-1} \Big(- \rprior(\omega_0) x^{(T-1)} + \rprior(\omega_1)  \Big) + \gamma^T \Big( -\rprior(\omega_0) x^{(T)} + \rprior(\omega_1) \Big) \\
        & \ge 0
    \end{align*}
    where the ``$\ge 0$'' is because $x$ satisfies \eqref{eq:instance-global-U-constraint-1}. 
    \item For $t = T-1$, we have 
    \begin{align*}
        & \gamma^t\frac{- \rprior(\omega_0) \tilde x^{(t)} + \rprior(\omega_1)}{\rprior(\omega_0) \tilde x^{(t)} + \rprior(\omega_1)} + \sum_{t' = t+1}^{T} \gamma^{t'} \Big(- \rprior(\omega_0) \tilde x^{(t')} + \rprior(\omega_1) \Big) \\
        & = \gamma^{T-1}\frac{- \rprior(\omega_0) \tilde x^{(T-1)} + \rprior(\omega_1)}{\rprior(\omega_0) \tilde x^{(T-1)} + \rprior(\omega_1)} + \gamma^T \cdot 0 \\
        & = \gamma^{T-1}\frac{- \rprior(\omega_0) \Big(x^{(T-1)} - \tfrac{\gamma^T}{\gamma^{T-1}} \big( \tfrac{\rprior(\omega_1)}{\rprior(\omega_0)} - x^{(T)} \big)\Big) + \rprior(\omega_1)}{\rprior(\omega_0) \tilde x^{(T-1)} + \rprior(\omega_1)} \\
        & = \frac{ \gamma^{T-1} \Big( - \rprior(\omega_0) x^{(T-1)} + \rprior(\omega_1) \Big)  + \gamma^T \Big(- \rprior(\omega_0) x^{(T)} + \rprior(\omega_1) \Big)}{\rprior(\omega_0) \tilde x^{(T-1)} + \rprior(\omega_1)} \\
        & \ge 0
    \end{align*}
    where the ``$\ge 0$'' is because $x$ satisfies \eqref{eq:instance-global-U-constraint-2}.
    \item For $t = T$, we have 
    \begin{align*}
        & \gamma^t\frac{- \rprior(\omega_0) \tilde x^{(T)} + \rprior(\omega_1)}{\rprior(\omega_0) \tilde x^{(T)} + \rprior(\omega_1)} = 0 \ge 0. 
    \end{align*}
\end{itemize}
Thus, $\tilde x$ satisfies the first constraint \eqref{eq:instance-global-U-constraint-1}. One can similarly verify that $\tilde x$ satisfies the second constraint \eqref{eq:instance-global-U-constraint-2}. We then show that $\tilde x$ has a weakly larger objective value than $x$: 
\begin{align*}
    \sum_{t=1}^T \tilde x^{(t)} & ~ = ~ \sum_{t=1}^{T-2} x^{(t)} + x^{(T-1)} - \tfrac{\gamma^T}{\gamma^{T-1}} \big( \tfrac{\rprior(\omega_1)}{\rprior(\omega_0)} - x^{(T)} \big) + \tfrac{\rprior(\omega_1)}{\rprior(\omega_0)} \\
    & ~ \ge ~ \sum_{t=1}^{T-2} x^{(t)} + x^{(T-1)} - \big( \tfrac{\rprior(\omega_1)}{\rprior(\omega_0)} - x^{(T)} \big) + \tfrac{\rprior(\omega_1)}{\rprior(\omega_0)} \\
    & ~ = ~ \sum_{t=1}^T x^{(t)}. 
\end{align*}

Since $\tilde x$ is feasible and weakly better than $x$, we can safely set $x^{(T)} = \frac{\rprior(\omega_1)}{\rprior(\omega_0)}$ and reduce the optimization problem to the $T-1$ period problem: 
\begin{align*} 
    \max_{0\le x^{(t)} \le 1} & \sum_{t=1}^{T-1} x^{(t)} \\
    \text{s.t.} ~~ 
    & \gamma^t\frac{- \rprior(\omega_0) x^{(t)} + \rprior(\omega_1)}{\rprior(\omega_0) x^{(t)} + \rprior(\omega_1)} + \sum_{t' = t+1}^{T-1} \gamma^{t'} \Big(- \rprior(\omega_0) x^{(t')} + \rprior(\omega_1) \Big) \ge 0,  \\
    & \text{and } \sum_{t' = t+1}^{T-1} \gamma^{t'} \Big(- \rprior(\omega_0) x^{(t')} + \rprior(\omega_1) \Big) \ge 0.
\end{align*}
By induction, the $T-1$ period problem has solution $\big(x^{(t)} = \frac{\rprior(\omega_1)}{\rprior(\omega_0)}\big)_{t=1}^{T-1}$. Thus, the $T$ period problem has solution $\big(x^{(t)} = \frac{\rprior(\omega_1)}{\rprior(\omega_0)}\big)_{t=1}^T$. 
\end{proof}

By Claim \ref{claim:instance-optimization-solution}, we have $x^{(t)} = \tfrac{\rprior(\omega_1)}{\rprior(\omega_0)}$ and thus 
\begin{align*}
     \mathcal U^{**}_T(\mathcal{I}) = \sum_{t=1}^T \Big( (1-p_0) \tfrac{\rprior(\omega_1)}{\rprior(\omega_0)} + p_0 \Big) = T \cdot \Big( (1-p_0) \tilde \mu_R + p_0 \Big) \stackrel{\text{by Claim \ref{claim:instance-BP-optimal}}}{=} T \cdot U^*_{\BP}(\mathcal I) = \mathcal U^*_T(\mathcal I)
\end{align*}
which proves Lemma \ref{lem:two-benchmark-equal}. 

% (1) Consider the average signaling scheme:
% \begin{align}
%     \overline \pi = \frac{1}{T} \sum_{t=1}^T \pi^{(t)}.  
% \end{align}
% Under $\overline \pi$, the receiver's obedient utility is
% \begin{align}
%     V(\overline \pi, \rho_\ob; \rprior) = \frac{1}{T} \sum_{t=1}^T V(\pi^{(t)}, \rho_\ob; \rprior) ??????? \ge V_\uninformed(\rprior).
% \end{align}

% (2) Consider the discounted weighted average of the signaling schemes
% \begin{align}
%     \overline \pi_\gamma = \frac{1}{T_\gamma} \sum_{t=1}^T \gamma^t \pi^{(t)}. 
% \end{align}
% Under $\overline \pi_\gamma$, the receiver's obedient utility is
% \begin{align}
%     V(\overline \pi_\gamma, \rho_\ob; \rprior) = \frac{1}{T_\gamma} \sum_{t=1}^T \gamma^t V(\pi^{(t)}, \rho_\ob; \rprior) = \frac{1}{T_\gamma} \mathcal V_T(\sigma, \phi_\ob; \rprior) \ge \frac{1}{T_\gamma} \sum_{t=1}^T \gamma^t V_\uninformed(\rprior) = V_\uninformed(\rprior). 
% \end{align}
% where the ``$\ge$'' is due to the dynamic persuasiveness of $\sigma$. 
% So, $\overline \pi_\gamma$ is a feasible solution to the optimization problem for $U^*_{\ge \uninformed}(\rprior)$. 
% The designer's utility satisfies
% \begin{align}
%     U^{**}_T(\mathcal I) ~ = ~ \sum_{t=1}^T U(\pi^{(t)}, \rho_\ob; \dprior) ????????? T\cdot U(\overline \pi_\gamma, \rho_\ob; \dprior) ~ \le ~ T\cdot U^*_{\ge \uninformed}(\rprior).  
% \end{align}

\subsection{Proof of Claim \ref{claim:icir}}
\label{proof:icir}

Under menu $\mathcal M$, the receiver's expected utility with prior $\rprior$ and a report $\mu$ in the single-period is
\begin{align*}
    V^{\sp}(\mathcal M, \rprior, \mu) & = \sum_{\omega \in \Omega} \rprior(\omega) \sum_{a\in A} \sigma_\mu(a|\omega) v(a, \omega) \\
    & = \E_{(\pi^{(t)}, \rho^{(t)}) \sim \Alg, \phi^*(\Alg, \mu)} \bigg[\frac{1}{T_\gamma}\sum_{t=1}^T \gamma^t \sum_{\omega \in \Omega} \rprior(\omega)  \sum_{s\in S}\pi^{(t)}(s|\omega) \sum_{a\in A}\rho^{(t)}(a|s) v(a, \omega) \bigg ] \\
    & \le \E_{(\pi^{(t)}, \rho^{(t)}) \sim \Alg, \phi^*(\Alg, \rprior)} \bigg[\frac{1}{T_\gamma}\sum_{t=1}^T \gamma^t \sum_{\omega \in \Omega} \rprior(\omega)  \sum_{s\in S}\pi^{(t)}(s|\omega) \sum_{a\in A}\rho^{(t)}(a|s) v(a, \omega) \bigg ] \\
    & = V^{\sp}(\mathcal M, \rprior, \rprior),
\end{align*}
where the $\le$ is because $\phi^*(\Alg, \rprior)$ is a best response to $\Alg$ for a strategic receiver with prior $\rprior$ in the $T$-period game. So, $\mathcal M$ is incentive compatible. 
By a similar argument, one can verify that $\mathcal M$ is individually rational. So the claim is proven.

%Then we want to show how the designer's single-period regret is linked to her T-period regret so as to compute the latter.
\subsection{Proof of Lemma \ref{lem:T-regret-single-period-regret}}
\label{proof:T-regret-single-period-regret}

% Then, we relate the designer's regret in the single-period game to her regret in the $T$-period game: 
We will prove $\Reg(\Alg; \dprior, \rprior) ~ \ge ~ T_\gamma \Reg^{\sp}(\mathcal M, \rprior)$. Consider: 
\begin{align*}
    & T_\gamma \Reg^{\sp}(\mathcal M, \rprior) \\
    & = T_\gamma \Big [ U^*_\sp(\rprior) -  U^{\sp}(\mathcal M, \rprior) \Big] \\
    & = T_\gamma \bigg [ U^*_\sp(\rprior) - \sum_{\omega \in \Omega} \dprior(\omega) \sum_{a\in A} \sigma_{\rprior}(a|\omega) u(a, \omega) \bigg] \\
    & = T_\gamma U^*_\sp(\rprior) - \E_{(\pi^{(t)}, \rho^{(t)}) \sim \Alg, \phi^*(\Alg, \rprior)} \bigg[ \sum_{t=1}^T \gamma^t \sum_{\omega \in \Omega} \dprior(\omega) \sum_{s\in S}\pi^{(t)}(s|\omega) \sum_{a\in A} \rho^{(t)}(a|s) u(a, \omega) \bigg] \\ 
    & = T U^*_\sp(\rprior) - (T-T_\gamma) U^*_\sp(\rprior) \\
    & \qquad - \E_{(\pi^{(t)}, \rho^{(t)}) \sim \Alg, \phi^*(\Alg, \rprior)} \bigg[ \sum_{t=1}^T (1 - 1 + \gamma^t) \sum_{\omega \in \Omega} \dprior(\omega) \sum_{s\in S}\pi^{(t)}(s|\omega) \sum_{a\in A} \rho^{(t)}(a|s) u(a, \omega) \bigg] \\ 
    & = T U^*_\sp(\rprior) - \E_{(\pi^{(t)}, \rho^{(t)}) \sim \Alg, \phi^*(\Alg, \rprior)} \bigg[ \sum_{t=1}^T \sum_{\omega \in \Omega} \dprior(\omega) \sum_{s\in S}\pi^{(t)}(s|\omega) \sum_{a\in A} \rho^{(t)}(a|s) u(a, \omega) \bigg]\\
    & \qquad - (T-T_\gamma) U^*_\sp(\rprior) + \E_{(\pi^{(t)}, \rho^{(t)}) \sim \Alg, \phi^*(\Alg, \rprior)} \bigg[ \sum_{t=1}^T ( 1 - \gamma^t)\sum_{\omega \in \Omega} \dprior(\omega) \sum_{s\in S}\pi^{(t)}(s|\omega) \sum_{a\in A} \rho^{(t)}(a|s) u(a, \omega) \bigg] \\ 
    & = \Reg(\Alg; \dprior, \rprior) \\
    & \qquad - \underbrace{\E_{(\pi^{(t)}, \rho^{(t)}) \sim \Alg, \phi^*(\Alg, \rprior)} \Big[ \sum_{t=1}^T ( 1 - \gamma^t) \Big(U^*_\sp(\rprior) - \sum_{\omega \in \Omega} \dprior(\omega) \sum_{s\in S}\pi^{(t)}(s|\omega) \sum_{a\in A} \rho^{(t)}(a|s) u(a, \omega) \Big) \bigg]}_{A}.  
\end{align*}
Then, we want to show $A \ge 0$. According to the definition of utility \eqref{eq:single-period-optimal-utility} \eqref{eq:sender-utility} in the two-state, two-action single-period game defined previously, 
\begin{align*}
    & U^*_\sp(\rprior) - \sum_{\omega \in \Omega} \dprior(\omega) \sum_{s\in S}\pi^{(t)}(s|\omega) \sum_{a\in A} \rho^{(t)}(a|s) u(a, \omega) \\
    & = (1-p_0) \tilde \mu_R + p_0 - (1-p_0) \sum_{s\in S} \pi^{(t)}(s|\omega_0) \rho^{(t)}(a_1|s) - p_0 \sum_{s\in S} \pi^{(t)}(s|\omega_1) \rho^{(t)}(a_1|s) \\
    & = (1-p_0) \Big( \tilde \mu_R  - \sum_{s\in S} \pi^{(t)}(s|\omega_0) \rho^{(t)}(a_1|s) \Big) + p_0 \Big( 1 - \sum_{s\in S} \pi^{(t)}(s|\omega_1) \rho^{(t)}(a_1|s) \Big) \\
    & \ge (1-p_0) (1+\tilde \mu_R) V^{\sp}(\pi^{(t)}, \rho^{(t)}) + 0,
    % & \qquad - p_0 \tilde \mu_R + p_0  + p_0 \sum_{s\in S} \pi^{(t)}(s|\omega_0) \rho^{(t)}(a_1|s) - p_0 \sum_{s\in S} \pi^{(t)}(s|\omega_1) \rho^{(t)}(a_1|s) \\
    % & \ge (1+\tilde \mu_R) V^{\sp}(\pi^{(t)}, \rho^{(t)}) - 2p_0. 
\end{align*}
where $V^{\sp}(\pi^{(t)}, \rho^{(t)}) = \frac{1}{1 + \tilde \mu_R} \Big(\tilde \mu_R \sum_{s\in S} \pi^{(t)}(s|\omega_1) \rho^{(t)}(a_1|s) - \sum_{s\in S} \pi^{(t)}(s|\omega_0) \rho^{(t)}(a_1|s)\Big)$ is the receiver's single-period utility under signaling scheme $\pi^{(t)}$ and responding strategy $\rho^{(t)}$. 
The following lemma shows that $\E[ \sum_{t=1}^T (1-\gamma^t) V^{\sp}(\pi^{(t)}, \rho^{(t)})] \ge 0$. Note that this lemma is not immediate because the receiver's utility can be negative in some periods. See Appendix \ref{proof:non-decreasing-discount} for the proof of this lemma.  

\begin{lemma}\label{lemma:non-decreasing-discount}
Assume that the receiver's discount factor sequence satisfies $1 \ge \gamma^t \ge 0$ and $\gamma^t \ge \gamma^{t+1}, \forall t$.  
% over time $\{\gamma^t\}$ for $t=1,...,T$ is positive and non-decreasing.
For any learning algorithm $\Alg$ of the information designer, with the receiver best responding by $\phi^*(\Alg, \rprior)$, we have $\E[ \sum_{t=1}^T (1-\gamma^t) V^{\sp}(\pi^{(t)}, \rho^{(t)})] \ge 0$. 
\end{lemma}
% 
% Denote by $\mathcal{R}^*(\Alg, \rprior)$ the {\color{red}{optimal} algorithm} that the receiver, who maximizes his total $T-$period utility with prior $\rprior$, uses to interact the designer who uses a learning algorithm $\Alg$.
% 
% \begin{lemma}\label{lemma:non-decreasing-discount}
%     Assume that the receiver's discount sequence over time $\{\gamma^t\}$ for $t=1,...,T$ is positive and non-decreasing. For any learning algorithm $\Alg$ of the information designer, the receiver's belief $\mu_R$, and the receiver's optimal algorithm $\mathcal{R}^*(\Alg, \rprior)$, $\E _{\{(\pi_{\rprior}^{(t)}, \phi^{(t)})\}_{t=1}^T \sim \mathcal{R}^*(\Alg, \rprior)} 
%  \sum_{t=1}^T \gamma^t \tfrac{1}{1+\tilde \mu_R}\Big( \tilde \mu_R \pi_\rprior(a_1|\omega_1) - \pi_\rprior(a_1|\omega_0) \Big) \leq \E _{\{(\pi_{\rprior}^{(t)}, \phi^{(t)})\}_{t=1}^T \sim \mathcal{R}^*(\Alg, \rprior)} 
%  \sum_{t=1}^T \tfrac{1}{1+\tilde \mu_R}\Big( \tilde \mu_R \pi_\rprior(a_1|\omega_1) - \pi_\rprior(a_1|\omega_0) \Big) $.
% \end{lemma}
So, we obtain 
\begin{equation*}
    A ~ \ge ~ (1-p_0)(1+\tilde \mu_R) \E\bigg[ \sum_{t=1}^T (1-\gamma^t) V^{\sp}(\pi^{(t)}, \rho^{(t)})\bigg] ~ \ge ~ 0,  
\end{equation*}
which implies
\begin{equation*}
    \Reg(\Alg; \dprior, \rprior) ~ \ge ~ T_\gamma \Reg^{\sp}(\mathcal M, \rprior). 
\end{equation*}
% By Lemma \ref{lemma:single-period-regret}, there exists an instance where $\Reg^{\sp}(\mathcal M, \rprior) \ge \frac{1}{16}$, so $\Reg(\Alg; \dprior, \rprior) \ge \frac{1}{16}T_\gamma$.
% \qed

\subsection{Proof of Lemma \ref{lemma:non-decreasing-discount}}
\label{proof:non-decreasing-discount}
In this proof, the expectation is over the stochastic process $\{ (\pi^{(t)}, \rho^{(t)}) \}_{t=1}^T$ generated by $\Alg$ and $\phi^*(\Alg, \rprior)$. 
Let
\begin{equation*}
    \E[V^{\sp}(\pi^{(t)}, \rho^{(t)})] = \E\bigg[ \frac{1}{1 + \tilde \mu_R} \Big(\tilde \mu_R \sum_{s\in S} \pi^{(t)}(s|\omega_1) \rho^{(t)}(a_1|\omega_1) - \sum_{s\in S} \pi^{(t)}(s|\omega_0) \rho^{(t)}(a_1|\omega_0)\Big) \bigg]
\end{equation*}
be the receiver's expected utility at period $t$, where the signaling scheme is $\pi^{(t)}$ and the receiver responds by strategy $\rho^{(t)}$. 
Denote the receiver's discounted and undiscounted total utility from period $t_1$ to $t_2$ by
\begin{equation*}
    DV(t_1, t_2) = \sum_{t=t_1}^{t_2} \gamma^t \E[V^{\sp}(\pi^{(t)}, \rho^{(t)})], \qquad \quad UV(t_1, t_2) = \sum_{t=t_1}^{t_2} \E[V^{\sp}(\pi^{(t)}, \rho^{(t)})]. 
\end{equation*}
% We want to prove $UV(1, T) \ge DV(1, T)$.

We will use an induction to prove $0 \le DV(t, T) \le \gamma^t UV(t, T)$ for every $t \in \{1, \ldots, T, T+1\}$. 
The base case where $t = T+1$ trivially holds. 
Consider any $t \le T$.  First, we have $DV(t, T) \ge \sum_{t'=1}^T \gamma^{t'} V_{\mathrm{uninformed}} \ge 0$ because the receiver is best-responding using $\phi^*(\Alg, \rprior)$ and he can always guarantee the uninformed utility by ignoring the signals. 
%of $\phi^*(\Alg, \rprior)$ is optimal for the receiver.
We then note that
\begin{align*}
    DV(t, T) & ~ = ~ DV(t, t) + DV(t+1, T) \\
    & ~ = ~ \gamma^t UV(t, t) + DV(t+1, T) \\
    & ~ \le ~ \gamma^t UV(t, t) + \gamma^{t+1} UV(t+1, T) \quad \text{by induction}. 
\end{align*}
Because $\gamma^t \ge \gamma^{t+1}$ and $UV(t+1, T) \ge \gamma^{t+1} DV(t+1, T) \ge 0$ by induction, we have 
\begin{align*}
    DV(t, T) ~ \le ~ \gamma^t UV(t, t) + \gamma^{t} UV(t+1, T)  ~ = ~ \gamma^t UV(t, T), 
\end{align*}
which proves the induction. 

For $t = 1$, $\gamma^t \le 1$, so $DV(1, T) \le \gamma^1 UV(1, T) \le UV(1, T)$, which implies
\begin{equation*}
    0 ~ \le ~ UV(1, T) - DV(1, T) ~ = ~ \E\Big[ \sum_{t=1}^T (1-\gamma^t) V^{\sp}(\pi^{(t)}, \rho^{(t)}) \Big]  
\end{equation*}
and proves the lemma.

\end{document}

%% file: notations.tex
\DeclareMathOperator*{\argmax}{arg\,max}
\DeclareMathOperator*{\argmin}{arg\,min}

\newtheorem{claim}{Claim}
\newtheorem{definition}{Definition}
\newtheorem{theorem}{Theorem}
\newtheorem{proposition}{Proposition}

\newtheorem{lemma}{Lemma}
\newtheorem{remark}{Remark}
\newtheorem{regularity assumption}{Regularity assumption}
\newtheorem{assumption}{Assumption}

\newtheorem{corollary}{Corollary}

\newcommand{\diss}{\iota}

\newcommand{\eps}{\varepsilon}

\newcommand{\E}{\mathbb{E}}
\newcommand{\Reg}{\mathrm{Reg}}

\newcommand{\rprior}{{\mu_R}}
\newcommand{\dprior}{{\mu_D}}

\newcommand{\Alg}{\mathcal{G}}

\newcommand{\uninformed}{\mathrm{uninformed}}

\newcommand{\ob}{\mathrm{ob}}

\newcommand{\full}{\mathrm{full}}
\newcommand{\terminate}{\tau}
\newcommand{\BP}{\mathrm{BP}}

\renewcommand{\sp}{\mathrm{sp}}

%% file: plots/designer-payoff-T=20.tex
\begin{tikzpicture}
\begin{axis}[
    width=12cm,
    height=9cm,
    title={Designer's payoff},
    xlabel={Period $t$},
    xmin=1, xmax=20,
    ymin=0.45, ymax=1.05,
    xtick={1,2,...,20},
    ytick={0.5,0.6,0.7,0.8,0.9,1.0},
    grid=both,
    grid style={gray!20},
    legend style={
        font=\small, 
        at={(0.57,0.012)},
        anchor=south west,
        draw=black,
        fill=white
    },
]

\addplot[
    red,
    thick,
    mark=*,
    mark size=2pt
] coordinates {
    (1,0.99649681)
    (2,0.70682326)
    (3,0.65813702)
    (4,0.55449705)
    (5,0.97215591)
    (6,0.65813702)
    (7,0.54798657)
    (8,0.99649681)
    (9,0.70682326)
    (10,0.65813702)
    (11,0.55449705)
    (12,0.97215591)
    (13,0.65813702)
    (14,0.55449705)
    (15,0.97215591)
    (16,0.65813702)
    (17,0.55449705)
    (18,0.97215591)
    (19,0.65813702)
    (20,0.54798657)
};
\addlegendentry{Dynamic benchmark}

\addplot[
    black,
    dashed,
    thick,
    domain=1:20,
    samples=2
] {0.7};
\addlegendentry{Static benchmark}

\end{axis}
\end{tikzpicture}

%% file: plots/receiver-payoff-T=20.tex
\begin{tikzpicture}
\begin{axis}[
    width=12cm,
    height=9cm,
    title={Receiver's payoff},
    xlabel={Period $t$},
    xmin=1, xmax=20,
    ymin=0.15, ymax=1.05,
    xtick={1,2,...,20},
    ytick={0.2,0.3,0.4,0.5,0.6,0.7,0.8,0.9,1.0},
    grid=both,
    grid style={gray!20},
    legend style={
        font=\small, 
        at={(0.388,0.937)},
        anchor=north west,
        draw=black,
        fill=white
    },
]

\addplot[
    blue,
    thick,
    mark=*,
    mark size=2pt
] coordinates {
    (1,0.20350318)
    (2,0.64317673)
    (3,0.69127517)
    (4,0.79418345)
    (5,0.22930649)
    (6,0.69127517)
    (7,0.80201342)
    (8,0.20350318)
    (9,0.64317673)
    (10,0.69127517)
    (11,0.79418345)
    (12,0.22930649)
    (13,0.69127517)
    (14,0.79418345)
    (15,0.22930649)
    (16,0.69127517)
    (17,0.79418345)
    (18,0.22930649)
    (19,0.69127517)
    (20,0.80201342)
};
\addlegendentry{Dynamic benchmark}

\addplot[
    black,
    dashed,
    thick,
    domain=1:20,
    samples=2
] {0.5};
\addlegendentry{Static benchmark (= $V_\uninformed$)}

\addplot[
    orange,
    thick,
    mark=asterisk,
    mark size=3pt, 
    domain=1:20,
    samples=2
] {1.0};
\addlegendentry{Full information payoff}

\end{axis}
\end{tikzpicture}